\documentclass[aps,prb,twocolumn,reprint,superscriptaddress,citeautoscript]{revtex4-2}

\usepackage{physics}
\usepackage{graphicx}
\usepackage{amsmath,amssymb}
\usepackage[normalem]{ulem}
\usepackage{bm}
\usepackage{upgreek}

\usepackage{xcolor}

\usepackage{hyperref}
\hypersetup{
	citecolor = blue,
	colorlinks = true,
	urlcolor = blue
}

\begin{document}
	
\title{Theory of the inverse Rashba-Edelstein effect induced by thermal spin injection}
	
\author{Kaiji Hosokawa}
\affiliation{Institute for Solid State Physics, University of Tokyo, Kashiwa 277-8581, Japan}
\author{Masaki Yama}
\affiliation{Institute for Solid State Physics, University of Tokyo, Kashiwa 277-8581, Japan}
\author{Mamoru Matsuo}
\affiliation{Kavli Institute for Theoretical Sciences, University of Chinese Academy of Sciences, Beijing, 100190, China.}
\affiliation{CAS Center for Excellence in Topological Quantum Computation, University of Chinese Academy of Sciences, Beijing 100190, China}
\affiliation{Advanced Science Research Center, Japan Atomic Energy Agency, Tokai, 319-1195, Japan}
\affiliation{RIKEN Center for Emergent Matter Science (CEMS), Wako, Saitama 351-0198, Japan}
\author{Takeo Kato}
\email[]{kato@issp.u-tokyo.ac.jp}
\affiliation{Institute for Solid State Physics, University of Tokyo, Kashiwa 277-8581, Japan}
	
\begin{abstract}
We theoretically consider a junction composed of a ferromagnetic insulator (FI) and a two-dimensional electron gas (2DEG) with Rashba- and Dresselhaus-type spin-orbit interactions. Using the Boltzmann equation, we calculate an electric current in 2DEG induced by the inverse Rashba-Edelstein effect when imposing the temperature difference between FI and 2DEG. We clarify how the induced current depends on the magnetization direction of FI, spin texture on the Fermi surface of 2DEG, and temperature. Our result provides an important foundation for an accurate analysis of the inverse Rashba-Edelstein effect induced by thermal spin injection.
\end{abstract}
	
\maketitle

\section{Introduction} 

Spin-charge conversion in two-dimensional electron gas (2DEG) is one of the key ingredients in modern spintronics technology.
In a system without spatial inversion symmetry, the spin polarization is generated by applying a charge current.
This phenomenon is called the Rashba-Edelstein effect (REE)~\cite{Aronov1989, Aronov1991, Edelstein1990, Inoue2003, Silsbee2004, Sinova2015, Soumyanarayanan2016,Bychkov1984, Rashba2015, Winkler2003, Manchon2015} or the inverse spin-galvanic effect~\cite{Gambardella2011, Manchon2019}.
In contrast, its inverse effect, that is, the generation of charge currents from spin polarization, is called the inverse Rashba-Edelstein effect (IREE)~\cite{Sanchez2013, Shen2014a, Soumyanarayanan2016} or the spin-galvanic effect~\cite{Ganichev2002, Ivchenko1989, Ivchenko1990, Ganichev2001, Ganichev2003b, Burkov2004, Sinova2015}. 
These spin-charge conversion phenomena in 2DEG are now becoming important in the field of semiconductor spintronics~\cite{Fabian2007, Awschalom2007, Manchon2015, Kohda2017, Dieny2020, VicenteArche21, Gupta22}.

In the last decade, spintronic devices that combine REE or IREE with standard methods of spintronics have been under intense investigation.
For example, spin pumping~\cite{Tserkovnyak2002, Tserkovnyak2005, Hellman2017} caused by ferromagnetic resonance (FMR) has been used to generate electron spins from an ferromagnet into an adjacent system.
This technique has been combined with IREE for spin-charge conversion in various materials~\cite{Sanchez2013, Nomura2015, Sangiao2015, Zhang2015, Matsushima2017,Lesne2016, Song2017, Vaz2019, Noel2020, Ohya2020, Bruneel2020, To2021, Trier2022,Shiomi2014, Sanchez2016, Wang2016, Song2016, Mendes2017, Sun2019, Singh2020, Dey2021, He2021, Zhang2016, Mendes2015, Dushenko2016, Mendes2019, Bangar2022, Chen2016, Oyarzun2016}. 

Recently, spin-charge conversion using IREE and thermal-gradient-induced magnon spin current has been demonstrated in 2DEG at the EuO--KTaO${}_3$ heterostructure~\cite{Zhang19}.
In this experiment, a large spin Seebeck coefficient was observed in comparison to the standard setup, i.e., the Pt/YIG heterostructure with the same thickness for a magnetic layer, suggesting a potential application to spintronics devices.
The current generation in this system is governed by a spin-momentum locking due to the strong Rashba spin-orbit interaction originating from the $5d$ atomic orbit at the tantalum atom~\cite{VicenteArche21, Gupta22}.

A similar spin-charge conversion due to thermal spin injection is expected to occur in semiconductor junctions such as the GaAs--Fe interface, in which two types of spin-orbit interactions, namely, Rashba and Dresselhaus spin-orbit interactions, coexist~\cite{Dresselhaus1955, Rocca1988, Winkler2003}.
Several studies on REE and IREE for 2DEG have been performed in semiconductor heterostructures~\cite{Ganichev2003a, Ganichev2003b, Ganichev2004, Giglberger2007, Belkov2008, Ganichev2008, Ivchenko2008, Ganichev2014, Sheikhabadi2017, Tao2021, Zhuravlev2022,Shytov2006, Raimondi2006, Trushin2007, Raichev2007, Engel2007, Gorini2010, Raimondi2012, Xintao2013, Shen2014b, Szolnoki2017, Gorini2017, Sheikhabadi2018, Tao2021, Tkach2021, Tkach2022, Suzuki2023, Yama2023a} and IREE combined with spin pumping has begun to be theoretically studied recently~\cite{Tolle2017,Dey2018,Fleury2023,Yama2023}.
However, a microscopic theory of IREE for thermal spin injection has not been provided so far.

\begin{figure}[tb]
    \centering
    \includegraphics[width=90mm]{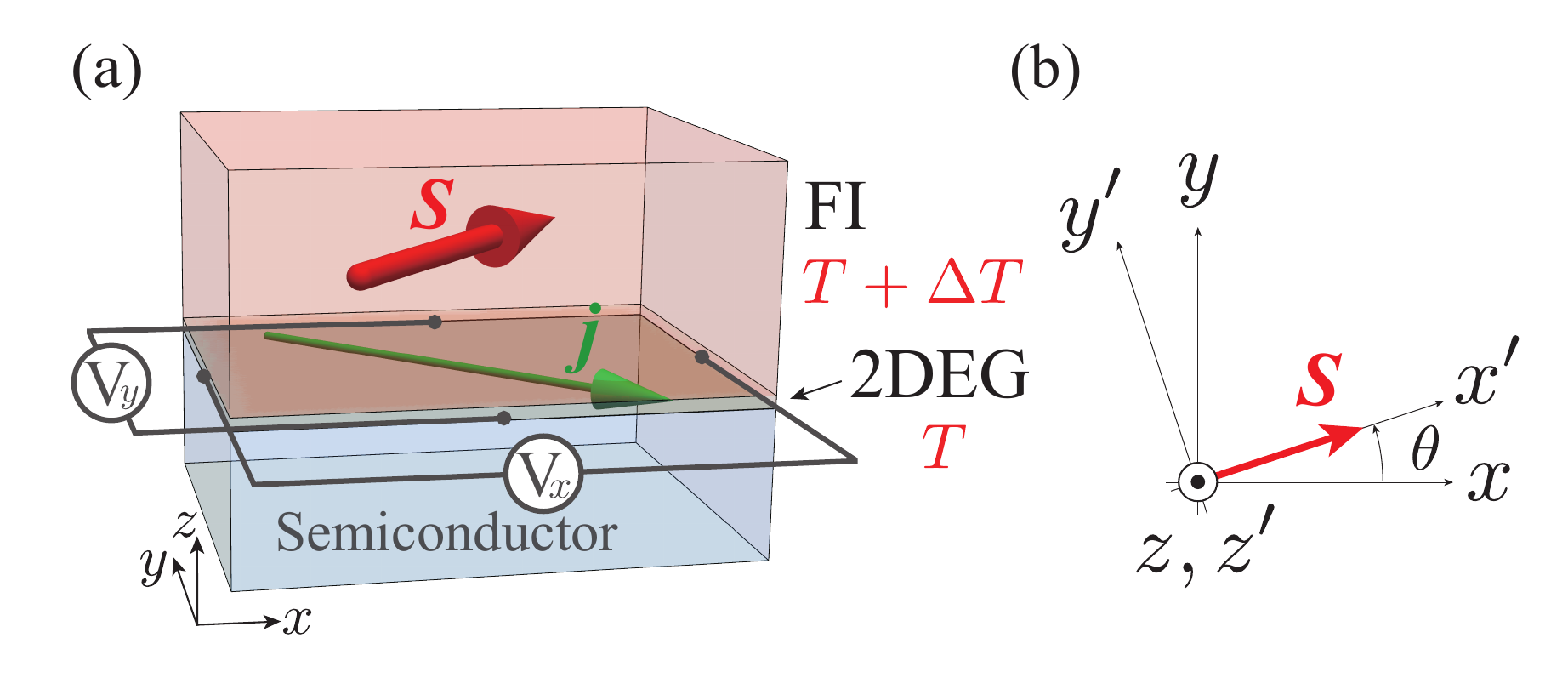}
    \caption{(a) A magnetic junction composed of a ferromagnetic insulator (FI) and two-dimensional electron gas (2DEG). (b) Coordinate transformation between the laboratory frame $(x,y,z)$ and the frame $(x',y',z')$ fixed to the spin orientation in the FI. We define $\theta$ as an angle between the spin orientation and the $x$ axis.}
    \label{fig:setup}
\end{figure}

In this paper, we theoretically discuss IREE induced by thermal spin injection.
We consider a magnetic junction composed of 2DEG and a ferromagnetic insulator (FI) as shown in Fig.~\ref{fig:setup}(a).
In our work, we set the temperatures of 2DEG and FI as $T$ and $T+\Delta T$ and explain IREE in the case of $\Delta T>0$ although our theory can also describe the opposite case.
We introduce the $x$-$y$ coordinate in the plane of 2DEG and denote the azimuth angle $\theta$ of the spin in the FI [see Fig.~\ref{fig:setup}(b)].
To clarify the effect of the spin-momentum locking, we consider two types of spin-orbit interactions, i.e. the Rashba and Dresselhaus spin-orbit interactions. 
For simplicity, we consider the case that the strength of these spin-orbit interactions is much larger than the temperature and energy broadening of electron scattering and much smaller than the Fermi energy.
We note that the effect of spin-momentum locking is most effective in this condition.
For describing non-equilibrium steady states of 2DEG, we derive the Boltzmann equation in which interfacial electron scattering at the interface evolving spin flipping is taken into account in terms of the collision term.
For this magnetic junction, we clarify how the induced current depends on the azimuth angle $\theta$ of the spin in the FI, the temperature, and the ratio of the Rashba and Dresselhaus interactions.
This study provides an important foundation for an accurate analysis of IREE induced by thermal spin injection into 2DEG.
 
\section{Model} 
\label{sec:model}

The Hamiltonian for a 2D electron system coupled with a ferromagnetic insulator is given by the following expression:
\begin{align}
H &= H_{\rm kin} + H_{\rm imp} + H_{\rm FI} + H_{\rm int},
\end{align}
where $H_{\rm kin}$ represents the kinetic energy and spin-orbit interaction of 2DEG, $H_{\rm imp}$ accounts for impurities in 2DEG, $H_{\rm FI}$ describes the FI, and $H_{\rm int}$ characterizes the interface between 2DEG and FI. 
Detailed explanations for each term are provided in the subsequent sections.
In the following, we use the laboratory coordinate $(x,y,z)$ so that the $xy$ plane is parallel to the 2DEG (see Fig.~\ref{fig:setup}).

\subsection{Two-dimensional electron gas (2DEG)}

The kinetic term $H_{\rm kin}$ of 2DEG is expressed as follows:
\begin{align}
& H_{\rm kin}
=\sum_{\bm k} \sum_{\sigma \sigma'} c_{{\bm k} \sigma}^{\dagger} (\hat{h}_{\bm k})_{\sigma\sigma'}
c_{{\bm k} \sigma'}, \\
& \hat{h}_{\bm k} =\xi_{\bm k} \hat{I}
+\alpha\left(k_y \sigma_x-k_x \sigma_y \right)
+\beta\left(k_x \sigma_x-k_y \sigma_y \right), \label{eq:hh}
\end{align}
where ${\bm k}=(k_x,k_y)$ is the two-dimensional wavenumber of the electrons, $\sigma, \sigma'$ ($=\uparrow,\downarrow$) indicates an electron spin, $\xi_{\bm k} = \hbar^2{\bm k}^2/2 m^*-\mu$ is the kinetic energy measured from the chemical potential $\mu$, $m^*$ is the effective mass, $I$ is a $2\times 2$ identity matrix, and ${\bm \sigma} = (\sigma_x,\sigma_y,\sigma_z)$ denotes Pauli matrices.
The strengths of the Rashba-type and Dresselhaus-type spin-orbit interactions are denoted by $\alpha$ and $\beta$, respectively.
The matrix $\hat{h}_{\bm k}$ is rewritten as
\begin{align}
\hat{h}_{\bm k} &=\xi_{\bm k} \hat{I} - {\bm h}_{\rm eff}({\bm k}) \cdot {\bm \sigma},
\end{align}
where ${\bm h}_{\rm eff}({\bm k})$ denotes an effective Zeeman field acting on the electrons:
\begin{align}
{\bm h}_{\rm eff}({\bm k}) &=k\left(\begin{array}{c} -\alpha\sin\varphi -\beta\cos\varphi \\
\alpha\cos\varphi +\beta\sin\varphi \\ 0 \end{array} \right) ,
\label{eq:heffkdep}
\end{align}
where we have introduced the polar representation ${\bm k}=(k\cos \varphi,k\sin \varphi)$ for the electron wavenumber.
By diagonalizaing the matrix $\hat{h}_{\bm k}$, the energy eigenvalue $E_{{\bm k}\gamma}$ and eigenvector ${\bm u}_{{\bm k}\gamma}$ are obtained as
\begin{align}
& E_{{\bm k}\gamma} = \xi_{\bm k} + \gamma |{\bm h}_{\rm eff}({\bm k})|, \\
& ({\bm u}_{{\bm k}\gamma})_\sigma = \frac{C(\varphi)}{\sqrt{2}} \delta_{\sigma,\uparrow} 
+ \frac{\gamma}{\sqrt{2}} \delta_{\sigma,\downarrow}, \label{eq:eigenvector}
\end{align}
where $\gamma = \pm 1$ assigns the spin-splitting band, 
$C(\varphi) = - \hat{h}_{{\rm eff},x}(\varphi) + i \hat{h}_{{\rm eff},y}(\varphi)$, and $\hat{\bm h}_{\rm eff}(\varphi) = {\bm h}_{\rm eff}({\bm k})/|{\bm h}_{\rm eff}({\bm k})|$ is a unit vector pointing to the effective Zeeman field.
We note that the spin of the state $\gamma = +1$ ($-1$) points in the opposite (the same) direction to $\hat{\bm h}_{\rm eff}$.

We consider impurities in 2DEG whose potential energy is described by the delta function as $v({\bm r}) = v_0 \delta({\bm r})$.
The Hamiltonian of the impurities is written as
\begin{align}
 H_{\rm imp} &= \frac{v_0}{{\cal A}}\sum_{{\bm k},{\bm q},\sigma} \rho_{\rm imp}({\bm q})c^\dagger_{{\bm k}+{\bm q}\sigma} c_{{\bm k}\sigma} , 
 \label{eq:Himp2d}
\end{align}
where ${\cal A}$ is an area of 2DEG,
$\rho_{\rm imp}({\bm q}) = \sum_i e^{-i{\bm q}\cdot {\bm R}_i}$ and ${\bm R}_i$ denotes the position of the impurity.

\subsection{Ferromagnetic insulator (FI)}

We assume that the spins in the FI
are aligned in the in-plane ($xy$) direction, as shown in Fig.~\ref{fig:setup}.
We define the azimuth angle of the spin measured from the $x$ axis as $\theta$ (see Fig.~\ref{fig:setup}(b)).
Then, the average of the localized spin in the FI is given as $\langle {\bm S}_i \rangle = (S_0\cos \theta, S_0 \sin \theta, 0)$, where $S_0$ denotes the amplitude of the localized spin.
To apply the spin-wave approximation, we introduce a new coordinate $(x',y',z')$ in which the $x'$ axis is taken in the direction of the spin in the FI.
Using this new coordinate, the Hamiltonian of the FI is given as
\begin{align}
H_{\rm FI} = \sum_{\langle i,j \rangle} J (S_i^{x'} S_j^{x'}
+ S_i^{y'} S_j^{y'}+ S_i^{z'} S_j^{z'}) -h_{\mathrm{dc}}\sum_{i}S_i^{x'},
\end{align}
where ${\bm S}_i = (S_i^{x'},S_i^{y'},S_i^{z'})$ is the spin operator of the FI, $\langle i, j \rangle$ indicates a pair of neighboring sites, $J$ is the exchange interaction, and $h_{\rm dc}$ is an external magnetic field.
Using the Holstein-Primakoff transformation, the spin operators can be expressed with annihilation and creation operators of the magnon as
\begin{align}
& S_i^{x'-} \simeq \sqrt{2S_0} b_i^\dagger,  \\
& S_i^{x'+} \simeq \sqrt{2S_0} b_i,  \\
& S_i^{x'} = S_0 - b_i^\dagger b_i. \label{eq:Sixprime}
\end{align}
The Hamiltonian is rewritten in the leading term with respect to $1/S_0$ as
\begin{align}
H_{\rm FI} &=\sum_{\bm q} \hbar \omega_{\bm q} b_{\bm q}^{\dagger} b_{\bm q},
\end{align}
where $b_{\bm q}$ is a Fourier transformation of $b_i$, $\hbar \omega_{\bm q}=h_{\mathrm{dc}}+{\cal D}{\bm q}^2$ is a magnon dispersion, and ${\cal D}$ is a spin stiffness.

\subsection{Interfacial exchange interaction}

The Hamiltonian for the interfacial exchange interaction is given as
\begin{align}
H_{\rm int} &=\sum_{{\bm q}_{\parallel},q_z}\left(T_{\bm q} S_{{\bm q}}^{x'+} s_{{\bm q}_\parallel}^{x'-}+T_{\bm q}^* S_{{\bm q}}^{x'-} s_{{\bm q}_\parallel}^{x'+}\right) ,
\label{Hintdef}
\end{align}
where ${\bm q}_{\parallel}=(q_x,q_y)$ is the in-plane component of the momentum transfer ${\bm q}=(q_x,q_y,q_z)$ and $T_{\bm q}$ represents the strength of the interfacial exchange coupling between the FI and 2DEG.
Here, we assumed conservation of the in-plane momentum, which is expected to hold for a clean interface~\footnote{We further assumed that an exchange bias due to the mean-field term, $\langle {\bm S}_i \rangle \cdot {\bm s}_j$, is sufficiently small.}, and $s_{\bm q}^{x'\pm}=s_{\pm \bm q}^{y'}\pm i s_{\pm \bm q}^{z'}$ is a Fourier transformation of spin ladder operators for electrons of 2DEG in the coordinate $(x',y',z')$:
\begin{align}
s_{\bm q}^{x'} &= \cos \theta s_{\bm q}^{x} + \sin \theta s_{\bm q}^{y}, \\
s_{\bm q}^{y'} &= -\sin \theta s_{\bm q}^{x} + \cos \theta s_{\bm q}^{y}, \\
s_{\bm q}^{z'} &= s_{\bm q}^{z}.
\end{align}
Here, $s_{\bm q}^{a}$ is the spin operators in the laboratory coordinate:
\begin{align}
s_{\bm q}^{a} = \frac12 \sum_{\sigma\sigma'}
\sum_{{\bm k}} c_{{\bm k}\sigma}^\dagger (\sigma_a)_{\sigma \sigma'}
c_{{\bm k}+{\bm q}\sigma'}, \quad (a=x,y,z).
\end{align}
Combining these equations, we obtain
\begin{align}
s_{\bm q}^{x'\pm} &= \frac12 \sum_{\sigma\sigma'}\sum_{\bm k} c_{{\bm k}\sigma}^\dagger (\hat{\sigma}^{x'\pm})_{\sigma \sigma'} c_{{\bm k}\pm{\bm q}\sigma'}, \\
\hat{\sigma}^{x'\pm} &= -\sin \theta \, \sigma_x + \cos \theta \, \sigma_y \pm i\sigma_z .
\end{align}
Therefore, the Hamiltoninan for the interface is given as
\begin{align}
H_{\rm int} 
&= \sum_{{\bm k},{\bm q},q_z} \sum_{\sigma \sigma'}\left(\frac{\sqrt{2S_0}T_{\bm q}}{2} 
b_{\bm q} c_{{\bm k}\sigma}^\dagger (\hat{\sigma}^{x'-})_{\sigma\sigma'} c_{{\bm k}-{\bm q}_{\parallel} \sigma'} + {\rm h.c.} \right).
\label{Hintdef2}
\end{align}
Here, we have omitted the static term proportional to $S_0$, which stems from the first term on the right-hand side of Eq.~(\ref{eq:Sixprime}). Generally, this omitted term corresponds to the exchange bias at the interface and can alter the spin-dependent energy dispersion of the 2DEG and its electronic states. However, such an effect can be ignored if the exchange coupling at the interface is much weaker than the spin splitting energy.

\section{Formulation}
\label{sec:formulation}

\subsection{Boltzmann equation}
\label{sec:Boltzmann}

We follow the method based on the Boltzmann equation in Ref.~\cite{Yama2023}.
We assume that the spin-orbit interactions are much larger than the temperature and energy broadening due to the impurity scattering rate, which is defined later. 
We also assume that the spin splitting energy is
much smaller than the chemical potential $\mu$.
Then, the electronic state of 2DEG is described by a distribution function $f({\bm k},\gamma)$ for a uniform steady state~\cite{Suzuki2023} and the Boltzmann equation contains only collision terms as
\begin{align}
0 = \frac{\partial f({\bm k},\gamma)}{\partial t}\biggl{|}_{\rm int}+\frac{\partial f({\bm k},\gamma)}{\partial t}\biggl{|}_{\rm imp} , \label{eq:0ecoll}
\end{align}
where the first term in r.h.s is an interfacial collision term due to spin injection from the FI and the second one is due to impurity scattering.
The 2DEG and FI temperatures are set as $T$ and $T+\Delta T$, respectively.
Within the linear response to the temperature difference $\Delta T$, we consider the modification of the distribution function in the form~\cite{Wilson1953,Ziman1960,Lundstrom2000}
\begin{align}
f(\bm{k},\gamma) \simeq f_0(E_{\bm{k}\gamma})
+ \frac{\partial f_0(E_{\bm{k}\gamma})}{\partial E_{\bm{k}\gamma}} \Phi(\varphi,\gamma),
\label{eq:fexpan}
\end{align}
where $f_0(\epsilon) =[\exp( \epsilon/k_{\rm B}T)+1]^{-1}$ is the Fermi distribution function
and $\Phi(\varphi,\gamma)$ denotes the shift of the chemical potential, which is proportional to the temperature gradient $\Delta T$.

\subsection{Impurity scattering}

The collision term due to impurity scattering is given as
\begin{align}
\left. \frac{\partial f({\bm k},\gamma)}{\partial t}\right|_{\rm imp} 
&= \Gamma_{({\bm k}'\gamma')\rightarrow({\bm k}\gamma)} f({\bm k}',\gamma') \notag \\
& - \Gamma_{({\bm k}\gamma)\rightarrow({\bm k}'\gamma')} f({\bm k},\gamma),
\end{align}
where the transition rate $\Gamma_{({\bm k}\gamma)\rightarrow({\bm k}'\gamma')}$ is given by Fermi's golden rule as
\begin{align}
&\Gamma_{({\bm k}\gamma)\rightarrow ({\bm k}'\gamma')} =
\frac{2\pi}{\hbar}| \langle {\bm k}'\gamma'|\hat{h}_{\rm imp}|{\bm k}\gamma \rangle  |^2  \delta(E_{{\bm k}'\gamma'}-E_{{\bm k}\gamma}) .
\end{align}
The matrix element $\langle {\bm k}'\gamma'|\hat{h}_{\rm imp}|{\bm k}\gamma \rangle$ is given from the Hamiltonian (\ref{eq:Himp2d}) as
\begin{align}
& \langle {\bm k}'\gamma'|h_{\rm imp}|{\bm k}\gamma \rangle  
= \frac{v_0}{\cal A} \rho_{\rm imp}({\bm k}'-{\bm k}) A_{\gamma'\gamma}, \\
& A_{\gamma'\gamma} =
\sum_{\sigma}  
({\bm u}_{{\bm k}'\gamma'})_\sigma^*
({\bm u}_{{\bm k}\gamma})_\sigma .
\end{align}
We can proceed in calculation, using the relations,
\begin{align}
\left|A_{\gamma'\gamma}\right|^2= 
\frac{1 + \gamma \gamma' \hat{\bm h}_{\rm eff}(\varphi) \cdot \hat{\bm h}_{\rm eff}(\varphi')}{2},
\end{align}
and $\langle |\rho_{\rm imp}({\bm k})|^2\rangle_{\rm imp}/{\cal A} = n_{\rm imp} $, where $n_{\rm imp}$ is an impurity density and $\langle \cdots \rangle_{\rm imp}$ indicates an average with respect to the impurity position ${\bm R}_i$.
Finally, we obtain
\begin{align}
\left. \frac{\partial f({\bm k},\gamma)}{\partial t}\right|_{\rm imp} 
&= \frac{2\pi v_0^2 n_{\rm imp}}{\hbar {\cal A}} 
\sum_{{\bm k}'\gamma'}
\frac{1+\gamma \gamma'\hat{\bm h}_{\rm eff}(\varphi) \cdot \hat{\bm h}_{\rm eff}(\varphi')}{2}
\notag \\
& \times (f({\bm k}', \gamma') -f({\bm k}, \gamma))
\delta(E_{{\bm k}'\gamma'}-E_{{\bm k}\gamma}).
\end{align}

\subsection{Interfacial scattering}

The collision term due to the interface scattering is a sum of magnon absorption ($\lambda = -$) and emission ($\lambda = +$) processes:
\begin{align}
& \frac{\partial f({\bm k},\gamma)}{\partial t}\biggl{|}_{{\rm int}} = \sum_{\lambda = \pm} [ \Gamma_{({\bm k}'\gamma')\rightarrow({\bm k}\gamma)}^\lambda f({\bm k}',\gamma') \notag \\
& \hspace{15mm} - \Gamma_{({\bm k}\gamma)\rightarrow({\bm k}'\gamma')}^\lambda f({\bm k},\gamma)].
\end{align}
Let us first consider the magnon absorption process.
The transition rate is calculated by Fermi's golden rule, after thermal average with respect to the magnon, as
\begin{align}
\Gamma^-_{({\bm k}\gamma)\rightarrow ({\bm k}'\gamma')} &=\frac{2\pi}{\hbar} |\langle {\bm k}'\gamma' |\hat{h}_{\rm int}^-|{\bm k}\gamma \rangle |^2 N_{{\bm q}} \notag \\
& \times \delta(E_{{\bm k}'\gamma'}-E_{{\bm k}\gamma}-\hbar \omega_{\bm q}) 
\delta_{{\bm q}_\parallel,{\bm k}'-{\bm k}},
\end{align}
where $N_{\bm q}$ is the Bose distribution function.
The matrix element $\langle {\bm k}'\gamma' |\hat{h}_{\rm int}^-|{\bm k}\gamma\rangle$ is determined from the first term of the Hamiltonian (\ref{Hintdef2}) as
\begin{align}
& \langle {\bm k}'\gamma'|\hat{h}_{\rm int}|{\bm k}\gamma \rangle  
= T_{\bm q} \frac{\sqrt{2 S_0}}{2} \delta_{{\bm q}_{\parallel}, {\bm k}-{\bm k}'} A_{\gamma'\gamma}^{-},\\
& A_{\gamma'\gamma}^{-} 
= \sum_{\sigma'\sigma} ({\bm u}_{{\bm k}'\gamma'})_{\sigma'}^* (\hat{\sigma}^{x'-})_{\sigma'\sigma} 
({\bm u}_{{\bm k}\gamma})_{\sigma}.
\end{align}
Using Eq.~(\ref{eq:eigenvector}), we obtain
\begin{align}
\left|A_{\gamma' \gamma}^{-}\right|^2 &= ( 1 + \gamma' \hat{h}_{\rm eff}(\varphi') \cdot \hat{\bm m})( 1 - \gamma \hat{h}_{\rm eff}(\varphi) \cdot \hat{\bm m} ) ,
\end{align}
where $\hat{\bm m}=(\cos \theta,\sin \theta,0)$ is a unit vector pointing to the direction of the ordered spin in FI.
Using this factor, the transition rate is written as
\begin{align}
& \Gamma^-_{({\bm k}\gamma)\rightarrow ({\bm k}'\gamma')} \notag \\
&= \frac{\pi S_0 |T_{\bm q}|^2}{\hbar} \left|A_{\gamma' \gamma}^{-}\right|^2 N_{\bm q} 
\delta_{{\bm q}_{\parallel}, {\bm k}'-{\bm k}}
\delta(E_{{\bm k}'\gamma'}-E_{{\bm k}\gamma}-\hbar \omega_{\bm q}).
\end{align}
Here, we note that the transition rate takes a maximum when $\gamma \hat{h}_{\rm eff}(\varphi) \cdot \hat{\bm m} = -\gamma' \hat{h}_{\rm eff}(\varphi') \cdot \hat{\bm m}=1$.
This is consistent with the fact that the magnon absorption induces spin flipping to 2DEG electrons in the opposite direction to the ordered spin.
In a similar way, the transition rate for the magnon emission process is calculated as
\begin{align}
& \Gamma^+_{({\bm k}\gamma)\rightarrow ({\bm k}'\gamma')} = \frac{\pi S_0 |T_{\bm q}|^2}{\hbar} 
\left|A_{\gamma' \gamma}^{+}\right|^2 (N_{\bm q}+1)
\delta_{{\bm q}_{\parallel}, {\bm k}-{\bm k}'}
\notag \\
& \hspace{10mm} \times \delta(E_{{\bm k}'\gamma'}-E_{{\bm k}\gamma}+\hbar \omega_{\bm q}), \\
& \left|A_{\gamma' \gamma}^{+}\right|^2 = ( 1 - \gamma' \hat{h}_{\rm eff}(\varphi') \cdot \hat{\bm m})( 1 + \gamma \hat{h}_{\rm eff}(\varphi) \cdot \hat{\bm m} ) ,
\end{align}
Thus, we finally obtain the collision term due to the interfacial scattering as
\begin{align}
& \left. \frac{\partial f({\bm k},\gamma)}{\partial t}\right|_{\rm int} 
= \frac{\pi S_0}{\hbar} \sum_{{\bm k}',{\bm q}, \gamma'} |T_{\bm q}|^2 \Bigl[
|A_{\gamma\gamma'}^{+}|^2 F_{\gamma\gamma'}({\bm q},{\bm k},{\bm k}')  \notag \\
& \hspace{33mm} - |A_{\gamma'\gamma}^{+}|^2 F_{\gamma'\gamma}({\bm q},{\bm k}',{\bm k})  \Bigr],
\label{coltmp} \\
& F_{\gamma\gamma'}({\bm q},{\bm k},{\bm k}') = 
[(N_{\bm q}+1)f({\bm k}',\gamma') (1-f({\bm k},\gamma) ) \notag \\
&\hspace{23mm} -N_{\bm q}
f({\bm k},\gamma) (1-f({\bm k}',\gamma'))] \notag \\
&\hspace{25mm}\times \delta(E_{{\bm k}\gamma}-E_{{\bm k}'\gamma'}+\hbar \omega_{\bm q}) \delta_{{\bm q}_{\parallel}, {\bm k}'-{\bm k}},
\end{align}
where we have used $(A_{\gamma' \gamma}^{\pm}(\varphi',\varphi))^{*} = A_{\gamma \gamma'}^{\mp}(\varphi,\varphi')$.

These collision terms reflect the energy conservation law through the delta function.
We note that the typical magnon energy relevant to spin injection into 2DEG is given by $k_{\rm B}T$, which is assumed to be much smaller than the spin-splitting energy.
Therefore, we can safely neglect the factor $\hbar\omega_{\bm q}$ in the delta function.
In the following, we employ this quasi-elastic approximation.

\subsection{Solution of the Boltzmann equation}

We first rewrite the collision terms with integrals, using
\begin{align}
\frac{1}{\cal A}\sum_{{\bm k}'} (\cdots) \rightarrow  
\frac{1}{2\pi} \int_0^{2\pi} \frac{d\varphi}{2\pi} \int dk' (\cdots).
\end{align}
Assuming that the spin splitting energy is sufficiently smaller than the Fermi energy, the $\delta$ function in the collision term is rewritten as
\begin{align}
&\delta(E_{{\bm k}'\gamma'}-E_{{\bm k}\gamma}) \notag \\
&= \frac{1}{\hbar v_{\rm F}} \delta(k'-k(\varphi',\gamma')-k+k(\varphi,\gamma)),
\label{deltafunccalc}
\end{align}
where $v_{\rm F}=\hbar k_{\rm F}/m^*$ is the Fermi velocity, $k_{\rm F}$ is the Fermi wavenumber in the absence of the spin-orbit interaction, and
\begin{align}
k(\varphi,\gamma) &=k_{\rm F} -\frac{\gamma k_{\rm F}}{\hbar v_{\rm F}} \sqrt{\alpha^2+\beta^2+2\alpha\beta\sin 2\varphi} \notag \\
&= k_{\rm F} -2\pi \gamma D(E_{\rm F}) \sqrt{\alpha^2+\beta^2+2\alpha\beta\sin 2\varphi},
\label{kvarphidep}
\end{align}
is the wavenumber of the Fermi surface in the direction of $\varphi$ for the band $\gamma$.
In the second equation of Eq.~(\ref{kvarphidep}), we have used the fact that the density of state per spin is given as $D(E_{\rm F}) = k_{\rm F}/2\pi\hbar v_{\rm F}$.

We note that the collision terms become zero for the thermal equilibrium states ($\Delta T = 0$).
The collision term due to impurities is calculated up to the first order of $\Delta T$ as
\begin{align}
& \left.\frac{\partial f({\bm k},\gamma)}{\partial t}\right|_{\rm imp} \notag = \frac{\Gamma}{\hbar}\sum_{\gamma'} \int \frac{d\varphi'}{2\pi} \, 
\frac{k-k(\varphi,\gamma)+k(\varphi',\gamma')}{k_{\rm F}} \notag \\
& \hspace{5mm}\times \frac{1+\gamma \gamma'\hat{\bm h}_{\rm eff}(\varphi) \cdot \hat{\bm h}_{\rm eff}(\varphi')}{2} \notag \\
& \hspace{5mm} \times f_0({\bm k},\gamma)(1-f_0({\bm k},\gamma))[\Phi(\varphi',\gamma') - \Phi(\varphi,\gamma)] ,
\label{colinteg_imp}
\end{align}
where $\Gamma=2\pi v_0^2 n_{\rm imp} D(E_{\rm F})$ is the energy broadening due to impurity scattering.
In a similar way, the collision term due to interfacial scattering is calculated up to the first order of $\Delta T$ as
\begin{align}
&\left.\frac{\partial f({\bm k},\gamma)}{\partial t}\right|_{\rm int} \notag \\
&= -\frac {\Delta T}{k_{\rm B}T^2} \frac{\Gamma_{\rm int}}{\hbar} \sum_{\gamma'} \int \frac{d\varphi'}{2\pi} \, \frac{k-k(\varphi,\gamma)+k(\varphi',\gamma')}{k_{\rm F}}
\notag \\
&\hspace{5mm} \times \frac{-\gamma \hat{h}_{\rm eff}(\varphi) \cdot \hat{\bm m} + \gamma' \hat{h}_{\rm eff}(\varphi') \cdot \hat{\bm m} }{2} \notag \\
&\hspace{5mm} \times  I(\varphi-\varphi') f_0({\bm k},\gamma) (1-f_0({\bm k},\gamma)) ,
\label{colinteg_int}
\end{align}
where the tunnel matrix element is assumed to be constant ($T_{\bm q}=\bar{T}$), the interfacial coupling strength is defined as
$\Gamma_{\rm int} = 2S_0d |\bar{T}|^2 {\cal A}D(E_{\rm F})/a$, and $a$ and $d$ denote the lattice spacing and the thickness of the FI, respectively.
The temperature-dependent factor $I(\varphi)$ is defined as
\begin{align}
& I(\varphi) = \int_{-\pi}^\pi d(q_z a) \, \hbar \omega(\varphi,q_z) n(\varphi,q_z), \\
& \hbar \omega(\varphi,q_z) = h_{\rm dc} + 4 {\cal D} k_{\rm F}^2 \sin^2\frac{\varphi}{2} +\mathcal{D}q_z^2 , \\
& n(\varphi,q_z) = \frac{1}{e^{\hbar \omega(\varphi,q_z)/k_{\rm B}T}-1}.
\end{align}
Combining Eqs.~(\ref{eq:0ecoll}) with (\ref{colinteg_imp}) and (\ref{colinteg_int}), we obtain analytic solution for $\Phi(\varphi,\gamma)$ as
\begin{align}
\Phi(\varphi,\gamma) &= \frac{\Delta T}{T} \frac{\Gamma_{\rm int}}{\Gamma}\Big[
2 \gamma I_1(T) \hat{\bm h}_{\rm eff}^T(\varphi)M \hat{\bm m} \notag \\
&\hspace{18mm}+ \frac{I_2(T)}{\hbar v_{\rm F}k_{\rm F}}{\bm h}_{\rm eff}(\varphi)\cdot\hat{\bm m}
\Big], \label{PhiResult}
\end{align}
where the superscript $T$ indicates vector transpose and $M$ is a $2\times 2$ matrix defined as
\begin{align}
M&= \frac{1}{1-D^2} \left(\begin{array}{cc}
1 & -D\\ -D & 1\end{array}\right), \\
D&=\frac{\alpha^2+\beta^2-|\alpha^2-\beta^2|}{2\alpha\beta}
= {\rm min}\, (\beta/\alpha, \alpha/\beta) ,
\label{DDef}
\end{align}
and the temperature-dependent factors are defined as
\begin{align}
I_1(T) &= \int_0^{2\pi} \! \frac{d\varphi'}{2\pi} \int_{-\pi}^{\pi} \! d(q_za)\, \hbar \omega(\varphi',q_z) n(\varphi',q_z), \label{eq:I1} \\
I_2(T) &= \int_0^{2\pi} \! \frac{d\varphi'}{2\pi} \int_{-\pi}^{\pi} \! d(q_za) \,  \hbar \omega(\varphi',q_z) n(\varphi',q_z) \cos \varphi' .\label{eq:I2}
\end{align}
For a detailed derivation, see Appendix~\ref{appA}.

\section{Induced Current}
\label{sec:current}

In this section, we calculate the current induced by IREE and thermal spin injection. For reference, we describe the result for spin accumulation and heat current in Appendices~\ref{appC} and \ref{appD}, respectively. 

\subsection{Analytic result}

The current in 2DEG induced by thermal spin injection is written with the distribution function as
\begin{align}
{\bm j} &= \frac{e}{{\cal A}}\sum_{{\bm k},\gamma} {\bm v}({\bm k},\gamma) f({\bm k},\gamma) ,
\end{align}
where ${\bm v}({\bm k},\gamma)$ is a velocity defined as
\begin{align}
{\bm v}({\bm k},\gamma) &\equiv \frac{1}{\hbar} \frac{\partial E_{{\bm k}\gamma}}{\partial {\bm k}}
= \frac{\hbar {\bm k}}{m^*} + \frac{\gamma}{\hbar} \frac{\partial h_{\rm eff}({\bm k})}{\partial {\bm k}} .
\end{align}
Replacing the sum with an integral and using $-\partial f_0({\bm k},\gamma)/\partial E_{{\bm k}\gamma} \simeq \delta(k-k(\varphi,\gamma))/\hbar v_{\rm F}$, the current is rewritten with $\Phi(\varphi,\gamma)$ as
\begin{align}
& {\bm j} = \frac{e k_{\rm F}}{2\pi \hbar v_{\rm F}} \sum_{\gamma} \int \frac{d\varphi}{2\pi} \frac{k(\varphi,\gamma)}{k_{\rm F}} 
{\bm v}(k(\varphi,\gamma),\varphi,\gamma)
\Phi(\varphi,\gamma) ,\label{currentPhi} \\
& {\bm v}(k(\varphi,\gamma),\varphi,\gamma)
= v_{\rm F}\, \hat{\bm k} + \frac{\gamma}{\hbar} \frac{2\alpha\beta\cos{2\varphi}}{\sqrt{\alpha^2+\beta^2+2\alpha\beta\sin{2\varphi}}} \hat{\boldsymbol{\varphi}} ,
\end{align}
where $\hat{\bm k} = (\cos \varphi,\sin \varphi)$ and $\hat{\boldsymbol{\varphi}} = (-\sin \varphi,\cos \varphi)$ are unit direction vectors.
Substituting Eq.~(\ref{PhiResult}), the current is calculated up to the first order of the spin-orbit interaction as
\begin{align}
{\bm j}
&= \frac{2e k_{\rm F}}{2\pi\hbar^2 v_{\rm F}}\frac{\Delta T}{T} \frac{\Gamma_{\rm int}}{\Gamma} \left(I_1(T)-\frac{I_2(T)}{2}\right) \left(\begin{array}{ll}
\beta & -\alpha \\ \alpha & -\beta\end{array}\right) \hat{\bm m} .
\label{eq:CurrentResult}
\end{align}
This is a main result of our work. For a detailed derivation, see Appendix~\ref{appB}.

Let us discuss the qualitative features of the induced current.
The current depends on the temperature only through the factor $I_1(T)-I_2(T)/2$ ($>0$) in Eq.~(\ref{eq:CurrentResult}), while it depends on the direction of the ordered spin, $\hat{\bm m}=(\cos \theta,\sin \theta)$, in the FI through the last part of Eq.~(\ref{eq:CurrentResult}).
The latter relation is rewritten as
\begin{align}
{\bm j} &\parallel 
\left(\begin{array}{ll}
\beta & -\alpha \\ \alpha & -\beta\end{array}\right) \hat{\bm m} \notag \\ 
&= \alpha \left(\begin{array}{c} \sin \theta \\ \cos \theta \end{array}\right) + \beta \left(\begin{array}{c} -\cos \theta \\ \sin \theta \end{array}\right).
\label{eq:CurrentAngleDep}
\end{align}
It is remarkable that this relation between the magnetization $\hat{\bm m}$ and the current ${\bm j}$ holds even for $\Gamma \gg k_{\rm F}\alpha, k_{\rm F}\beta$, which is the opposite condition to our calculation~\footnote{For example, see Eq.~(2) in Ref.~\cite{Ganichev2004}.
In previous work, two types of current were derived for $\Gamma \gg k_{\rm F}\alpha, k_{\rm F}\beta$ depending on the different spin relaxation mechanisms, that is, the Elliott-Yafet mechanism and the Dyakonov–Perel mechanism~\cite{Ivchenko2008}.
However, we should note that this separation by the spin relaxation mechanisms is not possible in our calculation, which treats the opposite condition, $\Gamma \ll k_{\rm F}\alpha, k_{\rm F}\beta$~\cite{Szolnoki2017,Suzuki2023}.}.
We also note that the current ${\bm j}(\alpha,\beta)$ has a symmetry relation
\begin{align}
j_x(\alpha,\beta) = j_y(\beta,\alpha).
\end{align}
This indicates that the induced current for a certain value of the ratio $\alpha/\beta$ can be related to that of its inverse, that is, $\beta/\alpha$.

\subsection{Spin-orientation dependence}
\label{sec:OrientationDep}

\begin{figure*}[tb]
    \centering
\includegraphics[width=160mm]{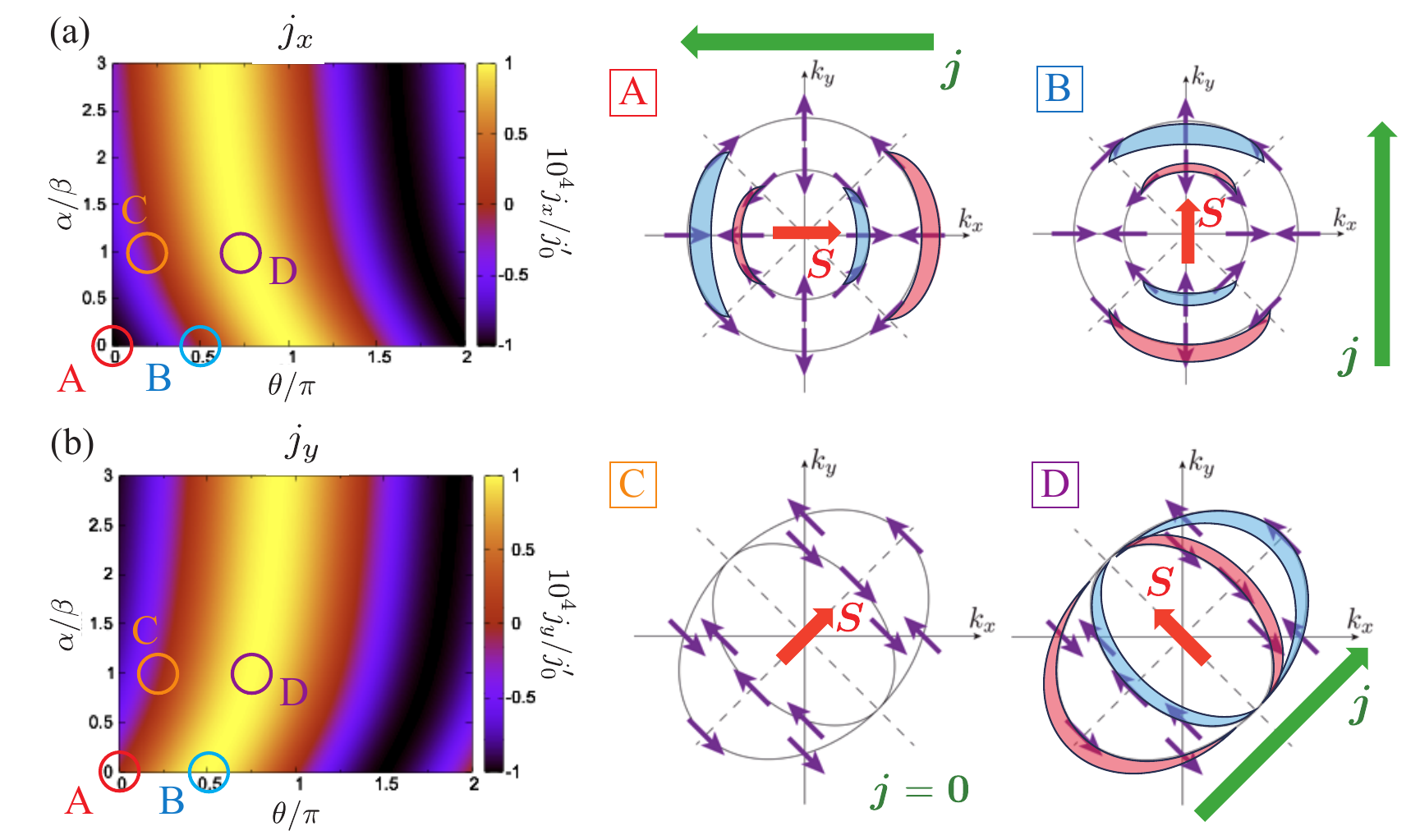}
    \caption{Contour plots of the current density ${\bm j}=(j_x,j_y)$ of 2DEG induced by thermal spin injection as a function of the azimuth angle $\theta$ of the spin in FI and $\alpha/\beta$: (a) $j_x$ and (b) $j_y$. The temperatures of FI and 2DEG are taken as $T+\Delta T$ and $T$, respectively. We set $k_{\rm B}T/4{\cal D}k_{\rm F}^2=1$ and $h_{\rm dc}=0$. The current density is normalized by $j_0'=j_0\Delta T/T$, where $j_0$ is defined in Eq.~(\ref{j0def}).
    The four right panels show schematic pictures of the Fermi surface modification by thermal spin injection at the four points, A, B, C, and D, in the contour plots.
    \label{fig:ContourJ}}
\end{figure*}

Next, we discuss how the induced current ${\bm j}=(j_x,j_y)$ depends on the orientation of the ordered spin ${\bm m}=(\cos\theta,\sin\theta)$.
Figure~\ref{fig:ContourJ}(a) and (b) respectively show $j_x$ and $j_y$ as a function of the ratio $\alpha/\beta$ and the azimuth angle $\theta$ of ${\bm m}$. The current is normalized by $j_0'=j_0\Delta T/T$, where
\begin{align}
j_0 = \frac{8|e|D(E_{\rm F}) {\cal D}k_{\rm F}^2 \sqrt{\alpha^2+\beta^2}}{\hbar} \frac{\Gamma_{\rm int}}{\Gamma} .
\label{j0def}
\end{align}
As indicated from Eq.~(\ref{eq:CurrentAngleDep}), the $x$- and $y$-component of the current is proportional to a trigonometric function of $\theta$ and takes a maximum (a minimum) at a specific value of $\theta$.
The position of the maximum (minimum) changes as the ratio $\alpha/\beta$ increases.

We first consider the case where only the Dresselhaus spin-orbit interaction exists ($\alpha/\beta=0$).
For $\theta = 0$ (indicated by A in the contour plot) the current flows in the $-x$ direction, while for $\theta = \pi/2$ (indicated by B) it flows in the $y$ direction.
To explain the physical mechanism of current generation at the points A and B, we show schematic pictures of the corresponding electron distribution functions in the upper two right panels of Fig.~\ref{fig:ContourJ}.
We first note that the magnon absorption process always becomes predominant over the magnon emission process, since the temperature of FI is assumed to be higher than that of 2DEG.
Because the magnon carries spin in the direction opposite to ${\bm S}$, the temperature gradient across the junction induces spin injection into 2DEG by flipping conduction electron spins in the direction opposite to the ordered spin in FI, $\langle {\bm S} \rangle$.
This spin flipping changes the distribution function of electrons, depending on the spin polarization of 2DEG electrons, which is depicted by arrows on the Fermi surface.
When the spin in the FI, $\langle {\bm S} \rangle$, points to the $+x$ direction ($\theta=0$, panel A), spin flipping toward the $-x$ direction occurs for 2DEG electrons by magnon absorption at the interface.
As a result, the distribution of electrons in the momentum space is modified as indicated by the red (blue) region at which the distribution function increases (decreases).
Since the shifts of the Fermi surface are opposite for the two spin-splitting bands, they are almost canceled.
However, the cancellation is not complete since the density of state is larger for the outer Fermi surface.
Therefore, the electrons flow in the $+x$ direction, resulting in the charge current in the $-x$ direction.
In the same way, we can explain the direction of the current when $\langle {\bm S} \rangle$ points to the $+y$ direction ($\theta=\pi/2$, panel B);
magnon absorption causes spin flipping of 2DEG in the $-y$ direction, leading to the Fermi surface modification shown by the red and blue regions in panel B.
This Fermi-surface modification produces the current in the $y$ direction.
We note that the direction of the current changes clockwise when the direction of $\langle {\bm S} \rangle$ rotates counterclockwise.

Next, we consider the case in which the two spin-orbit interactions compete ($\alpha/\beta=1$).
For $\theta = \pi/4$ (the point C in the contour plot), the spin polarization on the Fermi surface of 2DEG is always perpendicular to ${\bm S}$, leading to the vanishing current (see panel C in Fig.~\ref{fig:ContourJ}).
On the other hand, for $\theta = 3\pi/4$ (the point D), spin flipping of the conduction electrons toward the $-{\bm S}$ direction induces Fermi surface modification as indicated by the red and blue region in the panel D in Fig.~\ref{fig:ContourJ}, resulting in the charge current in the direction of the azimuth angle $\pi/4$.

Thus, the current generation due to the temperature gradient can be explained intuitively in terms of the Fermi surface modification depending on its spin polarization.

\subsection{Other dependence}

The temperature dependence of the current is determined only by the factor $I_1(T)-I_2(T)/2$, as seen in Eq.~(\ref{eq:CurrentResult}).
By changing the integral variable from $q_z$ to $x = \hbar \omega(\varphi,q_z)/k_{\rm B}T$ in Eqs.~(\ref{eq:I1}) and (\ref{eq:I2}), we can show that both $I_1(T)$ and $I_2(T)$ are proportional to $T^{3/2}$.
Therefore, when the temperature difference $\Delta T$ is fixed, the current becomes proportional to $T^{1/2}$.
The exponent of the temperature dependence depends on the dimension of the FI through the density of states of magnons.
For example, the current becomes independent of the temperature for a fixed $\Delta T$ if we consider a two-dimensional FI.

Regarding the dependence on $\alpha/\beta$, it is notable that the maximum of the induced current is always proportional to $\alpha^2 + \beta^2$. This means that the current is independent of the ratio $\alpha/\beta$ while keeping the amplitude $(\alpha^2 + \beta^2)^{1/2}$ constant.
This behavior is in contrast with the spin accumulation of 2DEG, which shows divergence at $\alpha/\beta = 1$ (see Appendix~\ref{appC}).

\section{Experimental Relevance}
\label{sec:estimate}

In comparison with experiments, we should be careful about how the temperature gradient is generated in a sample.
In particular, thermal conductivity due to phonons, which is not explicitly considered in our work, largely affects the current generation in 2DEG through the change of temperature distribution in a sample, which is difficult to measure.
Therefore, it will be difficult to experimentally observe the temperature dependence predicted in our work.
However, the prediction on the spin-orientation dependence will be tested experimentally since it is not affected by a detail of the temperature gradient.
We also mention that more information can be obtained by simultaneous measurement of the present phenomenon and IREE induced by spin pumping~\cite{Yama2023}.

To clarify the experimental relevance, we roughly estimate the current induced by thermal spin injection.
We first consider a junction composed of EuO and KTaO${}_3$~\cite{Zhang19}.
Using the electron density $n\simeq 10^{14} \, {\rm cm}^{-2}$, the mobility $\mu_e = 10^2 \, {\rm cm}^2/{\rm Vs}$, the effective mass $m^* = 0.52 m_e$ ($m_e$: electron mass) and the lattice spacing $a=4\,$\AA~\cite{Zhang19,Varotto2022}, we obtain $\Gamma=22\, {\rm meV}$.
When the interfacial exchange coupling is roughly estimated as $10\, {\rm meV}$, we obtain $\Gamma_{\rm int} = 3\times 10^{-2}\, {\rm meV}$.
Using the Rashba spin-orbit interaction $\alpha = 320 \, {\rm meV}$\AA~\cite{Varotto2022}, $S_0=7/2$, and ${\cal D}=10\,{\rm meV}$\AA${}^2$~\cite{Passell76,Hasegawa12}, we obtain $j_0 = 60 \, \mu{\rm A}/{\rm mm}$.
Setting $\Delta T = 1 \, {\rm K}$ and $k_{\rm B}T = 4{\cal D}k_{\rm F}^2 = 2.5 \, {\rm meV}$, we finally obtain the current $j=0.2 \, {\rm nA}/{\rm mm}$, which is comparable to the experimental value $j\sim 1 \, {\rm nA}/{\rm mm}$~\cite{Zhang19}.

As another example, we consider a GaAs--Fe junction in which the ratio between the Rashba and Dresselhaus spin-orbit interactions can be controlled. 
Using $n=1.1\times 10^{17} \, {\rm cm}^{-3}$, $\mu_e=3.5\times 10^3 \, {\rm cm}^2/{\rm Vs}$~\cite{Olejnik12}, $m^* = 0.067 m_e$, $S_0 \sim 2$~\cite{Bozorth1951}, and ${\cal D}=230\,{\rm meV}$\AA${}^2$~\cite{Lynn75}, we obtain $\Gamma = 5 \, {\rm meV}$ and $\Gamma_{\rm int} = 1.2 \, \mu{\rm eV}$, assuming the interfacial exchange coupling of $J = 10 \, {\rm meV}$.
When we set $\alpha=100 \, {\rm meV}$\AA~\cite{Chen2016}, $\beta = 0$, $\Delta T = 1 \, {\rm K}$, and $k_{\rm B}T = 4{\cal D}k_{\rm F}^2 = 0.2 \, {\rm meV}$ as a rough estimate, we obtain $j=1.4 \, {\rm pA}/{\rm mm}$, which is expected to be in a detectable range.

In this work, the spin splitting energy determined by $k_{\rm F}\alpha$ and $k_{\rm F}\beta$ is assumed to be much larger than the temperature ($k_{\rm B}T$), the energy broadening due to impurities ($\Gamma$), and the scattering rate at the interface ($\Gamma_{\rm int}$), while it is assumed to be much smaller than the chemical potential $\mu$.
The features of the induced current obtained in this work are expected to be observed most clearly under these conditions, for which the spin-momentum locking is most effective.
On the other hand, we expect the current to be induced even if some of the conditions are not well satisfied.
In fact, the theoretical description for the ordinary direct and inverse Rashba-Edelstein effects does not require such conditions.
We leave a detailed calculation that covers a wide range of parameters as a future problem. 

\section{Summary}
\label{sec:summary}

We theoretically examined current generation by the thermal spin injection into 2DEG with the Rashba and Dresselhaus spin-orbit interactions.
For a magnetic junction composed of 2DEG and FI, we formulated the electric current in 2DEG caused by the inverse Rashba-Edelstein effect under a temperature gradient between 2DEG and FI, employing the method of the Boltzmann equation.
In our formulation, a non-equilibrium steady state of 2DEG is realized by balancing impurity scattering and interfacial electron scattering accompanying spin flipping due to magnon absorption/emission.
In our work, we focused on the case in which the strength spin-orbit interactions are much larger than temperature and energy broadening due to electron scattering.
In this situation, the spin-momentum locking is most effective, and the direction of the generated current is largely affected by the spin texture on the Fermi surface, which can be controlled by the ratio between Rashba and Dresselhaus spin-orbit interactions.
We obtained an analytic formula for the current that can clarify the dependence of the magnetization of FI, temperature, and spin texture on the Fermi surface.
We also showed numerical results for the current as a function of the azimuth angle of the ordered spin in FI and the ratio between the two spin orbit interactions.
We found that the direction of the generated current is consistent with an intuitive explanation by spin flipping of the conduction electrons due to the magnon absorption(emission).

Our work will be helpful for an accurate analysis of the inverse Rashba-Edelstein effect induced by the temperature gradient of the sample.
Although we considered a simple 2DEG system with a circular Fermi surface and small spin-orbit interactions, it can be extended to more complex systems, including effective models obtained from first-principles calculations.
Details of such an extension will be discussed in subsequent papers.

\begin{acknowledgments}
The authors thank Y. Suzuki, Y. Kato, and M. Kohda for their helpful discussions. 
M. Y. was supported by JST SPRING (Grant No.~JPMJSP2108) and JSPS KAKENHI Grant Number JP24KJ0624. M. M. was supported by the National Natural Science Foundation of China (NSFC) under Grant No. 12374126, by the Priority Program of Chinese Academy of Sciences under Grant No.~XDB28000000, and by JSPS KAKENHI for Grants (No.~JP21H04565, No.~JP21H01800, No.~JP23H01839, and No.~24H00322) from MEXT, Japan.
T.K. was supported by JSPS KAKENHI Grant No.~JP24K06951.
\end{acknowledgments}

\appendix

\section{Detailed derivation of Eq.~(\ref{PhiResult})}
\label{appA}

We first define the part in the Boltzmann equation, which is independent of $\Phi(\varphi,\gamma)$, as
\begin{align}
F(\varphi,\gamma) &= -\frac{\Delta T}{T} \frac{\Gamma_{\rm int}}{\Gamma} \sum_{\gamma'} \int \frac{d\varphi'}{2\pi} \, \frac{k(\varphi',\gamma')}{k_{\rm F}}\notag \\
& \hspace{5mm} \times \frac{-\gamma \hat{\bm h}_{\rm eff}(\varphi) \cdot \hat{\bm m} + \gamma' \hat{\bm h}_{\rm eff}(\varphi') \cdot \hat{\bm m}}{2} \notag \\
& \hspace{5mm}  \times \hbar \omega(\varphi-\varphi') n(\varphi-\varphi').
\end{align}
We note that this function satisfies the following symmetry relation:
\begin{align}
F(\varphi+\pi,\gamma) &=-F(\varphi,\gamma) .
\end{align}
The Boltzmann equation is rewritten into the integral equation with $F(\varphi,\gamma)$ as
\begin{align}
& \Phi(\varphi,\gamma) =F(\varphi,\gamma) \notag \\
& + \sum_{\gamma'} \int \frac{d\varphi'}{2\pi} \, 
    \frac{k(\varphi',\gamma')}{k_{\rm F}} \frac{1+\gamma \gamma'\hat{\bm h}_{\rm eff}(\varphi) \cdot \hat{\bm h}_{\rm eff}(\varphi')}{2}\Phi(\varphi',\gamma') .
    \label{eq:IterativeEq}
\end{align} 
By iterative substitution of $\Phi(\varphi,\gamma)$ into the integral of the l.h.s of Eq.~(\ref{eq:IterativeEq}), we can obtain $\Phi(\varphi',\gamma')$ as a  series including multiple angle integrals of $F(\varphi,\gamma)$, which constitute a functional of $F(\varphi,\gamma)$.
To simplify this functional, we rewrite $F(\varphi,\gamma)$ as 
$F(\varphi,\gamma) = \gamma F_1(\varphi) + F_2(\varphi)$, where
\begin{align}
F_1(\varphi) &= \frac{\Delta T}{T}\frac{\Gamma_{\rm int}}{\Gamma} \hat{\bm h}_{\rm eff}(\varphi) \cdot \hat{\bm m} \notag \\ 
&\hspace{5mm} \times \int \frac{d\varphi'}{2\pi} \, \hbar \omega(\varphi-\varphi') n(\varphi-\varphi') ,\label{F1} \\
F_2(\varphi)&=\frac{\Delta T}{T} \frac{\Gamma_{\rm int}}{\Gamma} \int \frac{d\varphi'}{2\pi} \, g(\varphi) \hat{\bm h}_{\rm eff}(\varphi') \cdot \hat{\bm m} \notag \\
& \hspace{5mm} \times \hbar \omega(\varphi-\varphi') n(\varphi-\varphi') ,\label{F2}
\end{align}
where
\begin{align}
g(\varphi) &= \frac{1}{\hbar v_{\rm F}} \sqrt{\alpha^2+\beta^2+2\alpha\beta\sin 2\varphi}.
\end{align}
Then, $\Phi(\varphi,\gamma)$  can be expressed as 
\begin{align}
\Phi(\varphi,\gamma) &= \Phi[\gamma F_1(\varphi)]+\Phi[F_2(\varphi)] \notag \\
& \equiv \Phi_1(\varphi,\gamma) + \Phi_2(\varphi,\gamma).
\end{align} 
In the following, we separately calculate
$\Phi_1(\varphi,\gamma)$ and $\Phi_2(\varphi,\gamma)$.
By careful calculation for a series solution of Eq.~(\ref{eq:IterativeEq}), we obtain 
\begin{align}
& \Phi_1(\varphi,\gamma) =\gamma F_1(\varphi)\notag \\
&\hspace{5mm}+ \gamma \sum_{n=0}^{\infty} \int \frac{d\varphi'}{2\pi} \, \hat{\bm h}_{\rm eff}^T(\varphi)
A^{n}  \hat{\bm h}_{\rm eff} (\varphi') F_1(\varphi') , \label{Phi1}\\
& \Phi_2(\varphi,\gamma) =F_2(\varphi)\notag \\
&\hspace{5mm} + \gamma \sum_{n=0}^{\infty} \int \frac{d\varphi'}{2\pi} \, \hat{\bm h}_{\rm eff}^T(\varphi)
A^{n}  \hat{\bm h}_{\rm eff} (\varphi') g(\varphi')F_2(\varphi').
\label{Phi2}
\end{align} 
Here, $\Phi_2(\varphi,\gamma)$ can be approximated as $F_2(\varphi)$ up to the first order of the spin-orbit interaction since the second term of r.h.s of Eq.~(\ref{Phi2}) is of higher order.
The matrix $\hat{A}$ is defined as
\begin{align}
\hat{A} &= \int_0^{2\pi} \frac{d\varphi}{2\pi} \, \hat{\bm h}_{\rm eff}(\varphi)
\hat{\bm h}_{\rm eff}^T(\varphi), 
\end{align}
where
\begin{align}
{\bm a} \, {\bm a}^T
&= \left(\begin{array}{c} a_1 \\ a_2 \end{array}\right)
(a_1 \ a_2) = \left( \begin{array}{cc} a_1^2  & a_1 a_2 \\ a_1 a_1 & a_2^2 \end{array}\right) .
\end{align}
Straightforward calculation of $\hat{A}$ gives
\begin{align}
\hat{A} = \left( \begin{array}{cc} 1/2 & -D/2 \\ -D/2 & 1/2 \end{array} \right),
\end{align}
where $D$ is defined by Eq.~(\ref{DDef}).
Using
\begin{align}
\sum_{n=0}^{\infty} \hat{A}^n &= (\hat{I} - \hat{A})^{-1}= \frac{2}{1-D^2}\left(
\begin{array}{cc} 1 & -D \\ -D & 1 \end{array} \right).
\end{align}
Combining Eqs.~(\ref{Phi1}) and (\ref{Phi2}) with Eqs.~(\ref{F1}) and (\ref{F2}), we obtain Eq.~(\ref{PhiResult}).

\section{Spin Accumulation}
\label{appC}

In this appendix, we derive the analytic formula for the spin accumulation in 2DEG induced by thermal spin injection.
The spin density in 2DEG is defined as
\begin{align}
{\bm s} &= \frac{\hbar}{2{\cal A}} \sum_{{\bm k},\gamma}\langle{\bm k}\gamma|{\bm \sigma}|{\bm k}\gamma\rangle f({\bm k},\gamma) \notag \\
& \simeq -\hbar D(E_{\rm F})\int \frac{d\varphi}{2\pi} \hat{\bm h}_{\rm eff}(\varphi) \Phi_1(\varphi),
\end{align}
where we have used $\langle{\bm k}\gamma|{\bm \sigma}|{\bm k}\gamma\rangle=-\gamma \hat{\bm h}_{\rm eff}(\varphi)$ and $D(E_{\rm F}) = k_{\rm F}/(2\pi \hbar v_{\rm F})$.
In the second equation, the spin density is approximated up to the first order of the spin orbit interaction.
Using the result given in Appendix~\ref{appA}
and
\begin{equation}
\hat{A}(I-\hat{A})^{-1} = \frac{1}{1-D^2} \left(\begin{array}{ll}
1+D^2 & -2D\\ -2D & 1+D^2\end{array}\right),
\end{equation}
the spin density is calculated as
\begin{align}
{\bm s} &= -\hbar D(E_{\rm F}) \frac{\Delta T}{T} \frac{\Gamma_{\rm int}}{\Gamma}\frac{I_1(T)}{1-D^2} \left(\begin{array}{ll}
1+D^2 & -2D\\ -2D & 1+D^2\end{array}\right) \hat{\bm m}(\theta).
\end{align}
This is a general formula for the spin density in 2DEG.
We note that the spin density has the following symmetry relation:
\begin{align}
{\bm s}(\alpha,\beta) = {\bm s}(\beta,\alpha).
\label{ssymmetry}
\end{align}

\begin{figure}[tb]
    \centering
    \includegraphics[width=80mm]{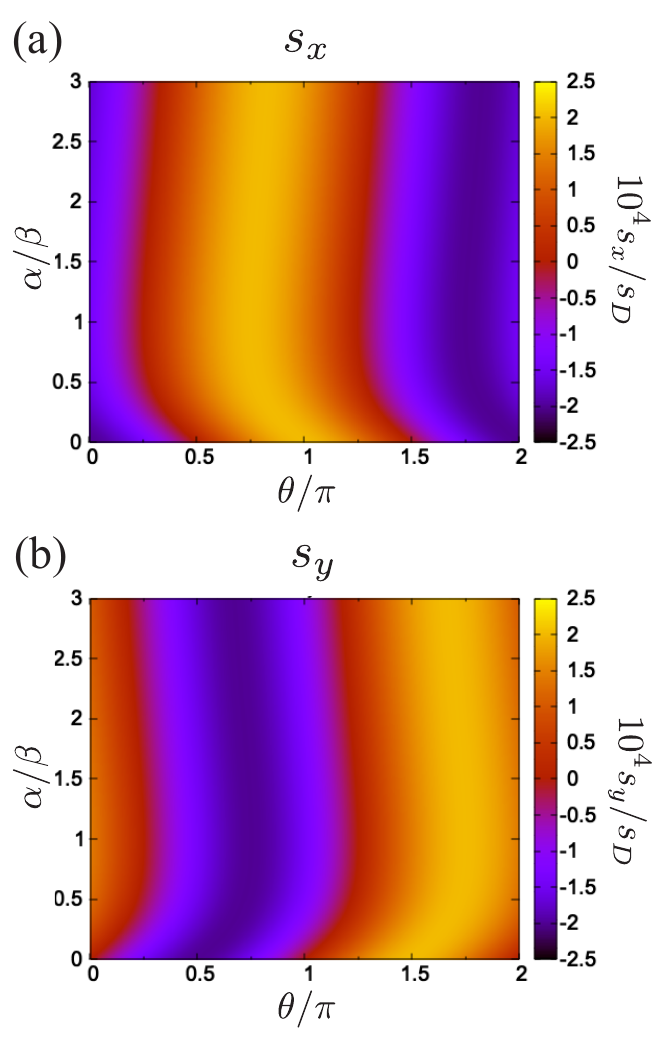}
    \caption{Contour plots of the spin density ${\bm s}=(s_x,s_y)$ of 2DEG induced by thermal spin injection as a function of the azimuth angle $\theta$ of the spin in FI and $\alpha/\beta$: (a) $s_x$ and (b) $s_y$. The temperatures of FI and 2DEG are taken as $T+\Delta T$ and $T$, respectively. We set $k_{\rm B}T/4{\cal D}k_{\rm F}^2=1$ and $h_{\rm dc}=0$. The spin density is normalized by $s_D$ defined in Eq.~(\ref{sD}).} 
    \label{fig:ContourS}
\end{figure}

Figure~\ref{fig:ContourS} shows the spin density of 2DEG induced by thermal spin injection as a function of $\theta$ and $\alpha/\beta$, where the azimuth angle of the spin in FI is defined as $\hat{\bm m}=(\cos\theta,\sin\theta)$ and the spin density is normalized by
\begin{align}
s_D &\equiv \frac{s_0 \Delta T}{T} \frac{\sqrt{1+6D^2+D^4}}{1-D^2},
\label{sD} \\
s_0 &\equiv \hbar D(E_{\rm F}) \hbar\omega_0 \frac{\Gamma_{\rm int}}{\Gamma}.
\end{align}
For $\alpha/\beta=0$, the direction of the accumulated spin becomes opposite to that of the spin in FI.
The same result is also obtained for $\alpha/\beta=\infty$ (see also Eq.~(\ref{ssymmetry})).
This result is consistent with an intuitive explanation that the magnon absorption at the interface induces spin flipping of electrons in 2DEG in the direction opposite to ${\bm m}$ (see also Sec.~\ref{sec:OrientationDep}).
On the other hand, for $\alpha/\beta \simeq 1$, the spin density points to the direction of $\varphi=7\pi/4$ when $\pi/4 < \theta < 5\pi/4$ while it points to that of $\varphi=3\pi/4$ otherwise.
This reflects the fact that the spin polarization on the Fermi surface aligns in the same direction as shown in panels, C and D, in Fig.~\ref{fig:ContourJ}.
In this situation, the electrons in 2DEG can receive the spin only in the direction of this spin polarization. 

Finally, let us discuss the dependence of $\alpha/\beta$.
The spin density depends on $\alpha/\beta$ through the factor $s_D$ given in Eq.~(\ref{sD}).
It is notable that $s_D$ diverges at $\alpha/\beta=1$.
This is because spin relaxation is never caused by
non-magnetic impurities when the spin polarization on the Fermi surface aligns in one direction.
This behavior is in contrast to that of the current; the current does not show any singularity at $\alpha/\beta = 1$.
This difference in behavior near $\alpha/\beta=1$ comes from the fact that the current is induced after a delicate cancellation between contributions from the inner and outer Fermi surfaces.

\section{Heat current}
\label{appD}

In our formalism, the heat current from the FI into the 2DEG across the junction can be expressed as
\begin{align}
I_h &= \sum_{{\bm k},\gamma} \sum_{{\bm k}',\gamma'} \left(E_{{\bm k}'\gamma'}-E_{{\bm k}\gamma}\right) 
\Gamma_{({\bm k},\gamma) \rightarrow ({\bm k}',\gamma')} \notag \\
& \hspace{15mm} \times f({\bm k},\gamma) (1-f({\bm k}',\gamma') ) \notag \\
&= \frac{2\pi}{\hbar} \frac{S_0}{2}|\bar{T}|^2 \sum_{{\bm k},\gamma,{\bm k}',\gamma',{\bm q}} (-\hbar\omega_{\bm q})
|A_{\gamma\gamma'}^{+}|^2 F_{\gamma\gamma'}({\bm q},{\bm k},{\bm k}').
\end{align}
Within the linear response to the temperature difference $\Delta T$, the heat current is calculated as
\begin{align}
I_h &= \frac{\Gamma_{\rm int}{\cal A}D(E_{\mathrm{F}})}{\hbar} \frac{\Delta T}{T} \, I_3(T), \\
I_3(T) &= \int_0^{2\pi} \! \frac{d\varphi}{2\pi} \int_{-\pi}^{\pi} \! d(q_za)\,
\left(\hbar\omega(\varphi,q_z)\right)^2 n(\varphi,q_z),
\end{align}
We note that $I_h$ is always positive for $\Delta T>0$ and is independent of the azimuth angle of $\hat{\bm m}$.

\section{Detailed derivation of Eq.~(\ref{eq:CurrentResult})}
\label{appB}

As expected from Eq.~(\ref{currentPhi}), the current is expressed by a sum of two contributions as ${\bm j}={\bm j}_1+{\bm j}_2$, where ${\bm j}_1$ and ${\bm j}_2$ depends on $\Phi_1(\varphi,\gamma)$ and  $\Phi_2(\varphi,\gamma)$, respectively 
Using the result given in Appendix~\ref{appA},
the current ${\bm j_2}$ is easily calculated as
\begin{align}
{\bm j_2}
&= \frac{2e k_{\rm F}}{2\pi\hbar}
\int \frac{d\varphi}{2\pi}\Phi_2(\varphi) \hat{\bm k}(\varphi) \notag \\
&= \frac{2e k_{\rm F}}{2\pi\hbar^2 v_{\rm F}}\frac{\Delta T}{T} \frac{\Gamma_{\rm int}}{\Gamma}
\frac{I_2(T)}{2}\left(\begin{array}{ll}
-\beta & \alpha\\ -\alpha & \beta\end{array}\right) \hat{\bm m}(\theta),
\label{j2}
\end{align}
where we have used
\begin{align}
\int_0^{2\pi} \frac{d\varphi}{2\pi} \frac{\hat{\bm k}(\varphi){\bm h}_{\rm eff}^{T}(\varphi)}{k_{\rm F}} &=\frac{1}{2}\left(\begin{array}{ll}
-\beta & \alpha\\ -\alpha & \beta\end{array}\right).
\label{appformula1}
\end{align}
On the other hand, ${\bm j}_1$ is calculated as
\begin{align}
{\bm j_1}
&= \frac{2e k_{\rm F}}{2\pi\hbar}
\int \frac{d\varphi}{2\pi}\Biggl[-g(\varphi)\hat{\bm k}(\varphi) \notag \\
&\hspace{5mm} + \frac{1}{\hbar v_{\rm F}} \frac{2\alpha \beta \cos 2\varphi}{\sqrt{\alpha^2+\beta^2+2\alpha \beta \sin 2\varphi}}\hat{\boldsymbol{\varphi}}(\varphi)\Biggr]\Phi_1(\varphi) \notag \\
&= \frac{2e k_{\rm F}}{2\pi\hbar^2 v_{\rm F}}\frac{\Delta T}{T} \frac{\Gamma_{\rm int}}{\Gamma} I_1(T) \left(\begin{array}{ll}
\beta & -\alpha \\ \alpha & -\beta\end{array}\right) \hat{\bm m}(\theta),
\label{j1}
\end{align}
where we have used Eq.~(\ref{appformula1}) and 
\begin{align}
&\int \frac{d\varphi}{2\pi} 
\frac{2\alpha \beta \cos 2\varphi}{\alpha^2+\beta^2+2\alpha \beta \sin 2\varphi} \frac{\hat{\boldsymbol{\varphi}}(\varphi) {\bm h}_{\rm eff}^T(\varphi)}{k_{\rm F}}
\notag \\
&\hspace{3mm} = \frac{D}{2} \left(\begin{array}{ll}
-\alpha & \beta\\ -\beta & \alpha\end{array}\right) .
\end{align}
By summing up Eqs.~(\ref{j2}) and (\ref{j1}), we obtain Eq.~(\ref{eq:CurrentResult}).

\bibliography{./reference}

\begin{thebibliography}{102}%
\makeatletter
\providecommand \@ifxundefined [1]{%
 \@ifx{#1\undefined}
}%
\providecommand \@ifnum [1]{%
 \ifnum #1\expandafter \@firstoftwo
 \else \expandafter \@secondoftwo
 \fi
}%
\providecommand \@ifx [1]{%
 \ifx #1\expandafter \@firstoftwo
 \else \expandafter \@secondoftwo
 \fi
}%
\providecommand \natexlab [1]{#1}%
\providecommand \enquote  [1]{``#1''}%
\providecommand \bibnamefont  [1]{#1}%
\providecommand \bibfnamefont [1]{#1}%
\providecommand \citenamefont [1]{#1}%
\providecommand \href@noop [0]{\@secondoftwo}%
\providecommand \href [0]{\begingroup \@sanitize@url \@href}%
\providecommand \@href[1]{\@@startlink{#1}\@@href}%
\providecommand \@@href[1]{\endgroup#1\@@endlink}%
\providecommand \@sanitize@url [0]{\catcode `\\12\catcode `\$12\catcode
  `\&12\catcode `\#12\catcode `\^12\catcode `\_12\catcode `\%12\relax}%
\providecommand \@@startlink[1]{}%
\providecommand \@@endlink[0]{}%
\providecommand \url  [0]{\begingroup\@sanitize@url \@url }%
\providecommand \@url [1]{\endgroup\@href {#1}{\urlprefix }}%
\providecommand \urlprefix  [0]{URL }%
\providecommand \Eprint [0]{\href }%
\providecommand \doibase [0]{https://doi.org/}%
\providecommand \selectlanguage [0]{\@gobble}%
\providecommand \bibinfo  [0]{\@secondoftwo}%
\providecommand \bibfield  [0]{\@secondoftwo}%
\providecommand \translation [1]{[#1]}%
\providecommand \BibitemOpen [0]{}%
\providecommand \bibitemStop [0]{}%
\providecommand \bibitemNoStop [0]{.\EOS\space}%
\providecommand \EOS [0]{\spacefactor3000\relax}%
\providecommand \BibitemShut  [1]{\csname bibitem#1\endcsname}%
\let\auto@bib@innerbib\@empty
\bibitem [{\citenamefont {Aronov}\ and\ \citenamefont
  {Lyanda-Geller}(1989)}]{Aronov1989}%
  \BibitemOpen
  \bibfield  {author} {\bibinfo {author} {\bibfnamefont {A.~G.}\ \bibnamefont
  {Aronov}}\ and\ \bibinfo {author} {\bibfnamefont {Y.~B.}\ \bibnamefont
  {Lyanda-Geller}},\ }\bibfield  {title} {\bibinfo {title} {Nuclear electric
  resonance and orientation of carrier spins by an electric field},\
  }\href@noop {} {\bibfield  {journal} {\bibinfo  {journal} {JETP Lett.}\
  }\textbf {\bibinfo {volume} {50}},\ \bibinfo {pages} {431} (\bibinfo {year}
  {1989})}\BibitemShut {NoStop}%
\bibitem [{\citenamefont {Aronov}\ \emph {et~al.}(1991)\citenamefont {Aronov},
  \citenamefont {Lyanda-Geller},\ and\ \citenamefont {Pikus}}]{Aronov1991}%
  \BibitemOpen
  \bibfield  {author} {\bibinfo {author} {\bibfnamefont {A.~G.}\ \bibnamefont
  {Aronov}}, \bibinfo {author} {\bibfnamefont {Y.~B.}\ \bibnamefont
  {Lyanda-Geller}},\ and\ \bibinfo {author} {\bibfnamefont {G.~E.}\
  \bibnamefont {Pikus}},\ }\bibfield  {title} {\bibinfo {title} {Spin
  polarization of electrons by an electric current},\ }\href@noop {} {\bibfield
   {journal} {\bibinfo  {journal} {Sov. Phys. JETP}\ }\textbf {\bibinfo
  {volume} {73}},\ \bibinfo {pages} {537} (\bibinfo {year} {1991})}\BibitemShut
  {NoStop}%
\bibitem [{\citenamefont {Edelstein}(1990)}]{Edelstein1990}%
  \BibitemOpen
  \bibfield  {author} {\bibinfo {author} {\bibfnamefont {V.}~\bibnamefont
  {Edelstein}},\ }\bibfield  {title} {\bibinfo {title} {Spin polarization of
  conduction electrons induced by electric current in two-dimensional
  asymmetric electron systems},\ }\href
  {https://doi.org/https://doi.org/10.1016/0038-1098(90)90963-C} {\bibfield
  {journal} {\bibinfo  {journal} {Solid State Commun.}\ }\textbf {\bibinfo
  {volume} {73}},\ \bibinfo {pages} {233} (\bibinfo {year} {1990})}\BibitemShut
  {NoStop}%
\bibitem [{\citenamefont {Inoue}\ \emph {et~al.}(2003)\citenamefont {Inoue},
  \citenamefont {Bauer},\ and\ \citenamefont {Molenkamp}}]{Inoue2003}%
  \BibitemOpen
  \bibfield  {author} {\bibinfo {author} {\bibfnamefont {J.-i.}\ \bibnamefont
  {Inoue}}, \bibinfo {author} {\bibfnamefont {G.~E.~W.}\ \bibnamefont
  {Bauer}},\ and\ \bibinfo {author} {\bibfnamefont {L.~W.}\ \bibnamefont
  {Molenkamp}},\ }\bibfield  {title} {\bibinfo {title} {Diffuse transport and
  spin accumulation in a rashba two-dimensional electron gas},\ }\href
  {https://doi.org/10.1103/PhysRevB.67.033104} {\bibfield  {journal} {\bibinfo
  {journal} {Phys. Rev. B}\ }\textbf {\bibinfo {volume} {67}},\ \bibinfo
  {pages} {033104} (\bibinfo {year} {2003})}\BibitemShut {NoStop}%
\bibitem [{\citenamefont {Silsbee}(2004)}]{Silsbee2004}%
  \BibitemOpen
  \bibfield  {author} {\bibinfo {author} {\bibfnamefont {R.~H.}\ \bibnamefont
  {Silsbee}},\ }\bibfield  {title} {\bibinfo {title} {Spin–orbit induced
  coupling of charge current and spin polarization},\ }\href
  {https://doi.org/10.1088/0953-8984/16/7/R02} {\bibfield  {journal} {\bibinfo
  {journal} {J. Phys. Condens. Matter}\ }\textbf {\bibinfo {volume} {16}},\
  \bibinfo {pages} {R179} (\bibinfo {year} {2004})}\BibitemShut {NoStop}%
\bibitem [{\citenamefont {Sinova}\ \emph {et~al.}(2015)\citenamefont {Sinova},
  \citenamefont {Valenzuela}, \citenamefont {Wunderlich}, \citenamefont
  {Back},\ and\ \citenamefont {Jungwirth}}]{Sinova2015}%
  \BibitemOpen
  \bibfield  {author} {\bibinfo {author} {\bibfnamefont {J.}~\bibnamefont
  {Sinova}}, \bibinfo {author} {\bibfnamefont {S.~O.}\ \bibnamefont
  {Valenzuela}}, \bibinfo {author} {\bibfnamefont {J.}~\bibnamefont
  {Wunderlich}}, \bibinfo {author} {\bibfnamefont {C.~H.}\ \bibnamefont
  {Back}},\ and\ \bibinfo {author} {\bibfnamefont {T.}~\bibnamefont
  {Jungwirth}},\ }\bibfield  {title} {\bibinfo {title} {Spin hall effects},\
  }\href {https://doi.org/10.1103/RevModPhys.87.1213} {\bibfield  {journal}
  {\bibinfo  {journal} {Rev. Mod. Phys.}\ }\textbf {\bibinfo {volume} {87}},\
  \bibinfo {pages} {1213} (\bibinfo {year} {2015})}\BibitemShut {NoStop}%
\bibitem [{\citenamefont {Soumyanarayanan}\ \emph {et~al.}(2016)\citenamefont
  {Soumyanarayanan}, \citenamefont {Reyren}, \citenamefont {Fert},\ and\
  \citenamefont {Panagopoulos}}]{Soumyanarayanan2016}%
  \BibitemOpen
  \bibfield  {author} {\bibinfo {author} {\bibfnamefont {A.}~\bibnamefont
  {Soumyanarayanan}}, \bibinfo {author} {\bibfnamefont {N.}~\bibnamefont
  {Reyren}}, \bibinfo {author} {\bibfnamefont {A.}~\bibnamefont {Fert}},\ and\
  \bibinfo {author} {\bibfnamefont {C.}~\bibnamefont {Panagopoulos}},\
  }\bibfield  {title} {\bibinfo {title} {Emergent phenomena induced by
  spin--orbit coupling at surfaces and interfaces},\ }\href
  {https://doi.org/10.1038/nature19820} {\bibfield  {journal} {\bibinfo
  {journal} {Nature}\ }\textbf {\bibinfo {volume} {539}},\ \bibinfo {pages}
  {509} (\bibinfo {year} {2016})}\BibitemShut {NoStop}%
\bibitem [{\citenamefont {Bychkov}\ and\ \citenamefont
  {Rashba}(1984)}]{Bychkov1984}%
  \BibitemOpen
  \bibfield  {author} {\bibinfo {author} {\bibfnamefont {Y.~A.}\ \bibnamefont
  {Bychkov}}\ and\ \bibinfo {author} {\bibfnamefont {E.~I.}\ \bibnamefont
  {Rashba}},\ }\bibfield  {title} {\bibinfo {title} {Oscillatory effects and
  the magnetic susceptibility of carriers in inversion layers},\ }\href
  {https://doi.org/10.1088/0022-3719/17/33/015} {\bibfield  {journal} {\bibinfo
   {journal} {J. Phys. C}\ }\textbf {\bibinfo {volume} {17}},\ \bibinfo {pages}
  {6039} (\bibinfo {year} {1984})}\BibitemShut {NoStop}%
\bibitem [{\citenamefont {Rashba}(2015)}]{Rashba2015}%
  \BibitemOpen
  \bibfield  {author} {\bibinfo {author} {\bibfnamefont {E.~I.}\ \bibnamefont
  {Rashba}},\ }\bibfield  {title} {\bibinfo {title} {{Semiconductors with a
  loop of extrema}},\ }\href {https://doi.org/10.1016/j.elspec.2014.10.002}
  {\bibfield  {journal} {\bibinfo  {journal} {J. Electron Spectros. Relat.
  Phenomena}\ }\textbf {\bibinfo {volume} {201}},\ \bibinfo {pages} {4}
  (\bibinfo {year} {2015})}\BibitemShut {NoStop}%
\bibitem [{\citenamefont {Winkler}(2003)}]{Winkler2003}%
  \BibitemOpen
  \bibfield  {author} {\bibinfo {author} {\bibfnamefont {R.}~\bibnamefont
  {Winkler}},\ }\href@noop {} {\emph {\bibinfo {title} {{Spin-Orbit Coupling
  Effects in Two-Dimensional Electron and Hole Systems}}}}\ (\bibinfo
  {publisher} {Springer Tracts Modern Physics Vol. 191 (Springer, Berlin,
  2003)},\ \bibinfo {year} {2003})\BibitemShut {NoStop}%
\bibitem [{\citenamefont {Manchon}\ \emph {et~al.}(2015)\citenamefont
  {Manchon}, \citenamefont {Koo}, \citenamefont {Nitta}, \citenamefont
  {Frolov},\ and\ \citenamefont {Duine}}]{Manchon2015}%
  \BibitemOpen
  \bibfield  {author} {\bibinfo {author} {\bibfnamefont {A.}~\bibnamefont
  {Manchon}}, \bibinfo {author} {\bibfnamefont {H.~C.}\ \bibnamefont {Koo}},
  \bibinfo {author} {\bibfnamefont {J.}~\bibnamefont {Nitta}}, \bibinfo
  {author} {\bibfnamefont {S.~M.}\ \bibnamefont {Frolov}},\ and\ \bibinfo
  {author} {\bibfnamefont {R.~A.}\ \bibnamefont {Duine}},\ }\bibfield  {title}
  {\bibinfo {title} {New perspectives for rashba spin-orbit coupling},\ }\href
  {https://doi.org/https://doi.org/10.1038/nmat4360} {\bibfield  {journal}
  {\bibinfo  {journal} {Nat. Mater.}\ }\textbf {\bibinfo {volume} {14}},\
  \bibinfo {pages} {871} (\bibinfo {year} {2015})}\BibitemShut {NoStop}%
\bibitem [{\citenamefont {Gambardella}\ and\ \citenamefont
  {Miron}(2011)}]{Gambardella2011}%
  \BibitemOpen
  \bibfield  {author} {\bibinfo {author} {\bibfnamefont {P.}~\bibnamefont
  {Gambardella}}\ and\ \bibinfo {author} {\bibfnamefont {I.~M.}\ \bibnamefont
  {Miron}},\ }\bibfield  {title} {\bibinfo {title} {Current-induced
  spin–orbit torques},\ }\href {https://doi.org/10.1098/rsta.2010.0336}
  {\bibfield  {journal} {\bibinfo  {journal} {Philos. Trans. R. Soc. London A}\
  }\textbf {\bibinfo {volume} {369}},\ \bibinfo {pages} {3175} (\bibinfo {year}
  {2011})}\BibitemShut {NoStop}%
\bibitem [{\citenamefont {Manchon}\ \emph {et~al.}(2019)\citenamefont
  {Manchon}, \citenamefont {\ifmmode~\check{Z}\else \v{Z}\fi{}elezn\'y},
  \citenamefont {Miron}, \citenamefont {Jungwirth}, \citenamefont {Sinova},
  \citenamefont {Thiaville}, \citenamefont {Garello},\ and\ \citenamefont
  {Gambardella}}]{Manchon2019}%
  \BibitemOpen
  \bibfield  {author} {\bibinfo {author} {\bibfnamefont {A.}~\bibnamefont
  {Manchon}}, \bibinfo {author} {\bibfnamefont {J.}~\bibnamefont
  {\ifmmode~\check{Z}\else \v{Z}\fi{}elezn\'y}}, \bibinfo {author}
  {\bibfnamefont {I.~M.}\ \bibnamefont {Miron}}, \bibinfo {author}
  {\bibfnamefont {T.}~\bibnamefont {Jungwirth}}, \bibinfo {author}
  {\bibfnamefont {J.}~\bibnamefont {Sinova}}, \bibinfo {author} {\bibfnamefont
  {A.}~\bibnamefont {Thiaville}}, \bibinfo {author} {\bibfnamefont
  {K.}~\bibnamefont {Garello}},\ and\ \bibinfo {author} {\bibfnamefont
  {P.}~\bibnamefont {Gambardella}},\ }\bibfield  {title} {\bibinfo {title}
  {Current-induced spin-orbit torques in ferromagnetic and antiferromagnetic
  systems},\ }\href {https://doi.org/10.1103/RevModPhys.91.035004} {\bibfield
  {journal} {\bibinfo  {journal} {Rev. Mod. Phys.}\ }\textbf {\bibinfo {volume}
  {91}},\ \bibinfo {pages} {035004} (\bibinfo {year} {2019})}\BibitemShut
  {NoStop}%
\bibitem [{\citenamefont {Rojas-S{\'a}nchez}\ \emph {et~al.}(2013)\citenamefont
  {Rojas-S{\'a}nchez}, \citenamefont {Vila}, \citenamefont {Desfonds},
  \citenamefont {Gambarelli}, \citenamefont {Attan{\'e}}, \citenamefont
  {De~Teresa}, \citenamefont {Mag{\'e}n},\ and\ \citenamefont
  {Fert}}]{Sanchez2013}%
  \BibitemOpen
  \bibfield  {author} {\bibinfo {author} {\bibfnamefont {J.-C.~R.}\
  \bibnamefont {Rojas-S{\'a}nchez}}, \bibinfo {author} {\bibfnamefont
  {L.}~\bibnamefont {Vila}}, \bibinfo {author} {\bibfnamefont {G.}~\bibnamefont
  {Desfonds}}, \bibinfo {author} {\bibfnamefont {S.}~\bibnamefont
  {Gambarelli}}, \bibinfo {author} {\bibfnamefont {J.~P.}\ \bibnamefont
  {Attan{\'e}}}, \bibinfo {author} {\bibfnamefont {J.~M.}\ \bibnamefont
  {De~Teresa}}, \bibinfo {author} {\bibfnamefont {C.}~\bibnamefont
  {Mag{\'e}n}},\ and\ \bibinfo {author} {\bibfnamefont {A.}~\bibnamefont
  {Fert}},\ }\bibfield  {title} {\bibinfo {title} {Spin-to-charge conversion
  using rashba coupling at the interface between non-magnetic materials},\
  }\href {https://doi.org/https://doi.org/10.1038/ncomms3944} {\bibfield
  {journal} {\bibinfo  {journal} {Nat. Commun.}\ }\textbf {\bibinfo {volume}
  {4}},\ \bibinfo {pages} {2944} (\bibinfo {year} {2013})}\BibitemShut
  {NoStop}%
\bibitem [{\citenamefont {Shen}\ \emph
  {et~al.}(2014{\natexlab{a}})\citenamefont {Shen}, \citenamefont {Vignale},\
  and\ \citenamefont {Raimondi}}]{Shen2014a}%
  \BibitemOpen
  \bibfield  {author} {\bibinfo {author} {\bibfnamefont {K.}~\bibnamefont
  {Shen}}, \bibinfo {author} {\bibfnamefont {G.}~\bibnamefont {Vignale}},\ and\
  \bibinfo {author} {\bibfnamefont {R.}~\bibnamefont {Raimondi}},\ }\bibfield
  {title} {\bibinfo {title} {Microscopic theory of the inverse edelstein
  effect},\ }\href {https://doi.org/10.1103/PhysRevLett.112.096601} {\bibfield
  {journal} {\bibinfo  {journal} {Phys. Rev. Lett.}\ }\textbf {\bibinfo
  {volume} {112}},\ \bibinfo {pages} {096601} (\bibinfo {year}
  {2014}{\natexlab{a}})}\BibitemShut {NoStop}%
\bibitem [{\citenamefont {Ganichev}\ \emph {et~al.}(2002)\citenamefont
  {Ganichev}, \citenamefont {Ivchenko}, \citenamefont {Bel'kov}, \citenamefont
  {Tarasenko}, \citenamefont {Sollinger}, \citenamefont {Weiss}, \citenamefont
  {Wegscheider},\ and\ \citenamefont {Prettl}}]{Ganichev2002}%
  \BibitemOpen
  \bibfield  {author} {\bibinfo {author} {\bibfnamefont {S.~D.}\ \bibnamefont
  {Ganichev}}, \bibinfo {author} {\bibfnamefont {E.~L.}\ \bibnamefont
  {Ivchenko}}, \bibinfo {author} {\bibfnamefont {V.~V.}\ \bibnamefont
  {Bel'kov}}, \bibinfo {author} {\bibfnamefont {S.~A.}\ \bibnamefont
  {Tarasenko}}, \bibinfo {author} {\bibfnamefont {M.}~\bibnamefont
  {Sollinger}}, \bibinfo {author} {\bibfnamefont {D.}~\bibnamefont {Weiss}},
  \bibinfo {author} {\bibfnamefont {W.}~\bibnamefont {Wegscheider}},\ and\
  \bibinfo {author} {\bibfnamefont {W.}~\bibnamefont {Prettl}},\ }\bibfield
  {title} {\bibinfo {title} {Spin-galvanic effect},\ }\href
  {https://doi.org/https://doi.org/10.1038/417153a} {\bibfield  {journal}
  {\bibinfo  {journal} {Nature (London)}\ }\textbf {\bibinfo {volume} {417}},\
  \bibinfo {pages} {153} (\bibinfo {year} {2002})}\BibitemShut {NoStop}%
\bibitem [{\citenamefont {Ivchenko}\ \emph {et~al.}(1989)\citenamefont
  {Ivchenko}, \citenamefont {Lyanda-Geller},\ and\ \citenamefont
  {Pikus}}]{Ivchenko1989}%
  \BibitemOpen
  \bibfield  {author} {\bibinfo {author} {\bibfnamefont {E.~L.}\ \bibnamefont
  {Ivchenko}}, \bibinfo {author} {\bibfnamefont {Y.~B.}\ \bibnamefont
  {Lyanda-Geller}},\ and\ \bibinfo {author} {\bibfnamefont {G.~E.}\
  \bibnamefont {Pikus}},\ }\bibfield  {title} {\bibinfo {title} {Photocurrent
  in structures with quantum wells with an optical orientation of free
  carriers},\ }\href@noop {} {\bibfield  {journal} {\bibinfo  {journal} {JETP
  Lett.}\ }\textbf {\bibinfo {volume} {50}},\ \bibinfo {pages} {175} (\bibinfo
  {year} {1989})}\BibitemShut {NoStop}%
\bibitem [{\citenamefont {Ivchenko}\ \emph {et~al.}(1990)\citenamefont
  {Ivchenko}, \citenamefont {Lyanda-Geller},\ and\ \citenamefont
  {Pikus}}]{Ivchenko1990}%
  \BibitemOpen
  \bibfield  {author} {\bibinfo {author} {\bibfnamefont {E.~L.}\ \bibnamefont
  {Ivchenko}}, \bibinfo {author} {\bibfnamefont {Y.~B.}\ \bibnamefont
  {Lyanda-Geller}},\ and\ \bibinfo {author} {\bibfnamefont {G.~E.}\
  \bibnamefont {Pikus}},\ }\bibfield  {title} {\bibinfo {title} {Current of
  thermalized spin-oriented photocarriers},\ }\href@noop {} {\bibfield
  {journal} {\bibinfo  {journal} {Sov. Phys. JETP}\ }\textbf {\bibinfo {volume}
  {71}},\ \bibinfo {pages} {550} (\bibinfo {year} {1990})}\BibitemShut
  {NoStop}%
\bibitem [{\citenamefont {Ganichev}\ \emph {et~al.}(2001)\citenamefont
  {Ganichev}, \citenamefont {Ivchenko}, \citenamefont {Danilov}, \citenamefont
  {Eroms}, \citenamefont {Wegscheider}, \citenamefont {Weiss},\ and\
  \citenamefont {Prettl}}]{Ganichev2001}%
  \BibitemOpen
  \bibfield  {author} {\bibinfo {author} {\bibfnamefont {S.~D.}\ \bibnamefont
  {Ganichev}}, \bibinfo {author} {\bibfnamefont {E.~L.}\ \bibnamefont
  {Ivchenko}}, \bibinfo {author} {\bibfnamefont {S.~N.}\ \bibnamefont
  {Danilov}}, \bibinfo {author} {\bibfnamefont {J.}~\bibnamefont {Eroms}},
  \bibinfo {author} {\bibfnamefont {W.}~\bibnamefont {Wegscheider}}, \bibinfo
  {author} {\bibfnamefont {D.}~\bibnamefont {Weiss}},\ and\ \bibinfo {author}
  {\bibfnamefont {W.}~\bibnamefont {Prettl}},\ }\bibfield  {title} {\bibinfo
  {title} {Conversion of spin into directed electric current in quantum
  wells},\ }\href {https://doi.org/10.1103/PhysRevLett.86.4358} {\bibfield
  {journal} {\bibinfo  {journal} {Phys. Rev. Lett.}\ }\textbf {\bibinfo
  {volume} {86}},\ \bibinfo {pages} {4358} (\bibinfo {year}
  {2001})}\BibitemShut {NoStop}%
\bibitem [{\citenamefont {Ganichev}\ and\ \citenamefont
  {Prettl}(2003)}]{Ganichev2003b}%
  \BibitemOpen
  \bibfield  {author} {\bibinfo {author} {\bibfnamefont {S.~D.}\ \bibnamefont
  {Ganichev}}\ and\ \bibinfo {author} {\bibfnamefont {W.}~\bibnamefont
  {Prettl}},\ }\bibfield  {title} {\bibinfo {title} {Spin photocurrents in
  quantum wells},\ }\href {https://doi.org/10.1088/0953-8984/15/20/204}
  {\bibfield  {journal} {\bibinfo  {journal} {J. Phys. Condens. Matter}\
  }\textbf {\bibinfo {volume} {15}},\ \bibinfo {pages} {R935} (\bibinfo {year}
  {2003})}\BibitemShut {NoStop}%
\bibitem [{\citenamefont {Burkov}\ \emph {et~al.}(2004)\citenamefont {Burkov},
  \citenamefont {N\'u\~nez},\ and\ \citenamefont {MacDonald}}]{Burkov2004}%
  \BibitemOpen
  \bibfield  {author} {\bibinfo {author} {\bibfnamefont {A.~A.}\ \bibnamefont
  {Burkov}}, \bibinfo {author} {\bibfnamefont {A.~S.}\ \bibnamefont
  {N\'u\~nez}},\ and\ \bibinfo {author} {\bibfnamefont {A.~H.}\ \bibnamefont
  {MacDonald}},\ }\bibfield  {title} {\bibinfo {title} {Theory of
  spin-charge-coupled transport in a two-dimensional electron gas with rashba
  spin-orbit interactions},\ }\href
  {https://doi.org/10.1103/PhysRevB.70.155308} {\bibfield  {journal} {\bibinfo
  {journal} {Phys. Rev. B}\ }\textbf {\bibinfo {volume} {70}},\ \bibinfo
  {pages} {155308} (\bibinfo {year} {2004})}\BibitemShut {NoStop}%
\bibitem [{\citenamefont {Fabian}\ \emph {et~al.}(2007)\citenamefont {Fabian},
  \citenamefont {Matos-Abiague}, \citenamefont {Ertler}, \citenamefont
  {Stano},\ and\ \citenamefont {\v{Z}uti{\'c}}}]{Fabian2007}%
  \BibitemOpen
  \bibfield  {author} {\bibinfo {author} {\bibfnamefont {J.}~\bibnamefont
  {Fabian}}, \bibinfo {author} {\bibfnamefont {A.}~\bibnamefont
  {Matos-Abiague}}, \bibinfo {author} {\bibfnamefont {C.}~\bibnamefont
  {Ertler}}, \bibinfo {author} {\bibfnamefont {P.}~\bibnamefont {Stano}},\ and\
  \bibinfo {author} {\bibfnamefont {I.}~\bibnamefont {\v{Z}uti{\'c}}},\
  }\bibfield  {title} {\bibinfo {title} {Semiconductor spintronics},\
  }\href@noop {} {\bibfield  {journal} {\bibinfo  {journal} {Acta Phys. Slov.}\
  }\textbf {\bibinfo {volume} {57}},\ \bibinfo {pages} {565} (\bibinfo {year}
  {2007})}\BibitemShut {NoStop}%
\bibitem [{\citenamefont {Awschalom}\ and\ \citenamefont
  {Flatt{\'e}}(2007)}]{Awschalom2007}%
  \BibitemOpen
  \bibfield  {author} {\bibinfo {author} {\bibfnamefont {D.~D.}\ \bibnamefont
  {Awschalom}}\ and\ \bibinfo {author} {\bibfnamefont {M.~E.}\ \bibnamefont
  {Flatt{\'e}}},\ }\bibfield  {title} {\bibinfo {title} {Challenges for
  semiconductor spintronics},\ }\href
  {https://doi.org/https://doi.org/10.1038/nphys551} {\bibfield  {journal}
  {\bibinfo  {journal} {Nat. Phys.}\ }\textbf {\bibinfo {volume} {3}},\
  \bibinfo {pages} {153} (\bibinfo {year} {2007})}\BibitemShut {NoStop}%
\bibitem [{\citenamefont {Kohda}\ and\ \citenamefont
  {Salis}(2017)}]{Kohda2017}%
  \BibitemOpen
  \bibfield  {author} {\bibinfo {author} {\bibfnamefont {M.}~\bibnamefont
  {Kohda}}\ and\ \bibinfo {author} {\bibfnamefont {G.}~\bibnamefont {Salis}},\
  }\bibfield  {title} {\bibinfo {title} {Physics and application of persistent
  spin helix state in semiconductor heterostructures},\ }\href
  {https://doi.org/10.1088/1361-6641/aa5dd6} {\bibfield  {journal} {\bibinfo
  {journal} {Semicond. Sci. Technol.}\ }\textbf {\bibinfo {volume} {32}},\
  \bibinfo {pages} {073002} (\bibinfo {year} {2017})}\BibitemShut {NoStop}%
\bibitem [{\citenamefont {Dieny}\ \emph {et~al.}(2020)\citenamefont {Dieny},
  \citenamefont {Prejbeanu}, \citenamefont {Garello}, \citenamefont
  {Gambardella}, \citenamefont {Freitas}, \citenamefont {Lehndorff},
  \citenamefont {Raberg}, \citenamefont {Ebels}, \citenamefont {Demokritov},
  \citenamefont {Akerman}, \citenamefont {Deac}, \citenamefont {Pirro},
  \citenamefont {Adelmann}, \citenamefont {Anane}, \citenamefont {Chumak},
  \citenamefont {Hirohata}, \citenamefont {Mangin}, \citenamefont {Valenzuela},
  \citenamefont {Cengiz~Onba\c{s}l{\i}}, \citenamefont {d'Aquino},
  \citenamefont {Prenat}, \citenamefont {Finocchio}, \citenamefont
  {Lopez-Diaz}, \citenamefont {Chantrell}, \citenamefont {Chubykalo-Fesenko},\
  and\ \citenamefont {Bortolotti}}]{Dieny2020}%
  \BibitemOpen
  \bibfield  {author} {\bibinfo {author} {\bibfnamefont {B.}~\bibnamefont
  {Dieny}}, \bibinfo {author} {\bibfnamefont {I.~L.}\ \bibnamefont
  {Prejbeanu}}, \bibinfo {author} {\bibfnamefont {K.}~\bibnamefont {Garello}},
  \bibinfo {author} {\bibfnamefont {P.}~\bibnamefont {Gambardella}}, \bibinfo
  {author} {\bibfnamefont {P.}~\bibnamefont {Freitas}}, \bibinfo {author}
  {\bibfnamefont {R.}~\bibnamefont {Lehndorff}}, \bibinfo {author}
  {\bibfnamefont {W.}~\bibnamefont {Raberg}}, \bibinfo {author} {\bibfnamefont
  {U.}~\bibnamefont {Ebels}}, \bibinfo {author} {\bibfnamefont {S.~O.}\
  \bibnamefont {Demokritov}}, \bibinfo {author} {\bibfnamefont
  {J.}~\bibnamefont {Akerman}}, \bibinfo {author} {\bibfnamefont
  {A.}~\bibnamefont {Deac}}, \bibinfo {author} {\bibfnamefont {P.}~\bibnamefont
  {Pirro}}, \bibinfo {author} {\bibfnamefont {C.}~\bibnamefont {Adelmann}},
  \bibinfo {author} {\bibfnamefont {A.}~\bibnamefont {Anane}}, \bibinfo
  {author} {\bibfnamefont {A.~V.}\ \bibnamefont {Chumak}}, \bibinfo {author}
  {\bibfnamefont {A.}~\bibnamefont {Hirohata}}, \bibinfo {author}
  {\bibfnamefont {S.}~\bibnamefont {Mangin}}, \bibinfo {author} {\bibfnamefont
  {S.~O.}\ \bibnamefont {Valenzuela}}, \bibinfo {author} {\bibfnamefont
  {M.}~\bibnamefont {Cengiz~Onba\c{s}l{\i}}}, \bibinfo {author} {\bibfnamefont
  {M.}~\bibnamefont {d'Aquino}}, \bibinfo {author} {\bibfnamefont
  {G.}~\bibnamefont {Prenat}}, \bibinfo {author} {\bibfnamefont
  {G.}~\bibnamefont {Finocchio}}, \bibinfo {author} {\bibfnamefont
  {L.}~\bibnamefont {Lopez-Diaz}}, \bibinfo {author} {\bibfnamefont
  {R.}~\bibnamefont {Chantrell}}, \bibinfo {author} {\bibfnamefont
  {O.}~\bibnamefont {Chubykalo-Fesenko}},\ and\ \bibinfo {author}
  {\bibfnamefont {P.}~\bibnamefont {Bortolotti}},\ }\bibfield  {title}
  {\bibinfo {title} {Opportunities and challenges for spintronics in the
  microelectronics industry},\ }\href
  {https://doi.org/https://doi.org/10.1038/s41928-020-0461-5} {\bibfield
  {journal} {\bibinfo  {journal} {Nature Electronics}\ }\textbf {\bibinfo
  {volume} {3}},\ \bibinfo {pages} {446} (\bibinfo {year} {2020})}\BibitemShut
  {NoStop}%
\bibitem [{\citenamefont {Vicente-Arche}\ \emph {et~al.}(2021)\citenamefont
  {Vicente-Arche}, \citenamefont {Bréhin}, \citenamefont {Varotto},
  \citenamefont {Cosset-Cheneau}, \citenamefont {Mallik}, \citenamefont
  {Salazar}, \citenamefont {Noël}, \citenamefont {Vaz}, \citenamefont {Trier},
  \citenamefont {Bhattacharya}, \citenamefont {Sander}, \citenamefont
  {Le~Fèvre}, \citenamefont {Bertran}, \citenamefont {Saiz}, \citenamefont
  {Ménard}, \citenamefont {Bergeal}, \citenamefont {Barthélémy},
  \citenamefont {Li}, \citenamefont {Lin}, \citenamefont {Nikonov},
  \citenamefont {Young}, \citenamefont {Rault}, \citenamefont {Vila},
  \citenamefont {Attané},\ and\ \citenamefont {Bibes}}]{VicenteArche21}%
  \BibitemOpen
  \bibfield  {author} {\bibinfo {author} {\bibfnamefont {L.~M.}\ \bibnamefont
  {Vicente-Arche}}, \bibinfo {author} {\bibfnamefont {J.}~\bibnamefont
  {Bréhin}}, \bibinfo {author} {\bibfnamefont {S.}~\bibnamefont {Varotto}},
  \bibinfo {author} {\bibfnamefont {M.}~\bibnamefont {Cosset-Cheneau}},
  \bibinfo {author} {\bibfnamefont {S.}~\bibnamefont {Mallik}}, \bibinfo
  {author} {\bibfnamefont {R.}~\bibnamefont {Salazar}}, \bibinfo {author}
  {\bibfnamefont {P.}~\bibnamefont {Noël}}, \bibinfo {author} {\bibfnamefont
  {D.~C.}\ \bibnamefont {Vaz}}, \bibinfo {author} {\bibfnamefont
  {F.}~\bibnamefont {Trier}}, \bibinfo {author} {\bibfnamefont
  {S.}~\bibnamefont {Bhattacharya}}, \bibinfo {author} {\bibfnamefont
  {A.}~\bibnamefont {Sander}}, \bibinfo {author} {\bibfnamefont
  {P.}~\bibnamefont {Le~Fèvre}}, \bibinfo {author} {\bibfnamefont
  {F.}~\bibnamefont {Bertran}}, \bibinfo {author} {\bibfnamefont
  {G.}~\bibnamefont {Saiz}}, \bibinfo {author} {\bibfnamefont {G.}~\bibnamefont
  {Ménard}}, \bibinfo {author} {\bibfnamefont {N.}~\bibnamefont {Bergeal}},
  \bibinfo {author} {\bibfnamefont {A.}~\bibnamefont {Barthélémy}}, \bibinfo
  {author} {\bibfnamefont {H.}~\bibnamefont {Li}}, \bibinfo {author}
  {\bibfnamefont {C.-C.}\ \bibnamefont {Lin}}, \bibinfo {author} {\bibfnamefont
  {D.~E.}\ \bibnamefont {Nikonov}}, \bibinfo {author} {\bibfnamefont {I.~A.}\
  \bibnamefont {Young}}, \bibinfo {author} {\bibfnamefont {J.~E.}\ \bibnamefont
  {Rault}}, \bibinfo {author} {\bibfnamefont {L.}~\bibnamefont {Vila}},
  \bibinfo {author} {\bibfnamefont {J.-P.}\ \bibnamefont {Attané}},\ and\
  \bibinfo {author} {\bibfnamefont {M.}~\bibnamefont {Bibes}},\ }\bibfield
  {title} {\bibinfo {title} {{Spin–Charge Interconversion in KTaO3 2D
  Electron Gases}},\ }\href
  {https://doi.org/https://doi.org/10.1002/adma.202102102} {\bibfield
  {journal} {\bibinfo  {journal} {Adv. Mater.}\ }\textbf {\bibinfo {volume}
  {33}},\ \bibinfo {pages} {2102102} (\bibinfo {year} {2021})}\BibitemShut
  {NoStop}%
\bibitem [{\citenamefont {Gupta}\ \emph {et~al.}(2022)\citenamefont {Gupta},
  \citenamefont {Silotia}, \citenamefont {Kumari}, \citenamefont {Dumen},
  \citenamefont {Goyal}, \citenamefont {Tomar}, \citenamefont {Wadehra},
  \citenamefont {Ayyub},\ and\ \citenamefont {Chakraverty}}]{Gupta22}%
  \BibitemOpen
  \bibfield  {author} {\bibinfo {author} {\bibfnamefont {A.}~\bibnamefont
  {Gupta}}, \bibinfo {author} {\bibfnamefont {H.}~\bibnamefont {Silotia}},
  \bibinfo {author} {\bibfnamefont {A.}~\bibnamefont {Kumari}}, \bibinfo
  {author} {\bibfnamefont {M.}~\bibnamefont {Dumen}}, \bibinfo {author}
  {\bibfnamefont {S.}~\bibnamefont {Goyal}}, \bibinfo {author} {\bibfnamefont
  {R.}~\bibnamefont {Tomar}}, \bibinfo {author} {\bibfnamefont
  {N.}~\bibnamefont {Wadehra}}, \bibinfo {author} {\bibfnamefont
  {P.}~\bibnamefont {Ayyub}},\ and\ \bibinfo {author} {\bibfnamefont
  {S.}~\bibnamefont {Chakraverty}},\ }\bibfield  {title} {\bibinfo {title}
  {{KTaO3--The New Kid on the Spintronics Block}},\ }\href
  {https://doi.org/https://doi.org/10.1002/adma.202106481} {\bibfield
  {journal} {\bibinfo  {journal} {Adv. Mater.}\ }\textbf {\bibinfo {volume}
  {34}},\ \bibinfo {pages} {2106481} (\bibinfo {year} {2022})}\BibitemShut
  {NoStop}%
\bibitem [{\citenamefont {Tserkovnyak}\ \emph {et~al.}(2002)\citenamefont
  {Tserkovnyak}, \citenamefont {Brataas},\ and\ \citenamefont
  {Bauer}}]{Tserkovnyak2002}%
  \BibitemOpen
  \bibfield  {author} {\bibinfo {author} {\bibfnamefont {Y.}~\bibnamefont
  {Tserkovnyak}}, \bibinfo {author} {\bibfnamefont {A.}~\bibnamefont
  {Brataas}},\ and\ \bibinfo {author} {\bibfnamefont {G.~E.~W.}\ \bibnamefont
  {Bauer}},\ }\bibfield  {title} {\bibinfo {title} {Enhanced gilbert damping in
  thin ferromagnetic films},\ }\href
  {https://doi.org/10.1103/PhysRevLett.88.117601} {\bibfield  {journal}
  {\bibinfo  {journal} {Phys. Rev. Lett.}\ }\textbf {\bibinfo {volume} {88}},\
  \bibinfo {pages} {117601} (\bibinfo {year} {2002})}\BibitemShut {NoStop}%
\bibitem [{\citenamefont {Tserkovnyak}\ \emph {et~al.}(2005)\citenamefont
  {Tserkovnyak}, \citenamefont {Brataas}, \citenamefont {Bauer},\ and\
  \citenamefont {Halperin}}]{Tserkovnyak2005}%
  \BibitemOpen
  \bibfield  {author} {\bibinfo {author} {\bibfnamefont {Y.}~\bibnamefont
  {Tserkovnyak}}, \bibinfo {author} {\bibfnamefont {A.}~\bibnamefont
  {Brataas}}, \bibinfo {author} {\bibfnamefont {G.~E.~W.}\ \bibnamefont
  {Bauer}},\ and\ \bibinfo {author} {\bibfnamefont {B.~I.}\ \bibnamefont
  {Halperin}},\ }\bibfield  {title} {\bibinfo {title} {Nonlocal magnetization
  dynamics in ferromagnetic heterostructures},\ }\href
  {https://doi.org/10.1103/RevModPhys.77.1375} {\bibfield  {journal} {\bibinfo
  {journal} {Rev. Mod. Phys.}\ }\textbf {\bibinfo {volume} {77}},\ \bibinfo
  {pages} {1375} (\bibinfo {year} {2005})}\BibitemShut {NoStop}%
\bibitem [{\citenamefont {Hellman}\ \emph {et~al.}(2017)\citenamefont
  {Hellman}, \citenamefont {Hoffmann}, \citenamefont {Tserkovnyak},
  \citenamefont {Beach}, \citenamefont {Fullerton}, \citenamefont {Leighton},
  \citenamefont {MacDonald}, \citenamefont {Ralph}, \citenamefont {Arena},
  \citenamefont {D\"urr}, \citenamefont {Fischer}, \citenamefont {Grollier},
  \citenamefont {Heremans}, \citenamefont {Jungwirth}, \citenamefont {Kimel},
  \citenamefont {Koopmans}, \citenamefont {Krivorotov}, \citenamefont {May},
  \citenamefont {Petford-Long}, \citenamefont {Rondinelli}, \citenamefont
  {Samarth}, \citenamefont {Schuller}, \citenamefont {Slavin}, \citenamefont
  {Stiles}, \citenamefont {Tchernyshyov}, \citenamefont {Thiaville},\ and\
  \citenamefont {Zink}}]{Hellman2017}%
  \BibitemOpen
  \bibfield  {author} {\bibinfo {author} {\bibfnamefont {F.}~\bibnamefont
  {Hellman}}, \bibinfo {author} {\bibfnamefont {A.}~\bibnamefont {Hoffmann}},
  \bibinfo {author} {\bibfnamefont {Y.}~\bibnamefont {Tserkovnyak}}, \bibinfo
  {author} {\bibfnamefont {G.~S.~D.}\ \bibnamefont {Beach}}, \bibinfo {author}
  {\bibfnamefont {E.~E.}\ \bibnamefont {Fullerton}}, \bibinfo {author}
  {\bibfnamefont {C.}~\bibnamefont {Leighton}}, \bibinfo {author}
  {\bibfnamefont {A.~H.}\ \bibnamefont {MacDonald}}, \bibinfo {author}
  {\bibfnamefont {D.~C.}\ \bibnamefont {Ralph}}, \bibinfo {author}
  {\bibfnamefont {D.~A.}\ \bibnamefont {Arena}}, \bibinfo {author}
  {\bibfnamefont {H.~A.}\ \bibnamefont {D\"urr}}, \bibinfo {author}
  {\bibfnamefont {P.}~\bibnamefont {Fischer}}, \bibinfo {author} {\bibfnamefont
  {J.}~\bibnamefont {Grollier}}, \bibinfo {author} {\bibfnamefont {J.~P.}\
  \bibnamefont {Heremans}}, \bibinfo {author} {\bibfnamefont {T.}~\bibnamefont
  {Jungwirth}}, \bibinfo {author} {\bibfnamefont {A.~V.}\ \bibnamefont
  {Kimel}}, \bibinfo {author} {\bibfnamefont {B.}~\bibnamefont {Koopmans}},
  \bibinfo {author} {\bibfnamefont {I.~N.}\ \bibnamefont {Krivorotov}},
  \bibinfo {author} {\bibfnamefont {S.~J.}\ \bibnamefont {May}}, \bibinfo
  {author} {\bibfnamefont {A.~K.}\ \bibnamefont {Petford-Long}}, \bibinfo
  {author} {\bibfnamefont {J.~M.}\ \bibnamefont {Rondinelli}}, \bibinfo
  {author} {\bibfnamefont {N.}~\bibnamefont {Samarth}}, \bibinfo {author}
  {\bibfnamefont {I.~K.}\ \bibnamefont {Schuller}}, \bibinfo {author}
  {\bibfnamefont {A.~N.}\ \bibnamefont {Slavin}}, \bibinfo {author}
  {\bibfnamefont {M.~D.}\ \bibnamefont {Stiles}}, \bibinfo {author}
  {\bibfnamefont {O.}~\bibnamefont {Tchernyshyov}}, \bibinfo {author}
  {\bibfnamefont {A.}~\bibnamefont {Thiaville}},\ and\ \bibinfo {author}
  {\bibfnamefont {B.~L.}\ \bibnamefont {Zink}},\ }\bibfield  {title} {\bibinfo
  {title} {Interface-induced phenomena in magnetism},\ }\href
  {https://doi.org/10.1103/RevModPhys.89.025006} {\bibfield  {journal}
  {\bibinfo  {journal} {Rev. Mod. Phys.}\ }\textbf {\bibinfo {volume} {89}},\
  \bibinfo {pages} {025006} (\bibinfo {year} {2017})}\BibitemShut {NoStop}%
\bibitem [{\citenamefont {Nomura}\ \emph {et~al.}(2015)\citenamefont {Nomura},
  \citenamefont {Tashiro}, \citenamefont {Nakayama},\ and\ \citenamefont
  {Ando}}]{Nomura2015}%
  \BibitemOpen
  \bibfield  {author} {\bibinfo {author} {\bibfnamefont {A.}~\bibnamefont
  {Nomura}}, \bibinfo {author} {\bibfnamefont {T.}~\bibnamefont {Tashiro}},
  \bibinfo {author} {\bibfnamefont {H.}~\bibnamefont {Nakayama}},\ and\
  \bibinfo {author} {\bibfnamefont {K.}~\bibnamefont {Ando}},\ }\bibfield
  {title} {\bibinfo {title} {{Temperature dependence of inverse
  Rashba-Edelstein effect at metallic interface}},\ }\href
  {https://doi.org/10.1063/1.4921765} {\bibfield  {journal} {\bibinfo
  {journal} {Appl. Phys. Lett.}\ }\textbf {\bibinfo {volume} {106}},\ \bibinfo
  {pages} {212403} (\bibinfo {year} {2015})}\BibitemShut {NoStop}%
\bibitem [{\citenamefont {Sangiao}\ \emph {et~al.}(2015)\citenamefont
  {Sangiao}, \citenamefont {De~Teresa}, \citenamefont {Morellon}, \citenamefont
  {Lucas}, \citenamefont {Mart{\'\i}nez-Velarte},\ and\ \citenamefont
  {Viret}}]{Sangiao2015}%
  \BibitemOpen
  \bibfield  {author} {\bibinfo {author} {\bibfnamefont {S.}~\bibnamefont
  {Sangiao}}, \bibinfo {author} {\bibfnamefont {J.~M.}\ \bibnamefont
  {De~Teresa}}, \bibinfo {author} {\bibfnamefont {L.}~\bibnamefont {Morellon}},
  \bibinfo {author} {\bibfnamefont {I.}~\bibnamefont {Lucas}}, \bibinfo
  {author} {\bibfnamefont {M.~C.}\ \bibnamefont {Mart{\'\i}nez-Velarte}},\ and\
  \bibinfo {author} {\bibfnamefont {M.}~\bibnamefont {Viret}},\ }\bibfield
  {title} {\bibinfo {title} {Control of the spin to charge conversion using the
  inverse rashba-edelstein effect},\ }\href
  {https://doi.org/https://doi.org/10.1063/1.4919129} {\bibfield  {journal}
  {\bibinfo  {journal} {Appl. Phys. Lett.}\ }\textbf {\bibinfo {volume}
  {106}},\ \bibinfo {pages} {172403} (\bibinfo {year} {2015})}\BibitemShut
  {NoStop}%
\bibitem [{\citenamefont {Zhang}\ \emph {et~al.}(2015)\citenamefont {Zhang},
  \citenamefont {Jungfleisch}, \citenamefont {Jiang}, \citenamefont {Pearson},\
  and\ \citenamefont {Hoffmann}}]{Zhang2015}%
  \BibitemOpen
  \bibfield  {author} {\bibinfo {author} {\bibfnamefont {W.}~\bibnamefont
  {Zhang}}, \bibinfo {author} {\bibfnamefont {M.~B.}\ \bibnamefont
  {Jungfleisch}}, \bibinfo {author} {\bibfnamefont {W.}~\bibnamefont {Jiang}},
  \bibinfo {author} {\bibfnamefont {J.~E.}\ \bibnamefont {Pearson}},\ and\
  \bibinfo {author} {\bibfnamefont {A.}~\bibnamefont {Hoffmann}},\ }\bibfield
  {title} {\bibinfo {title} {{Spin pumping and inverse Rashba-Edelstein effect
  in NiFe/Ag/Bi and NiFe/Ag/Sb}},\ }\href {https://doi.org/10.1063/1.4915479}
  {\bibfield  {journal} {\bibinfo  {journal} {J. Appl. Phys.}\ }\textbf
  {\bibinfo {volume} {117}},\ \bibinfo {pages} {17C727} (\bibinfo {year}
  {2015})}\BibitemShut {NoStop}%
\bibitem [{\citenamefont {Matsushima}\ \emph {et~al.}(2017)\citenamefont
  {Matsushima}, \citenamefont {Ando}, \citenamefont {Dushenko}, \citenamefont
  {Ohshima}, \citenamefont {Kumamoto}, \citenamefont {Shinjo},\ and\
  \citenamefont {Shiraishi}}]{Matsushima2017}%
  \BibitemOpen
  \bibfield  {author} {\bibinfo {author} {\bibfnamefont {M.}~\bibnamefont
  {Matsushima}}, \bibinfo {author} {\bibfnamefont {Y.}~\bibnamefont {Ando}},
  \bibinfo {author} {\bibfnamefont {S.}~\bibnamefont {Dushenko}}, \bibinfo
  {author} {\bibfnamefont {R.}~\bibnamefont {Ohshima}}, \bibinfo {author}
  {\bibfnamefont {R.}~\bibnamefont {Kumamoto}}, \bibinfo {author}
  {\bibfnamefont {T.}~\bibnamefont {Shinjo}},\ and\ \bibinfo {author}
  {\bibfnamefont {M.}~\bibnamefont {Shiraishi}},\ }\bibfield  {title} {\bibinfo
  {title} {{Quantitative investigation of the inverse Rashba-Edelstein effect
  in Bi/Ag and Ag/Bi on YIG}},\ }\href {https://doi.org/10.1063/1.4976691}
  {\bibfield  {journal} {\bibinfo  {journal} {Appl. Phys. Lett.}\ }\textbf
  {\bibinfo {volume} {110}},\ \bibinfo {pages} {072404} (\bibinfo {year}
  {2017})}\BibitemShut {NoStop}%
\bibitem [{\citenamefont {Lesne}\ \emph {et~al.}(2016)\citenamefont {Lesne},
  \citenamefont {Fu}, \citenamefont {Oyarzun}, \citenamefont
  {Rojas-S{\'a}nchez}, \citenamefont {Vaz}, \citenamefont {Naganuma},
  \citenamefont {Sicoli}, \citenamefont {Attan{\'e}}, \citenamefont {Jamet},
  \citenamefont {Jacquet}, \citenamefont {George}, \citenamefont
  {Barth\'{e}l\'{e}my}, \citenamefont {Jaffr\`{e}s}, \citenamefont {Fert},
  \citenamefont {Bibes},\ and\ \citenamefont {Vila}}]{Lesne2016}%
  \BibitemOpen
  \bibfield  {author} {\bibinfo {author} {\bibfnamefont {E.}~\bibnamefont
  {Lesne}}, \bibinfo {author} {\bibfnamefont {Y.}~\bibnamefont {Fu}}, \bibinfo
  {author} {\bibfnamefont {S.}~\bibnamefont {Oyarzun}}, \bibinfo {author}
  {\bibfnamefont {J.~C.}\ \bibnamefont {Rojas-S{\'a}nchez}}, \bibinfo {author}
  {\bibfnamefont {D.~C.}\ \bibnamefont {Vaz}}, \bibinfo {author} {\bibfnamefont
  {H.}~\bibnamefont {Naganuma}}, \bibinfo {author} {\bibfnamefont
  {G.}~\bibnamefont {Sicoli}}, \bibinfo {author} {\bibfnamefont {J.-P.}\
  \bibnamefont {Attan{\'e}}}, \bibinfo {author} {\bibfnamefont
  {M.}~\bibnamefont {Jamet}}, \bibinfo {author} {\bibfnamefont
  {E.}~\bibnamefont {Jacquet}}, \bibinfo {author} {\bibfnamefont {J.-M.}\
  \bibnamefont {George}}, \bibinfo {author} {\bibfnamefont {A.}~\bibnamefont
  {Barth\'{e}l\'{e}my}}, \bibinfo {author} {\bibfnamefont {H.}~\bibnamefont
  {Jaffr\`{e}s}}, \bibinfo {author} {\bibfnamefont {A.}~\bibnamefont {Fert}},
  \bibinfo {author} {\bibfnamefont {M.}~\bibnamefont {Bibes}},\ and\ \bibinfo
  {author} {\bibfnamefont {L.}~\bibnamefont {Vila}},\ }\bibfield  {title}
  {\bibinfo {title} {Highly efficient and tunable spin-to-charge conversion
  through rashba coupling at oxide interfaces},\ }\href
  {https://doi.org/https://doi.org/10.1038/nmat4726} {\bibfield  {journal}
  {\bibinfo  {journal} {Nat. Mater.}\ }\textbf {\bibinfo {volume} {15}},\
  \bibinfo {pages} {1261} (\bibinfo {year} {2016})}\BibitemShut {NoStop}%
\bibitem [{\citenamefont {Song}\ \emph {et~al.}(2017)\citenamefont {Song},
  \citenamefont {Zhang}, \citenamefont {Su}, \citenamefont {Yuan},
  \citenamefont {Chen}, \citenamefont {Xing}, \citenamefont {Shi},
  \citenamefont {Sun},\ and\ \citenamefont {Han}}]{Song2017}%
  \BibitemOpen
  \bibfield  {author} {\bibinfo {author} {\bibfnamefont {Q.}~\bibnamefont
  {Song}}, \bibinfo {author} {\bibfnamefont {H.}~\bibnamefont {Zhang}},
  \bibinfo {author} {\bibfnamefont {T.}~\bibnamefont {Su}}, \bibinfo {author}
  {\bibfnamefont {W.}~\bibnamefont {Yuan}}, \bibinfo {author} {\bibfnamefont
  {Y.}~\bibnamefont {Chen}}, \bibinfo {author} {\bibfnamefont {W.}~\bibnamefont
  {Xing}}, \bibinfo {author} {\bibfnamefont {J.}~\bibnamefont {Shi}}, \bibinfo
  {author} {\bibfnamefont {J.}~\bibnamefont {Sun}},\ and\ \bibinfo {author}
  {\bibfnamefont {W.}~\bibnamefont {Han}},\ }\bibfield  {title} {\bibinfo
  {title} {Observation of inverse edelstein effect in rashba-split 2deg between
  srtio3 and laalo3 at room temperature},\ }\href
  {https://doi.org/10.1126/sciadv.1602312} {\bibfield  {journal} {\bibinfo
  {journal} {Sci. Adv.}\ }\textbf {\bibinfo {volume} {3}},\ \bibinfo {pages}
  {e1602312} (\bibinfo {year} {2017})}\BibitemShut {NoStop}%
\bibitem [{\citenamefont {Vaz}\ \emph {et~al.}(2019)\citenamefont {Vaz},
  \citenamefont {No{\"e}l}, \citenamefont {Johansson}, \citenamefont
  {G{\"o}bel}, \citenamefont {Bruno}, \citenamefont {Singh}, \citenamefont
  {Mckeown-Walker}, \citenamefont {Trier}, \citenamefont {Vicente-Arche},
  \citenamefont {Sander}, \citenamefont {Valencia}, \citenamefont {Bruneel},
  \citenamefont {Vivek}, \citenamefont {Gabay}, \citenamefont {Bergeal},
  \citenamefont {Baumberger}, \citenamefont {Okuno}, \citenamefont
  {Barth\'{e}l\'{e}my}, \citenamefont {Fert}, \citenamefont {Vila},
  \citenamefont {Mertig}, \citenamefont {Attan\'{e}},\ and\ \citenamefont
  {Bibes}}]{Vaz2019}%
  \BibitemOpen
  \bibfield  {author} {\bibinfo {author} {\bibfnamefont {D.~C.}\ \bibnamefont
  {Vaz}}, \bibinfo {author} {\bibfnamefont {P.}~\bibnamefont {No{\"e}l}},
  \bibinfo {author} {\bibfnamefont {A.}~\bibnamefont {Johansson}}, \bibinfo
  {author} {\bibfnamefont {B.}~\bibnamefont {G{\"o}bel}}, \bibinfo {author}
  {\bibfnamefont {F.~Y.}\ \bibnamefont {Bruno}}, \bibinfo {author}
  {\bibfnamefont {G.}~\bibnamefont {Singh}}, \bibinfo {author} {\bibfnamefont
  {S.}~\bibnamefont {Mckeown-Walker}}, \bibinfo {author} {\bibfnamefont
  {F.}~\bibnamefont {Trier}}, \bibinfo {author} {\bibfnamefont {L.~M.}\
  \bibnamefont {Vicente-Arche}}, \bibinfo {author} {\bibfnamefont
  {A.}~\bibnamefont {Sander}}, \bibinfo {author} {\bibfnamefont
  {S.}~\bibnamefont {Valencia}}, \bibinfo {author} {\bibfnamefont
  {P.}~\bibnamefont {Bruneel}}, \bibinfo {author} {\bibfnamefont
  {M.}~\bibnamefont {Vivek}}, \bibinfo {author} {\bibfnamefont
  {M.}~\bibnamefont {Gabay}}, \bibinfo {author} {\bibfnamefont
  {N.}~\bibnamefont {Bergeal}}, \bibinfo {author} {\bibfnamefont
  {F.}~\bibnamefont {Baumberger}}, \bibinfo {author} {\bibfnamefont
  {H.}~\bibnamefont {Okuno}}, \bibinfo {author} {\bibfnamefont
  {A.}~\bibnamefont {Barth\'{e}l\'{e}my}}, \bibinfo {author} {\bibfnamefont
  {A.}~\bibnamefont {Fert}}, \bibinfo {author} {\bibfnamefont {L.}~\bibnamefont
  {Vila}}, \bibinfo {author} {\bibfnamefont {I.}~\bibnamefont {Mertig}},
  \bibinfo {author} {\bibfnamefont {J.-P.}\ \bibnamefont {Attan\'{e}}},\ and\
  \bibinfo {author} {\bibfnamefont {M.}~\bibnamefont {Bibes}},\ }\bibfield
  {title} {\bibinfo {title} {Mapping spin-charge conversion to the band
  structure in a topological oxide two-dimensional electron gas},\ }\href
  {https://doi.org/https://doi.org/10.1038/s41563-019-0467-4} {\bibfield
  {journal} {\bibinfo  {journal} {Nat. Mater.}\ }\textbf {\bibinfo {volume}
  {18}},\ \bibinfo {pages} {1187} (\bibinfo {year} {2019})}\BibitemShut
  {NoStop}%
\bibitem [{\citenamefont {No{\"e}l}\ \emph {et~al.}(2020)\citenamefont
  {No{\"e}l}, \citenamefont {Trier}, \citenamefont {Vicente~Arche},
  \citenamefont {Br{\'e}hin}, \citenamefont {Vaz}, \citenamefont {Garcia},
  \citenamefont {Fusil}, \citenamefont {Barth{\'e}l{\'e}my}, \citenamefont
  {Vila}, \citenamefont {Bibes},\ and\ \citenamefont {Attan{\'e}}}]{Noel2020}%
  \BibitemOpen
  \bibfield  {author} {\bibinfo {author} {\bibfnamefont {P.}~\bibnamefont
  {No{\"e}l}}, \bibinfo {author} {\bibfnamefont {F.}~\bibnamefont {Trier}},
  \bibinfo {author} {\bibfnamefont {L.~M.}\ \bibnamefont {Vicente~Arche}},
  \bibinfo {author} {\bibfnamefont {J.}~\bibnamefont {Br{\'e}hin}}, \bibinfo
  {author} {\bibfnamefont {D.~C.}\ \bibnamefont {Vaz}}, \bibinfo {author}
  {\bibfnamefont {V.}~\bibnamefont {Garcia}}, \bibinfo {author} {\bibfnamefont
  {S.}~\bibnamefont {Fusil}}, \bibinfo {author} {\bibfnamefont
  {A.}~\bibnamefont {Barth{\'e}l{\'e}my}}, \bibinfo {author} {\bibfnamefont
  {L.}~\bibnamefont {Vila}}, \bibinfo {author} {\bibfnamefont {M.}~\bibnamefont
  {Bibes}},\ and\ \bibinfo {author} {\bibfnamefont {J.-P.}\ \bibnamefont
  {Attan{\'e}}},\ }\bibfield  {title} {\bibinfo {title} {Non-volatile electric
  control of spin-charge conversion in a srtio3 rashba system},\ }\href
  {https://doi.org/10.1038/s41586-020-2197-9} {\bibfield  {journal} {\bibinfo
  {journal} {Nature}\ }\textbf {\bibinfo {volume} {580}},\ \bibinfo {pages}
  {483} (\bibinfo {year} {2020})}\BibitemShut {NoStop}%
\bibitem [{\citenamefont {Ohya}\ \emph {et~al.}(2020)\citenamefont {Ohya},
  \citenamefont {Araki}, \citenamefont {Anh}, \citenamefont {Kaneta},
  \citenamefont {Seki}, \citenamefont {Tabata},\ and\ \citenamefont
  {Tanaka}}]{Ohya2020}%
  \BibitemOpen
  \bibfield  {author} {\bibinfo {author} {\bibfnamefont {S.}~\bibnamefont
  {Ohya}}, \bibinfo {author} {\bibfnamefont {D.}~\bibnamefont {Araki}},
  \bibinfo {author} {\bibfnamefont {L.~D.}\ \bibnamefont {Anh}}, \bibinfo
  {author} {\bibfnamefont {S.}~\bibnamefont {Kaneta}}, \bibinfo {author}
  {\bibfnamefont {M.}~\bibnamefont {Seki}}, \bibinfo {author} {\bibfnamefont
  {H.}~\bibnamefont {Tabata}},\ and\ \bibinfo {author} {\bibfnamefont
  {M.}~\bibnamefont {Tanaka}},\ }\bibfield  {title} {\bibinfo {title}
  {Efficient intrinsic spin-to-charge current conversion in an all-epitaxial
  single-crystal perovskite-oxide heterostructure of
  $\mathrm{L}{\mathrm{a}}_{0.67}\mathrm{S}{\mathrm{r}}_{0.33}\mathrm{Mn}{\mathrm{o}}_{3}/\mathrm{LaAl}{\mathrm{o}}_{3}/\mathrm{SrTi}{\mathrm{o}}_{3}$},\
  }\href {https://doi.org/10.1103/PhysRevResearch.2.012014} {\bibfield
  {journal} {\bibinfo  {journal} {Phys. Rev. Res.}\ }\textbf {\bibinfo {volume}
  {2}},\ \bibinfo {pages} {012014(R)} (\bibinfo {year} {2020})}\BibitemShut
  {NoStop}%
\bibitem [{\citenamefont {Bruneel}\ and\ \citenamefont
  {Gabay}(2020)}]{Bruneel2020}%
  \BibitemOpen
  \bibfield  {author} {\bibinfo {author} {\bibfnamefont {P.}~\bibnamefont
  {Bruneel}}\ and\ \bibinfo {author} {\bibfnamefont {M.}~\bibnamefont
  {Gabay}},\ }\bibfield  {title} {\bibinfo {title} {Spin texture driven
  spintronic enhancement at the
  $\mathrm{La}\mathrm{Al}{\mathrm{o}}_{3}/\mathrm{Sr}\mathrm{Ti}{\mathrm{o}}_{3}$
  interface},\ }\href {https://doi.org/10.1103/PhysRevB.102.144407} {\bibfield
  {journal} {\bibinfo  {journal} {Phys. Rev. B}\ }\textbf {\bibinfo {volume}
  {102}},\ \bibinfo {pages} {144407} (\bibinfo {year} {2020})}\BibitemShut
  {NoStop}%
\bibitem [{\citenamefont {To}\ \emph {et~al.}(2021)\citenamefont {To},
  \citenamefont {Dang}, \citenamefont {Vila}, \citenamefont {Attan\'e},
  \citenamefont {Bibes},\ and\ \citenamefont {Jaffr\`es}}]{To2021}%
  \BibitemOpen
  \bibfield  {author} {\bibinfo {author} {\bibfnamefont {D.~Q.}\ \bibnamefont
  {To}}, \bibinfo {author} {\bibfnamefont {T.~H.}\ \bibnamefont {Dang}},
  \bibinfo {author} {\bibfnamefont {L.}~\bibnamefont {Vila}}, \bibinfo {author}
  {\bibfnamefont {J.~P.}\ \bibnamefont {Attan\'e}}, \bibinfo {author}
  {\bibfnamefont {M.}~\bibnamefont {Bibes}},\ and\ \bibinfo {author}
  {\bibfnamefont {H.}~\bibnamefont {Jaffr\`es}},\ }\bibfield  {title} {\bibinfo
  {title} {Spin to charge conversion at rashba-split ${\mathbf{srtio}}_{3}$
  interfaces from resonant tunneling},\ }\href
  {https://doi.org/10.1103/PhysRevResearch.3.043170} {\bibfield  {journal}
  {\bibinfo  {journal} {Phys. Rev. Res.}\ }\textbf {\bibinfo {volume} {3}},\
  \bibinfo {pages} {043170} (\bibinfo {year} {2021})}\BibitemShut {NoStop}%
\bibitem [{\citenamefont {Trier}\ \emph {et~al.}(2022)\citenamefont {Trier},
  \citenamefont {No{\"e}l}, \citenamefont {Kim}, \citenamefont {Attan{\'e}},
  \citenamefont {Vila},\ and\ \citenamefont {Bibes}}]{Trier2022}%
  \BibitemOpen
  \bibfield  {author} {\bibinfo {author} {\bibfnamefont {F.}~\bibnamefont
  {Trier}}, \bibinfo {author} {\bibfnamefont {P.}~\bibnamefont {No{\"e}l}},
  \bibinfo {author} {\bibfnamefont {J.-V.}\ \bibnamefont {Kim}}, \bibinfo
  {author} {\bibfnamefont {J.-P.}\ \bibnamefont {Attan{\'e}}}, \bibinfo
  {author} {\bibfnamefont {L.}~\bibnamefont {Vila}},\ and\ \bibinfo {author}
  {\bibfnamefont {M.}~\bibnamefont {Bibes}},\ }\bibfield  {title} {\bibinfo
  {title} {Oxide spin-orbitronics: spin-charge interconversion and topological
  spin textures},\ }\href {https://doi.org/10.1038/s41578-021-00395-9}
  {\bibfield  {journal} {\bibinfo  {journal} {Nat. Rev. Mater.}\ }\textbf
  {\bibinfo {volume} {7}},\ \bibinfo {pages} {258} (\bibinfo {year}
  {2022})}\BibitemShut {NoStop}%
\bibitem [{\citenamefont {Shiomi}\ \emph {et~al.}(2014)\citenamefont {Shiomi},
  \citenamefont {Nomura}, \citenamefont {Kajiwara}, \citenamefont {Eto},
  \citenamefont {Novak}, \citenamefont {Segawa}, \citenamefont {Ando},\ and\
  \citenamefont {Saitoh}}]{Shiomi2014}%
  \BibitemOpen
  \bibfield  {author} {\bibinfo {author} {\bibfnamefont {Y.}~\bibnamefont
  {Shiomi}}, \bibinfo {author} {\bibfnamefont {K.}~\bibnamefont {Nomura}},
  \bibinfo {author} {\bibfnamefont {Y.}~\bibnamefont {Kajiwara}}, \bibinfo
  {author} {\bibfnamefont {K.}~\bibnamefont {Eto}}, \bibinfo {author}
  {\bibfnamefont {M.}~\bibnamefont {Novak}}, \bibinfo {author} {\bibfnamefont
  {K.}~\bibnamefont {Segawa}}, \bibinfo {author} {\bibfnamefont
  {Y.}~\bibnamefont {Ando}},\ and\ \bibinfo {author} {\bibfnamefont
  {E.}~\bibnamefont {Saitoh}},\ }\bibfield  {title} {\bibinfo {title}
  {Spin-electricity conversion induced by spin injection into topological
  insulators},\ }\href {https://doi.org/10.1103/PhysRevLett.113.196601}
  {\bibfield  {journal} {\bibinfo  {journal} {Phys. Rev. Lett.}\ }\textbf
  {\bibinfo {volume} {113}},\ \bibinfo {pages} {196601} (\bibinfo {year}
  {2014})}\BibitemShut {NoStop}%
\bibitem [{\citenamefont {Rojas-S\'anchez}\ \emph {et~al.}(2016)\citenamefont
  {Rojas-S\'anchez}, \citenamefont {Oyarz\'un}, \citenamefont {Fu},
  \citenamefont {Marty}, \citenamefont {Vergnaud}, \citenamefont {Gambarelli},
  \citenamefont {Vila}, \citenamefont {Jamet}, \citenamefont {Ohtsubo},
  \citenamefont {Taleb-Ibrahimi}, \citenamefont {Le~F\`evre}, \citenamefont
  {Bertran}, \citenamefont {Reyren}, \citenamefont {George},\ and\
  \citenamefont {Fert}}]{Sanchez2016}%
  \BibitemOpen
  \bibfield  {author} {\bibinfo {author} {\bibfnamefont {J.-C.}\ \bibnamefont
  {Rojas-S\'anchez}}, \bibinfo {author} {\bibfnamefont {S.}~\bibnamefont
  {Oyarz\'un}}, \bibinfo {author} {\bibfnamefont {Y.}~\bibnamefont {Fu}},
  \bibinfo {author} {\bibfnamefont {A.}~\bibnamefont {Marty}}, \bibinfo
  {author} {\bibfnamefont {C.}~\bibnamefont {Vergnaud}}, \bibinfo {author}
  {\bibfnamefont {S.}~\bibnamefont {Gambarelli}}, \bibinfo {author}
  {\bibfnamefont {L.}~\bibnamefont {Vila}}, \bibinfo {author} {\bibfnamefont
  {M.}~\bibnamefont {Jamet}}, \bibinfo {author} {\bibfnamefont
  {Y.}~\bibnamefont {Ohtsubo}}, \bibinfo {author} {\bibfnamefont
  {A.}~\bibnamefont {Taleb-Ibrahimi}}, \bibinfo {author} {\bibfnamefont
  {P.}~\bibnamefont {Le~F\`evre}}, \bibinfo {author} {\bibfnamefont
  {F.}~\bibnamefont {Bertran}}, \bibinfo {author} {\bibfnamefont
  {N.}~\bibnamefont {Reyren}}, \bibinfo {author} {\bibfnamefont {J.-M.}\
  \bibnamefont {George}},\ and\ \bibinfo {author} {\bibfnamefont
  {A.}~\bibnamefont {Fert}},\ }\bibfield  {title} {\bibinfo {title} {Spin to
  charge conversion at room temperature by spin pumping into a new type of
  topological insulator: $\ensuremath{\alpha}$-sn films},\ }\href
  {https://doi.org/10.1103/PhysRevLett.116.096602} {\bibfield  {journal}
  {\bibinfo  {journal} {Phys. Rev. Lett.}\ }\textbf {\bibinfo {volume} {116}},\
  \bibinfo {pages} {096602} (\bibinfo {year} {2016})}\BibitemShut {NoStop}%
\bibitem [{\citenamefont {Wang}\ \emph {et~al.}(2016)\citenamefont {Wang},
  \citenamefont {Kally}, \citenamefont {Lee}, \citenamefont {Liu},
  \citenamefont {Chang}, \citenamefont {Hickey}, \citenamefont {Mkhoyan},
  \citenamefont {Wu}, \citenamefont {Richardella},\ and\ \citenamefont
  {Samarth}}]{Wang2016}%
  \BibitemOpen
  \bibfield  {author} {\bibinfo {author} {\bibfnamefont {H.}~\bibnamefont
  {Wang}}, \bibinfo {author} {\bibfnamefont {J.}~\bibnamefont {Kally}},
  \bibinfo {author} {\bibfnamefont {J.~S.}\ \bibnamefont {Lee}}, \bibinfo
  {author} {\bibfnamefont {T.}~\bibnamefont {Liu}}, \bibinfo {author}
  {\bibfnamefont {H.}~\bibnamefont {Chang}}, \bibinfo {author} {\bibfnamefont
  {D.~R.}\ \bibnamefont {Hickey}}, \bibinfo {author} {\bibfnamefont {K.~A.}\
  \bibnamefont {Mkhoyan}}, \bibinfo {author} {\bibfnamefont {M.}~\bibnamefont
  {Wu}}, \bibinfo {author} {\bibfnamefont {A.}~\bibnamefont {Richardella}},\
  and\ \bibinfo {author} {\bibfnamefont {N.}~\bibnamefont {Samarth}},\
  }\bibfield  {title} {\bibinfo {title} {Surface-state-dominated spin-charge
  current conversion in topological-insulator--ferromagnetic-insulator
  heterostructures},\ }\href {https://doi.org/10.1103/PhysRevLett.117.076601}
  {\bibfield  {journal} {\bibinfo  {journal} {Phys. Rev. Lett.}\ }\textbf
  {\bibinfo {volume} {117}},\ \bibinfo {pages} {076601} (\bibinfo {year}
  {2016})}\BibitemShut {NoStop}%
\bibitem [{\citenamefont {Song}\ \emph {et~al.}(2016)\citenamefont {Song},
  \citenamefont {Mi}, \citenamefont {Zhao}, \citenamefont {Su}, \citenamefont
  {Yuan}, \citenamefont {Xing}, \citenamefont {Chen}, \citenamefont {Wang},
  \citenamefont {Wu}, \citenamefont {Chen}, \citenamefont {Xie}, \citenamefont
  {Zhang}, \citenamefont {Shi},\ and\ \citenamefont {Han}}]{Song2016}%
  \BibitemOpen
  \bibfield  {author} {\bibinfo {author} {\bibfnamefont {Q.}~\bibnamefont
  {Song}}, \bibinfo {author} {\bibfnamefont {J.}~\bibnamefont {Mi}}, \bibinfo
  {author} {\bibfnamefont {D.}~\bibnamefont {Zhao}}, \bibinfo {author}
  {\bibfnamefont {T.}~\bibnamefont {Su}}, \bibinfo {author} {\bibfnamefont
  {W.}~\bibnamefont {Yuan}}, \bibinfo {author} {\bibfnamefont {W.}~\bibnamefont
  {Xing}}, \bibinfo {author} {\bibfnamefont {Y.}~\bibnamefont {Chen}}, \bibinfo
  {author} {\bibfnamefont {T.}~\bibnamefont {Wang}}, \bibinfo {author}
  {\bibfnamefont {T.}~\bibnamefont {Wu}}, \bibinfo {author} {\bibfnamefont
  {X.~H.}\ \bibnamefont {Chen}}, \bibinfo {author} {\bibfnamefont {X.~C.}\
  \bibnamefont {Xie}}, \bibinfo {author} {\bibfnamefont {C.}~\bibnamefont
  {Zhang}}, \bibinfo {author} {\bibfnamefont {J.}~\bibnamefont {Shi}},\ and\
  \bibinfo {author} {\bibfnamefont {W.}~\bibnamefont {Han}},\ }\bibfield
  {title} {\bibinfo {title} {Spin injection and inverse edelstein effect in the
  surface states of topological kondo insulator smb6},\ }\href
  {https://doi.org/10.1038/ncomms13485} {\bibfield  {journal} {\bibinfo
  {journal} {Nat. Commun.}\ }\textbf {\bibinfo {volume} {7}},\ \bibinfo {pages}
  {13485} (\bibinfo {year} {2016})}\BibitemShut {NoStop}%
\bibitem [{\citenamefont {Mendes}\ \emph {et~al.}(2017)\citenamefont {Mendes},
  \citenamefont {Alves~Santos}, \citenamefont {Holanda}, \citenamefont
  {Loreto}, \citenamefont {de~Araujo}, \citenamefont {Chang}, \citenamefont
  {Moodera}, \citenamefont {Azevedo},\ and\ \citenamefont
  {Rezende}}]{Mendes2017}%
  \BibitemOpen
  \bibfield  {author} {\bibinfo {author} {\bibfnamefont {J.~B.~S.}\
  \bibnamefont {Mendes}}, \bibinfo {author} {\bibfnamefont {O.}~\bibnamefont
  {Alves~Santos}}, \bibinfo {author} {\bibfnamefont {J.}~\bibnamefont
  {Holanda}}, \bibinfo {author} {\bibfnamefont {R.~P.}\ \bibnamefont {Loreto}},
  \bibinfo {author} {\bibfnamefont {C.~I.~L.}\ \bibnamefont {de~Araujo}},
  \bibinfo {author} {\bibfnamefont {C.-Z.}\ \bibnamefont {Chang}}, \bibinfo
  {author} {\bibfnamefont {J.~S.}\ \bibnamefont {Moodera}}, \bibinfo {author}
  {\bibfnamefont {A.}~\bibnamefont {Azevedo}},\ and\ \bibinfo {author}
  {\bibfnamefont {S.~M.}\ \bibnamefont {Rezende}},\ }\bibfield  {title}
  {\bibinfo {title} {Dirac-surface-state-dominated spin to charge current
  conversion in the topological insulator
  (${\mathrm{bi}}_{0.22}{\mathrm{sb}}_{0.78}){}_{2}{\mathrm{te}}_{3}$ films at
  room temperature},\ }\href {https://doi.org/10.1103/PhysRevB.96.180415}
  {\bibfield  {journal} {\bibinfo  {journal} {Phys. Rev. B}\ }\textbf {\bibinfo
  {volume} {96}},\ \bibinfo {pages} {180415(R)} (\bibinfo {year}
  {2017})}\BibitemShut {NoStop}%
\bibitem [{\citenamefont {Sun}\ \emph {et~al.}(2019)\citenamefont {Sun},
  \citenamefont {Yang}, \citenamefont {Yang}, \citenamefont {Vetter},
  \citenamefont {Sun}, \citenamefont {Li}, \citenamefont {Su}, \citenamefont
  {Li}, \citenamefont {Li}, \citenamefont {Gong}, \citenamefont {Xie},
  \citenamefont {Hou}, \citenamefont {Gul}, \citenamefont {He}, \citenamefont
  {Zhang},\ and\ \citenamefont {Cheng}}]{Sun2019}%
  \BibitemOpen
  \bibfield  {author} {\bibinfo {author} {\bibfnamefont {R.}~\bibnamefont
  {Sun}}, \bibinfo {author} {\bibfnamefont {S.}~\bibnamefont {Yang}}, \bibinfo
  {author} {\bibfnamefont {X.}~\bibnamefont {Yang}}, \bibinfo {author}
  {\bibfnamefont {E.}~\bibnamefont {Vetter}}, \bibinfo {author} {\bibfnamefont
  {D.}~\bibnamefont {Sun}}, \bibinfo {author} {\bibfnamefont {N.}~\bibnamefont
  {Li}}, \bibinfo {author} {\bibfnamefont {L.}~\bibnamefont {Su}}, \bibinfo
  {author} {\bibfnamefont {Y.}~\bibnamefont {Li}}, \bibinfo {author}
  {\bibfnamefont {Y.}~\bibnamefont {Li}}, \bibinfo {author} {\bibfnamefont
  {Z.-z.}\ \bibnamefont {Gong}}, \bibinfo {author} {\bibfnamefont {Z.-k.}\
  \bibnamefont {Xie}}, \bibinfo {author} {\bibfnamefont {K.-y.}\ \bibnamefont
  {Hou}}, \bibinfo {author} {\bibfnamefont {Q.}~\bibnamefont {Gul}}, \bibinfo
  {author} {\bibfnamefont {W.}~\bibnamefont {He}}, \bibinfo {author}
  {\bibfnamefont {X.-q.}\ \bibnamefont {Zhang}},\ and\ \bibinfo {author}
  {\bibfnamefont {Z.-h.}\ \bibnamefont {Cheng}},\ }\bibfield  {title} {\bibinfo
  {title} {Large tunable spin-to-charge conversion induced by hybrid rashba and
  dirac surface states in topological insulator heterostructures},\ }\href
  {https://doi.org/10.1021/acs.nanolett.9b01151} {\bibfield  {journal}
  {\bibinfo  {journal} {Nano Lett.}\ }\textbf {\bibinfo {volume} {19}},\
  \bibinfo {pages} {4420} (\bibinfo {year} {2019})}\BibitemShut {NoStop}%
\bibitem [{\citenamefont {Singh}\ \emph {et~al.}(2020)\citenamefont {Singh},
  \citenamefont {Jena}, \citenamefont {Samanta}, \citenamefont {Biswas},\ and\
  \citenamefont {Bedanta}}]{Singh2020}%
  \BibitemOpen
  \bibfield  {author} {\bibinfo {author} {\bibfnamefont {B.~B.}\ \bibnamefont
  {Singh}}, \bibinfo {author} {\bibfnamefont {S.~K.}\ \bibnamefont {Jena}},
  \bibinfo {author} {\bibfnamefont {M.}~\bibnamefont {Samanta}}, \bibinfo
  {author} {\bibfnamefont {K.}~\bibnamefont {Biswas}},\ and\ \bibinfo {author}
  {\bibfnamefont {S.}~\bibnamefont {Bedanta}},\ }\bibfield  {title} {\bibinfo
  {title} {High spin to charge conversion efficiency in electron
  beam-evaporated topological insulator bi2se3},\ }\href
  {https://doi.org/10.1021/acsami.0c13540} {\bibfield  {journal} {\bibinfo
  {journal} {ACS Appl. Mater. Interfaces}\ }\textbf {\bibinfo {volume} {12}},\
  \bibinfo {pages} {53409} (\bibinfo {year} {2020})}\BibitemShut {NoStop}%
\bibitem [{\citenamefont {Dey}\ \emph {et~al.}(2021)\citenamefont {Dey},
  \citenamefont {Roy}, \citenamefont {Register},\ and\ \citenamefont
  {Banerjee}}]{Dey2021}%
  \BibitemOpen
  \bibfield  {author} {\bibinfo {author} {\bibfnamefont {R.}~\bibnamefont
  {Dey}}, \bibinfo {author} {\bibfnamefont {A.}~\bibnamefont {Roy}}, \bibinfo
  {author} {\bibfnamefont {L.~F.}\ \bibnamefont {Register}},\ and\ \bibinfo
  {author} {\bibfnamefont {S.~K.}\ \bibnamefont {Banerjee}},\ }\bibfield
  {title} {\bibinfo {title} {{Recent progress on measurement of spin–charge
  interconversion in topological insulators using ferromagnetic resonance}},\
  }\href {https://doi.org/10.1063/5.0049887} {\bibfield  {journal} {\bibinfo
  {journal} {APL Mater.}\ }\textbf {\bibinfo {volume} {9}},\ \bibinfo {pages}
  {060702} (\bibinfo {year} {2021})}\BibitemShut {NoStop}%
\bibitem [{\citenamefont {He}\ \emph {et~al.}(2021)\citenamefont {He},
  \citenamefont {Tai}, \citenamefont {Wu}, \citenamefont {Wu}, \citenamefont
  {Razavi}, \citenamefont {Gosavi}, \citenamefont {Walker}, \citenamefont
  {Oguz}, \citenamefont {Lin}, \citenamefont {Wong}, \citenamefont {Liu},
  \citenamefont {Dai},\ and\ \citenamefont {Wang}}]{He2021}%
  \BibitemOpen
  \bibfield  {author} {\bibinfo {author} {\bibfnamefont {H.}~\bibnamefont
  {He}}, \bibinfo {author} {\bibfnamefont {L.}~\bibnamefont {Tai}}, \bibinfo
  {author} {\bibfnamefont {H.}~\bibnamefont {Wu}}, \bibinfo {author}
  {\bibfnamefont {D.}~\bibnamefont {Wu}}, \bibinfo {author} {\bibfnamefont
  {A.}~\bibnamefont {Razavi}}, \bibinfo {author} {\bibfnamefont {T.~A.}\
  \bibnamefont {Gosavi}}, \bibinfo {author} {\bibfnamefont {E.~S.}\
  \bibnamefont {Walker}}, \bibinfo {author} {\bibfnamefont {K.}~\bibnamefont
  {Oguz}}, \bibinfo {author} {\bibfnamefont {C.-C.}\ \bibnamefont {Lin}},
  \bibinfo {author} {\bibfnamefont {K.}~\bibnamefont {Wong}}, \bibinfo {author}
  {\bibfnamefont {Y.}~\bibnamefont {Liu}}, \bibinfo {author} {\bibfnamefont
  {B.}~\bibnamefont {Dai}},\ and\ \bibinfo {author} {\bibfnamefont {K.~L.}\
  \bibnamefont {Wang}},\ }\bibfield  {title} {\bibinfo {title} {Conversion
  between spin and charge currents in topological-insulator/nonmagnetic-metal
  systems},\ }\href {https://doi.org/10.1103/PhysRevB.104.L220407} {\bibfield
  {journal} {\bibinfo  {journal} {Phys. Rev. B}\ }\textbf {\bibinfo {volume}
  {104}},\ \bibinfo {pages} {L220407} (\bibinfo {year} {2021})}\BibitemShut
  {NoStop}%
\bibitem [{\citenamefont {Zhang}\ and\ \citenamefont {Fert}(2016)}]{Zhang2016}%
  \BibitemOpen
  \bibfield  {author} {\bibinfo {author} {\bibfnamefont {S.}~\bibnamefont
  {Zhang}}\ and\ \bibinfo {author} {\bibfnamefont {A.}~\bibnamefont {Fert}},\
  }\bibfield  {title} {\bibinfo {title} {Conversion between spin and charge
  currents with topological insulators},\ }\href
  {https://doi.org/10.1103/PhysRevB.94.184423} {\bibfield  {journal} {\bibinfo
  {journal} {Phys. Rev. B}\ }\textbf {\bibinfo {volume} {94}},\ \bibinfo
  {pages} {184423} (\bibinfo {year} {2016})}\BibitemShut {NoStop}%
\bibitem [{\citenamefont {Mendes}\ \emph {et~al.}(2015)\citenamefont {Mendes},
  \citenamefont {Alves~Santos}, \citenamefont {Meireles}, \citenamefont
  {Lacerda}, \citenamefont {Vilela-Le\~ao}, \citenamefont {Machado},
  \citenamefont {Rodr\'{\i}guez-Su\'arez}, \citenamefont {Azevedo},\ and\
  \citenamefont {Rezende}}]{Mendes2015}%
  \BibitemOpen
  \bibfield  {author} {\bibinfo {author} {\bibfnamefont {J.~B.~S.}\
  \bibnamefont {Mendes}}, \bibinfo {author} {\bibfnamefont {O.}~\bibnamefont
  {Alves~Santos}}, \bibinfo {author} {\bibfnamefont {L.~M.}\ \bibnamefont
  {Meireles}}, \bibinfo {author} {\bibfnamefont {R.~G.}\ \bibnamefont
  {Lacerda}}, \bibinfo {author} {\bibfnamefont {L.~H.}\ \bibnamefont
  {Vilela-Le\~ao}}, \bibinfo {author} {\bibfnamefont {F.~L.~A.}\ \bibnamefont
  {Machado}}, \bibinfo {author} {\bibfnamefont {R.~L.}\ \bibnamefont
  {Rodr\'{\i}guez-Su\'arez}}, \bibinfo {author} {\bibfnamefont
  {A.}~\bibnamefont {Azevedo}},\ and\ \bibinfo {author} {\bibfnamefont {S.~M.}\
  \bibnamefont {Rezende}},\ }\bibfield  {title} {\bibinfo {title} {Spin-current
  to charge-current conversion and magnetoresistance in a hybrid structure of
  graphene and yttrium iron garnet},\ }\href
  {https://doi.org/10.1103/PhysRevLett.115.226601} {\bibfield  {journal}
  {\bibinfo  {journal} {Phys. Rev. Lett.}\ }\textbf {\bibinfo {volume} {115}},\
  \bibinfo {pages} {226601} (\bibinfo {year} {2015})}\BibitemShut {NoStop}%
\bibitem [{\citenamefont {Dushenko}\ \emph {et~al.}(2016)\citenamefont
  {Dushenko}, \citenamefont {Ago}, \citenamefont {Kawahara}, \citenamefont
  {Tsuda}, \citenamefont {Kuwabata}, \citenamefont {Takenobu}, \citenamefont
  {Shinjo}, \citenamefont {Ando},\ and\ \citenamefont
  {Shiraishi}}]{Dushenko2016}%
  \BibitemOpen
  \bibfield  {author} {\bibinfo {author} {\bibfnamefont {S.}~\bibnamefont
  {Dushenko}}, \bibinfo {author} {\bibfnamefont {H.}~\bibnamefont {Ago}},
  \bibinfo {author} {\bibfnamefont {K.}~\bibnamefont {Kawahara}}, \bibinfo
  {author} {\bibfnamefont {T.}~\bibnamefont {Tsuda}}, \bibinfo {author}
  {\bibfnamefont {S.}~\bibnamefont {Kuwabata}}, \bibinfo {author}
  {\bibfnamefont {T.}~\bibnamefont {Takenobu}}, \bibinfo {author}
  {\bibfnamefont {T.}~\bibnamefont {Shinjo}}, \bibinfo {author} {\bibfnamefont
  {Y.}~\bibnamefont {Ando}},\ and\ \bibinfo {author} {\bibfnamefont
  {M.}~\bibnamefont {Shiraishi}},\ }\bibfield  {title} {\bibinfo {title}
  {Gate-tunable spin-charge conversion and the role of spin-orbit interaction
  in graphene},\ }\href {https://doi.org/10.1103/PhysRevLett.116.166102}
  {\bibfield  {journal} {\bibinfo  {journal} {Phys. Rev. Lett.}\ }\textbf
  {\bibinfo {volume} {116}},\ \bibinfo {pages} {166102} (\bibinfo {year}
  {2016})}\BibitemShut {NoStop}%
\bibitem [{\citenamefont {Mendes}\ \emph {et~al.}(2019)\citenamefont {Mendes},
  \citenamefont {Alves~Santos}, \citenamefont {Chagas}, \citenamefont
  {Magalh\~aes Paniago}, \citenamefont {Mori}, \citenamefont {Holanda},
  \citenamefont {Meireles}, \citenamefont {Lacerda}, \citenamefont {Azevedo},\
  and\ \citenamefont {Rezende}}]{Mendes2019}%
  \BibitemOpen
  \bibfield  {author} {\bibinfo {author} {\bibfnamefont {J.~B.~S.}\
  \bibnamefont {Mendes}}, \bibinfo {author} {\bibfnamefont {O.}~\bibnamefont
  {Alves~Santos}}, \bibinfo {author} {\bibfnamefont {T.}~\bibnamefont
  {Chagas}}, \bibinfo {author} {\bibfnamefont {R.}~\bibnamefont {Magalh\~aes
  Paniago}}, \bibinfo {author} {\bibfnamefont {T.~J.~A.}\ \bibnamefont {Mori}},
  \bibinfo {author} {\bibfnamefont {J.}~\bibnamefont {Holanda}}, \bibinfo
  {author} {\bibfnamefont {L.~M.}\ \bibnamefont {Meireles}}, \bibinfo {author}
  {\bibfnamefont {R.~G.}\ \bibnamefont {Lacerda}}, \bibinfo {author}
  {\bibfnamefont {A.}~\bibnamefont {Azevedo}},\ and\ \bibinfo {author}
  {\bibfnamefont {S.~M.}\ \bibnamefont {Rezende}},\ }\bibfield  {title}
  {\bibinfo {title} {Direct detection of induced magnetic moment and efficient
  spin-to-charge conversion in graphene/ferromagnetic structures},\ }\href
  {https://doi.org/10.1103/PhysRevB.99.214446} {\bibfield  {journal} {\bibinfo
  {journal} {Phys. Rev. B}\ }\textbf {\bibinfo {volume} {99}},\ \bibinfo
  {pages} {214446} (\bibinfo {year} {2019})}\BibitemShut {NoStop}%
\bibitem [{\citenamefont {Bangar}\ \emph {et~al.}(2022)\citenamefont {Bangar},
  \citenamefont {Kumar}, \citenamefont {Chowdhury}, \citenamefont {Mudgal},
  \citenamefont {Gupta}, \citenamefont {Yadav}, \citenamefont {Das},\ and\
  \citenamefont {Muduli}}]{Bangar2022}%
  \BibitemOpen
  \bibfield  {author} {\bibinfo {author} {\bibfnamefont {H.}~\bibnamefont
  {Bangar}}, \bibinfo {author} {\bibfnamefont {A.}~\bibnamefont {Kumar}},
  \bibinfo {author} {\bibfnamefont {N.}~\bibnamefont {Chowdhury}}, \bibinfo
  {author} {\bibfnamefont {R.}~\bibnamefont {Mudgal}}, \bibinfo {author}
  {\bibfnamefont {P.}~\bibnamefont {Gupta}}, \bibinfo {author} {\bibfnamefont
  {R.~S.}\ \bibnamefont {Yadav}}, \bibinfo {author} {\bibfnamefont
  {S.}~\bibnamefont {Das}},\ and\ \bibinfo {author} {\bibfnamefont {P.~K.}\
  \bibnamefont {Muduli}},\ }\bibfield  {title} {\bibinfo {title} {{Large
  Spin-To-Charge Conversion at the Two-Dimensional Interface of
  Transition-Metal Dichalcogenides and Permalloy}},\ }\href
  {https://doi.org/10.1021/acsami.2c11162} {\bibfield  {journal} {\bibinfo
  {journal} {ACS Appl. Mater. Interfaces}\ }\textbf {\bibinfo {volume} {14}},\
  \bibinfo {pages} {41598} (\bibinfo {year} {2022})}\BibitemShut {NoStop}%
\bibitem [{\citenamefont {Chen}\ \emph {et~al.}(2016)\citenamefont {Chen},
  \citenamefont {Decker}, \citenamefont {Kronseder}, \citenamefont {Islinger},
  \citenamefont {Gmitra}, \citenamefont {Schuh}, \citenamefont {Bougeard},
  \citenamefont {Fabian}, \citenamefont {Weiss},\ and\ \citenamefont
  {Back}}]{Chen2016}%
  \BibitemOpen
  \bibfield  {author} {\bibinfo {author} {\bibfnamefont {L.}~\bibnamefont
  {Chen}}, \bibinfo {author} {\bibfnamefont {M.}~\bibnamefont {Decker}},
  \bibinfo {author} {\bibfnamefont {M.}~\bibnamefont {Kronseder}}, \bibinfo
  {author} {\bibfnamefont {R.}~\bibnamefont {Islinger}}, \bibinfo {author}
  {\bibfnamefont {M.}~\bibnamefont {Gmitra}}, \bibinfo {author} {\bibfnamefont
  {D.}~\bibnamefont {Schuh}}, \bibinfo {author} {\bibfnamefont
  {D.}~\bibnamefont {Bougeard}}, \bibinfo {author} {\bibfnamefont
  {J.}~\bibnamefont {Fabian}}, \bibinfo {author} {\bibfnamefont
  {D.}~\bibnamefont {Weiss}},\ and\ \bibinfo {author} {\bibfnamefont {C.~H.}\
  \bibnamefont {Back}},\ }\bibfield  {title} {\bibinfo {title} {{Robust
  spin-orbit torque and spin-galvanic effect at the Fe/GaAs (001) interface at
  room temperature}},\ }\href {https://doi.org/10.1038/ncomms13802} {\bibfield
  {journal} {\bibinfo  {journal} {Nat. Commun.}\ }\textbf {\bibinfo {volume}
  {7}},\ \bibinfo {pages} {13802} (\bibinfo {year} {2016})}\BibitemShut
  {NoStop}%
\bibitem [{\citenamefont {Oyarz{\'u}n}\ \emph {et~al.}(2016)\citenamefont
  {Oyarz{\'u}n}, \citenamefont {Nandy}, \citenamefont {Rortais}, \citenamefont
  {Rojas-S{\'a}nchez}, \citenamefont {Dau}, \citenamefont {No{\"e}l},
  \citenamefont {Laczkowski}, \citenamefont {Pouget}, \citenamefont {Okuno},
  \citenamefont {Vila}, \citenamefont {Vergnaud}, \citenamefont {Beign{\'e}},
  \citenamefont {Marty}, \citenamefont {Attan{\'e}}, \citenamefont
  {Gambarelli}, \citenamefont {George}, \citenamefont {Jaffr\`{e}s},
  \citenamefont {Bl{\"u}gel},\ and\ \citenamefont {Jamet}}]{Oyarzun2016}%
  \BibitemOpen
  \bibfield  {author} {\bibinfo {author} {\bibfnamefont {S.}~\bibnamefont
  {Oyarz{\'u}n}}, \bibinfo {author} {\bibfnamefont {A.}~\bibnamefont {Nandy}},
  \bibinfo {author} {\bibfnamefont {F.}~\bibnamefont {Rortais}}, \bibinfo
  {author} {\bibfnamefont {J.-C.}\ \bibnamefont {Rojas-S{\'a}nchez}}, \bibinfo
  {author} {\bibfnamefont {M.-T.}\ \bibnamefont {Dau}}, \bibinfo {author}
  {\bibfnamefont {P.}~\bibnamefont {No{\"e}l}}, \bibinfo {author}
  {\bibfnamefont {P.}~\bibnamefont {Laczkowski}}, \bibinfo {author}
  {\bibfnamefont {S.}~\bibnamefont {Pouget}}, \bibinfo {author} {\bibfnamefont
  {H.}~\bibnamefont {Okuno}}, \bibinfo {author} {\bibfnamefont
  {L.}~\bibnamefont {Vila}}, \bibinfo {author} {\bibfnamefont {C.}~\bibnamefont
  {Vergnaud}}, \bibinfo {author} {\bibfnamefont {C.}~\bibnamefont
  {Beign{\'e}}}, \bibinfo {author} {\bibfnamefont {A.}~\bibnamefont {Marty}},
  \bibinfo {author} {\bibfnamefont {J.-P.}\ \bibnamefont {Attan{\'e}}},
  \bibinfo {author} {\bibfnamefont {S.}~\bibnamefont {Gambarelli}}, \bibinfo
  {author} {\bibfnamefont {J.-M.}\ \bibnamefont {George}}, \bibinfo {author}
  {\bibfnamefont {H.}~\bibnamefont {Jaffr\`{e}s}}, \bibinfo {author}
  {\bibfnamefont {S.}~\bibnamefont {Bl{\"u}gel}},\ and\ \bibinfo {author}
  {\bibfnamefont {M.}~\bibnamefont {Jamet}},\ }\bibfield  {title} {\bibinfo
  {title} {Evidence for spin-to-charge conversion by rashba coupling in
  metallic states at the fe/ge (111) interface},\ }\href
  {https://doi.org/10.1038/ncomms13857} {\bibfield  {journal} {\bibinfo
  {journal} {Nat. Commun.}\ }\textbf {\bibinfo {volume} {7}},\ \bibinfo {pages}
  {13857} (\bibinfo {year} {2016})}\BibitemShut {NoStop}%
\bibitem [{\citenamefont {Zhang}\ \emph {et~al.}(2019)\citenamefont {Zhang},
  \citenamefont {Ma}, \citenamefont {Zhang}, \citenamefont {Chen},
  \citenamefont {Wang}, \citenamefont {Li}, \citenamefont {Yun}, \citenamefont
  {Yan}, \citenamefont {Chen}, \citenamefont {Hu}, \citenamefont {Cai},
  \citenamefont {Shen}, \citenamefont {Han},\ and\ \citenamefont
  {Sun}}]{Zhang19}%
  \BibitemOpen
  \bibfield  {author} {\bibinfo {author} {\bibfnamefont {H.}~\bibnamefont
  {Zhang}}, \bibinfo {author} {\bibfnamefont {Y.}~\bibnamefont {Ma}}, \bibinfo
  {author} {\bibfnamefont {H.}~\bibnamefont {Zhang}}, \bibinfo {author}
  {\bibfnamefont {X.}~\bibnamefont {Chen}}, \bibinfo {author} {\bibfnamefont
  {S.}~\bibnamefont {Wang}}, \bibinfo {author} {\bibfnamefont {G.}~\bibnamefont
  {Li}}, \bibinfo {author} {\bibfnamefont {Y.}~\bibnamefont {Yun}}, \bibinfo
  {author} {\bibfnamefont {X.}~\bibnamefont {Yan}}, \bibinfo {author}
  {\bibfnamefont {Y.}~\bibnamefont {Chen}}, \bibinfo {author} {\bibfnamefont
  {F.}~\bibnamefont {Hu}}, \bibinfo {author} {\bibfnamefont {J.}~\bibnamefont
  {Cai}}, \bibinfo {author} {\bibfnamefont {B.}~\bibnamefont {Shen}}, \bibinfo
  {author} {\bibfnamefont {W.}~\bibnamefont {Han}},\ and\ \bibinfo {author}
  {\bibfnamefont {J.}~\bibnamefont {Sun}},\ }\bibfield  {title} {\bibinfo
  {title} {{Thermal Spin Injection and Inverse Edelstein Effect of the
  Two-Dimensional Electron Gas at EuO--KTaO${}_3$ Interfaces}},\ }\href
  {https://doi.org/10.1021/acs.nanolett.8b04509} {\bibfield  {journal}
  {\bibinfo  {journal} {Nano Lett.}\ }\textbf {\bibinfo {volume} {19}},\
  \bibinfo {pages} {1605} (\bibinfo {year} {2019})}\BibitemShut {NoStop}%
\bibitem [{\citenamefont {Dresselhaus}(1955)}]{Dresselhaus1955}%
  \BibitemOpen
  \bibfield  {author} {\bibinfo {author} {\bibfnamefont {G.}~\bibnamefont
  {Dresselhaus}},\ }\bibfield  {title} {\bibinfo {title} {Spin-orbit coupling
  effects in zinc blende structures},\ }\href
  {https://doi.org/10.1103/PhysRev.100.580} {\bibfield  {journal} {\bibinfo
  {journal} {Phys. Rev.}\ }\textbf {\bibinfo {volume} {100}},\ \bibinfo {pages}
  {580} (\bibinfo {year} {1955})}\BibitemShut {NoStop}%
\bibitem [{\citenamefont {La~Rocca}\ \emph {et~al.}(1988)\citenamefont
  {La~Rocca}, \citenamefont {Kim},\ and\ \citenamefont
  {Rodriguez}}]{Rocca1988}%
  \BibitemOpen
  \bibfield  {author} {\bibinfo {author} {\bibfnamefont {G.~C.}\ \bibnamefont
  {La~Rocca}}, \bibinfo {author} {\bibfnamefont {N.}~\bibnamefont {Kim}},\ and\
  \bibinfo {author} {\bibfnamefont {S.}~\bibnamefont {Rodriguez}},\ }\bibfield
  {title} {\bibinfo {title} {Effect of uniaxial stress on the electron spin
  resonance in zinc-blende semiconductors},\ }\href
  {https://doi.org/10.1103/PhysRevB.38.7595} {\bibfield  {journal} {\bibinfo
  {journal} {Phys. Rev. B}\ }\textbf {\bibinfo {volume} {38}},\ \bibinfo
  {pages} {7595} (\bibinfo {year} {1988})}\BibitemShut {NoStop}%
\bibitem [{\citenamefont {Ganichev}\ \emph {et~al.}(2003)\citenamefont
  {Ganichev}, \citenamefont {Schneider}, \citenamefont {Bel'kov}, \citenamefont
  {Ivchenko}, \citenamefont {Tarasenko}, \citenamefont {Wegscheider},
  \citenamefont {Weiss}, \citenamefont {Schuh}, \citenamefont {Murdin},
  \citenamefont {Phillips}, \citenamefont {Pidgeon}, \citenamefont {Clarke},
  \citenamefont {Merrick}, \citenamefont {Murzyn}, \citenamefont {Beregulin},\
  and\ \citenamefont {Prettl}}]{Ganichev2003a}%
  \BibitemOpen
  \bibfield  {author} {\bibinfo {author} {\bibfnamefont {S.~D.}\ \bibnamefont
  {Ganichev}}, \bibinfo {author} {\bibfnamefont {P.}~\bibnamefont {Schneider}},
  \bibinfo {author} {\bibfnamefont {V.~V.}\ \bibnamefont {Bel'kov}}, \bibinfo
  {author} {\bibfnamefont {E.~L.}\ \bibnamefont {Ivchenko}}, \bibinfo {author}
  {\bibfnamefont {S.~A.}\ \bibnamefont {Tarasenko}}, \bibinfo {author}
  {\bibfnamefont {W.}~\bibnamefont {Wegscheider}}, \bibinfo {author}
  {\bibfnamefont {D.}~\bibnamefont {Weiss}}, \bibinfo {author} {\bibfnamefont
  {D.}~\bibnamefont {Schuh}}, \bibinfo {author} {\bibfnamefont {B.~N.}\
  \bibnamefont {Murdin}}, \bibinfo {author} {\bibfnamefont {P.~J.}\
  \bibnamefont {Phillips}}, \bibinfo {author} {\bibfnamefont {C.~R.}\
  \bibnamefont {Pidgeon}}, \bibinfo {author} {\bibfnamefont {D.~G.}\
  \bibnamefont {Clarke}}, \bibinfo {author} {\bibfnamefont {M.}~\bibnamefont
  {Merrick}}, \bibinfo {author} {\bibfnamefont {P.}~\bibnamefont {Murzyn}},
  \bibinfo {author} {\bibfnamefont {E.~V.}\ \bibnamefont {Beregulin}},\ and\
  \bibinfo {author} {\bibfnamefont {W.}~\bibnamefont {Prettl}},\ }\bibfield
  {title} {\bibinfo {title} {Spin-galvanic effect due to optical spin
  orientation in n-type gaas quantum well structures},\ }\href
  {https://doi.org/10.1103/PhysRevB.68.081302} {\bibfield  {journal} {\bibinfo
  {journal} {Phys. Rev. B}\ }\textbf {\bibinfo {volume} {68}},\ \bibinfo
  {pages} {081302} (\bibinfo {year} {2003})}\BibitemShut {NoStop}%
\bibitem [{\citenamefont {Ganichev}\ \emph {et~al.}(2004)\citenamefont
  {Ganichev}, \citenamefont {Bel'kov}, \citenamefont {Golub}, \citenamefont
  {Ivchenko}, \citenamefont {Schneider}, \citenamefont {Giglberger},
  \citenamefont {Eroms}, \citenamefont {De~Boeck}, \citenamefont {Borghs},
  \citenamefont {Wegscheider}, \citenamefont {Weiss},\ and\ \citenamefont
  {Prettl}}]{Ganichev2004}%
  \BibitemOpen
  \bibfield  {author} {\bibinfo {author} {\bibfnamefont {S.~D.}\ \bibnamefont
  {Ganichev}}, \bibinfo {author} {\bibfnamefont {V.~V.}\ \bibnamefont
  {Bel'kov}}, \bibinfo {author} {\bibfnamefont {L.~E.}\ \bibnamefont {Golub}},
  \bibinfo {author} {\bibfnamefont {E.~L.}\ \bibnamefont {Ivchenko}}, \bibinfo
  {author} {\bibfnamefont {P.}~\bibnamefont {Schneider}}, \bibinfo {author}
  {\bibfnamefont {S.}~\bibnamefont {Giglberger}}, \bibinfo {author}
  {\bibfnamefont {J.}~\bibnamefont {Eroms}}, \bibinfo {author} {\bibfnamefont
  {J.}~\bibnamefont {De~Boeck}}, \bibinfo {author} {\bibfnamefont
  {G.}~\bibnamefont {Borghs}}, \bibinfo {author} {\bibfnamefont
  {W.}~\bibnamefont {Wegscheider}}, \bibinfo {author} {\bibfnamefont
  {D.}~\bibnamefont {Weiss}},\ and\ \bibinfo {author} {\bibfnamefont
  {W.}~\bibnamefont {Prettl}},\ }\bibfield  {title} {\bibinfo {title}
  {Experimental separation of rashba and dresselhaus spin splittings in
  semiconductor quantum wells},\ }\href
  {https://doi.org/10.1103/PhysRevLett.92.256601} {\bibfield  {journal}
  {\bibinfo  {journal} {Phys. Rev. Lett.}\ }\textbf {\bibinfo {volume} {92}},\
  \bibinfo {pages} {256601} (\bibinfo {year} {2004})}\BibitemShut {NoStop}%
\bibitem [{\citenamefont {Giglberger}\ \emph {et~al.}(2007)\citenamefont
  {Giglberger}, \citenamefont {Golub}, \citenamefont {Bel'kov}, \citenamefont
  {Danilov}, \citenamefont {Schuh}, \citenamefont {Gerl}, \citenamefont
  {Rohlfing}, \citenamefont {Stahl}, \citenamefont {Wegscheider}, \citenamefont
  {Weiss}, \citenamefont {Prettl},\ and\ \citenamefont
  {Ganichev}}]{Giglberger2007}%
  \BibitemOpen
  \bibfield  {author} {\bibinfo {author} {\bibfnamefont {S.}~\bibnamefont
  {Giglberger}}, \bibinfo {author} {\bibfnamefont {L.~E.}\ \bibnamefont
  {Golub}}, \bibinfo {author} {\bibfnamefont {V.~V.}\ \bibnamefont {Bel'kov}},
  \bibinfo {author} {\bibfnamefont {S.~N.}\ \bibnamefont {Danilov}}, \bibinfo
  {author} {\bibfnamefont {D.}~\bibnamefont {Schuh}}, \bibinfo {author}
  {\bibfnamefont {C.}~\bibnamefont {Gerl}}, \bibinfo {author} {\bibfnamefont
  {F.}~\bibnamefont {Rohlfing}}, \bibinfo {author} {\bibfnamefont
  {J.}~\bibnamefont {Stahl}}, \bibinfo {author} {\bibfnamefont
  {W.}~\bibnamefont {Wegscheider}}, \bibinfo {author} {\bibfnamefont
  {D.}~\bibnamefont {Weiss}}, \bibinfo {author} {\bibfnamefont
  {W.}~\bibnamefont {Prettl}},\ and\ \bibinfo {author} {\bibfnamefont {S.~D.}\
  \bibnamefont {Ganichev}},\ }\bibfield  {title} {\bibinfo {title} {Rashba and
  dresselhaus spin splittings in semiconductor quantum wells measured by spin
  photocurrents},\ }\href {https://doi.org/10.1103/PhysRevB.75.035327}
  {\bibfield  {journal} {\bibinfo  {journal} {Phys. Rev. B}\ }\textbf {\bibinfo
  {volume} {75}},\ \bibinfo {pages} {035327} (\bibinfo {year}
  {2007})}\BibitemShut {NoStop}%
\bibitem [{\citenamefont {Bel'kov}\ and\ \citenamefont
  {Ganichev}(2008)}]{Belkov2008}%
  \BibitemOpen
  \bibfield  {author} {\bibinfo {author} {\bibfnamefont {V.~V.}\ \bibnamefont
  {Bel'kov}}\ and\ \bibinfo {author} {\bibfnamefont {S.~D.}\ \bibnamefont
  {Ganichev}},\ }\bibfield  {title} {\bibinfo {title} {Magneto-gyrotropic
  effects in semiconductor quantum wells},\ }\href
  {https://doi.org/10.1088/0268-1242/23/11/114003} {\bibfield  {journal}
  {\bibinfo  {journal} {Semicond. Sci. Technol.}\ }\textbf {\bibinfo {volume}
  {23}},\ \bibinfo {pages} {114003} (\bibinfo {year} {2008})}\BibitemShut
  {NoStop}%
\bibitem [{\citenamefont {Ganichev}(2008)}]{Ganichev2008}%
  \BibitemOpen
  \bibfield  {author} {\bibinfo {author} {\bibfnamefont {S.~D.}\ \bibnamefont
  {Ganichev}},\ }\bibfield  {title} {\bibinfo {title} {Spin-galvanic effect and
  spin orientation by current in non-magnetic semiconductors},\ }\href
  {https://doi.org/10.1142/S0217979208046001} {\bibfield  {journal} {\bibinfo
  {journal} {Int. J. Mod. Phys. B}\ }\textbf {\bibinfo {volume} {22}},\
  \bibinfo {pages} {1} (\bibinfo {year} {2008})}\BibitemShut {NoStop}%
\bibitem [{\citenamefont {Ivchenko}\ and\ \citenamefont
  {Ganichev}(2008)}]{Ivchenko2008}%
  \BibitemOpen
  \bibfield  {author} {\bibinfo {author} {\bibfnamefont {E.~L.}\ \bibnamefont
  {Ivchenko}}\ and\ \bibinfo {author} {\bibfnamefont {S.~D.}\ \bibnamefont
  {Ganichev}},\ }\bibinfo {title} {Spin-photogalvanics},\ in\ \href
  {https://doi.org/10.1007/978-3-540-78820-1_9} {\emph {\bibinfo {booktitle}
  {Spin Physics in Semiconductors}}},\ \bibinfo {editor} {edited by\ \bibinfo
  {editor} {\bibfnamefont {M.~I.}\ \bibnamefont {Dyakonov}}}\ (\bibinfo
  {publisher} {Springer Berlin Heidelberg},\ \bibinfo {address} {Berlin,
  Heidelberg},\ \bibinfo {year} {2008})\ pp.\ \bibinfo {pages}
  {245--277}\BibitemShut {NoStop}%
\bibitem [{\citenamefont {Ganichev}\ and\ \citenamefont
  {Golub}(2014)}]{Ganichev2014}%
  \BibitemOpen
  \bibfield  {author} {\bibinfo {author} {\bibfnamefont {S.~D.}\ \bibnamefont
  {Ganichev}}\ and\ \bibinfo {author} {\bibfnamefont {L.~E.}\ \bibnamefont
  {Golub}},\ }\bibfield  {title} {\bibinfo {title} {Interplay of
  rashba/dresselhaus spin splittings probed by photogalvanic spectroscopy –a
  review},\ }\href {https://doi.org/https://doi.org/10.1002/pssb.201350261}
  {\bibfield  {journal} {\bibinfo  {journal} {Phys. Status Solidi B}\ }\textbf
  {\bibinfo {volume} {251}},\ \bibinfo {pages} {1801} (\bibinfo {year}
  {2014})}\BibitemShut {NoStop}%
\bibitem [{\citenamefont {Sheikhabadi}\ and\ \citenamefont
  {Raimondi}(2017)}]{Sheikhabadi2017}%
  \BibitemOpen
  \bibfield  {author} {\bibinfo {author} {\bibfnamefont {A.~M.}\ \bibnamefont
  {Sheikhabadi}}\ and\ \bibinfo {author} {\bibfnamefont {R.}~\bibnamefont
  {Raimondi}},\ }\bibfield  {title} {\bibinfo {title} {Inverse spin galvanic
  effect in the presence of impurity spin-orbit scattering: A diagrammatic
  approach},\ }\href {https://doi.org/10.3390/condmat2020017} {\bibfield
  {journal} {\bibinfo  {journal} {Condens. Matter}\ }\textbf {\bibinfo {volume}
  {2}},\ \bibinfo {pages} {17} (\bibinfo {year} {2017})}\BibitemShut {NoStop}%
\bibitem [{\citenamefont {Tao}\ and\ \citenamefont {Tsymbal}(2021)}]{Tao2021}%
  \BibitemOpen
  \bibfield  {author} {\bibinfo {author} {\bibfnamefont {L.~L.}\ \bibnamefont
  {Tao}}\ and\ \bibinfo {author} {\bibfnamefont {E.~Y.}\ \bibnamefont
  {Tsymbal}},\ }\bibfield  {title} {\bibinfo {title} {Spin-orbit dependence of
  anisotropic current-induced spin polarization},\ }\href
  {https://doi.org/10.1103/PhysRevB.104.085438} {\bibfield  {journal} {\bibinfo
   {journal} {Phys. Rev. B}\ }\textbf {\bibinfo {volume} {104}},\ \bibinfo
  {pages} {085438} (\bibinfo {year} {2021})}\BibitemShut {NoStop}%
\bibitem [{\citenamefont {Zhuravlev}\ \emph {et~al.}(2022)\citenamefont
  {Zhuravlev}, \citenamefont {Alexandrov},\ and\ \citenamefont
  {Vedyayev}}]{Zhuravlev2022}%
  \BibitemOpen
  \bibfield  {author} {\bibinfo {author} {\bibfnamefont {M.}~\bibnamefont
  {Zhuravlev}}, \bibinfo {author} {\bibfnamefont {A.}~\bibnamefont
  {Alexandrov}},\ and\ \bibinfo {author} {\bibfnamefont {A.}~\bibnamefont
  {Vedyayev}},\ }\bibfield  {title} {\bibinfo {title} {Spin accumulation and
  spin hall effect in a two-layer system with a thin ferromagnetic layer},\
  }\href {https://doi.org/10.1088/1361-648X/ac4c65} {\bibfield  {journal}
  {\bibinfo  {journal} {J. Phys. Condens. Matter}\ }\textbf {\bibinfo {volume}
  {34}},\ \bibinfo {pages} {145301} (\bibinfo {year} {2022})}\BibitemShut
  {NoStop}%
\bibitem [{\citenamefont {Shytov}\ \emph {et~al.}(2006)\citenamefont {Shytov},
  \citenamefont {Mishchenko}, \citenamefont {Engel},\ and\ \citenamefont
  {Halperin}}]{Shytov2006}%
  \BibitemOpen
  \bibfield  {author} {\bibinfo {author} {\bibfnamefont {A.~V.}\ \bibnamefont
  {Shytov}}, \bibinfo {author} {\bibfnamefont {E.~G.}\ \bibnamefont
  {Mishchenko}}, \bibinfo {author} {\bibfnamefont {H.-A.}\ \bibnamefont
  {Engel}},\ and\ \bibinfo {author} {\bibfnamefont {B.~I.}\ \bibnamefont
  {Halperin}},\ }\bibfield  {title} {\bibinfo {title} {Small-angle impurity
  scattering and the spin hall conductivity in two-dimensional semiconductor
  systems},\ }\href {https://doi.org/10.1103/PhysRevB.73.075316} {\bibfield
  {journal} {\bibinfo  {journal} {Phys. Rev. B}\ }\textbf {\bibinfo {volume}
  {73}},\ \bibinfo {pages} {075316} (\bibinfo {year} {2006})}\BibitemShut
  {NoStop}%
\bibitem [{\citenamefont {Raimondi}\ \emph {et~al.}(2006)\citenamefont
  {Raimondi}, \citenamefont {Gorini}, \citenamefont {Schwab},\ and\
  \citenamefont {Dzierzawa}}]{Raimondi2006}%
  \BibitemOpen
  \bibfield  {author} {\bibinfo {author} {\bibfnamefont {R.}~\bibnamefont
  {Raimondi}}, \bibinfo {author} {\bibfnamefont {C.}~\bibnamefont {Gorini}},
  \bibinfo {author} {\bibfnamefont {P.}~\bibnamefont {Schwab}},\ and\ \bibinfo
  {author} {\bibfnamefont {M.}~\bibnamefont {Dzierzawa}},\ }\bibfield  {title}
  {\bibinfo {title} {Quasiclassical approach to the spin hall effect in the
  two-dimensional electron gas},\ }\href
  {https://doi.org/10.1103/PhysRevB.74.035340} {\bibfield  {journal} {\bibinfo
  {journal} {Phys. Rev. B}\ }\textbf {\bibinfo {volume} {74}},\ \bibinfo
  {pages} {035340} (\bibinfo {year} {2006})}\BibitemShut {NoStop}%
\bibitem [{\citenamefont {Trushin}\ and\ \citenamefont
  {Schliemann}(2007)}]{Trushin2007}%
  \BibitemOpen
  \bibfield  {author} {\bibinfo {author} {\bibfnamefont {M.}~\bibnamefont
  {Trushin}}\ and\ \bibinfo {author} {\bibfnamefont {J.}~\bibnamefont
  {Schliemann}},\ }\bibfield  {title} {\bibinfo {title} {Anisotropic
  current-induced spin accumulation in the two-dimensional electron gas with
  spin-orbit coupling},\ }\href {https://doi.org/10.1103/PhysRevB.75.155323}
  {\bibfield  {journal} {\bibinfo  {journal} {Phys. Rev. B}\ }\textbf {\bibinfo
  {volume} {75}},\ \bibinfo {pages} {155323} (\bibinfo {year}
  {2007})}\BibitemShut {NoStop}%
\bibitem [{\citenamefont {Raichev}(2007)}]{Raichev2007}%
  \BibitemOpen
  \bibfield  {author} {\bibinfo {author} {\bibfnamefont {O.~E.}\ \bibnamefont
  {Raichev}},\ }\bibfield  {title} {\bibinfo {title} {Frequency dependence of
  induced spin polarization and spin current in quantum wells},\ }\href
  {https://doi.org/10.1103/PhysRevB.75.205340} {\bibfield  {journal} {\bibinfo
  {journal} {Phys. Rev. B}\ }\textbf {\bibinfo {volume} {75}},\ \bibinfo
  {pages} {205340} (\bibinfo {year} {2007})}\BibitemShut {NoStop}%
\bibitem [{\citenamefont {Engel}\ \emph {et~al.}(2007)\citenamefont {Engel},
  \citenamefont {Rashba},\ and\ \citenamefont {Halperin}}]{Engel2007}%
  \BibitemOpen
  \bibfield  {author} {\bibinfo {author} {\bibfnamefont {H.-A.}\ \bibnamefont
  {Engel}}, \bibinfo {author} {\bibfnamefont {E.~I.}\ \bibnamefont {Rashba}},\
  and\ \bibinfo {author} {\bibfnamefont {B.~I.}\ \bibnamefont {Halperin}},\
  }\bibfield  {title} {\bibinfo {title} {Out-of-plane spin polarization from
  in-plane electric and magnetic fields},\ }\href
  {https://doi.org/10.1103/PhysRevLett.98.036602} {\bibfield  {journal}
  {\bibinfo  {journal} {Phys. Rev. Lett.}\ }\textbf {\bibinfo {volume} {98}},\
  \bibinfo {pages} {036602} (\bibinfo {year} {2007})}\BibitemShut {NoStop}%
\bibitem [{\citenamefont {Gorini}\ \emph {et~al.}(2010)\citenamefont {Gorini},
  \citenamefont {Schwab}, \citenamefont {Raimondi},\ and\ \citenamefont
  {Shelankov}}]{Gorini2010}%
  \BibitemOpen
  \bibfield  {author} {\bibinfo {author} {\bibfnamefont {C.}~\bibnamefont
  {Gorini}}, \bibinfo {author} {\bibfnamefont {P.}~\bibnamefont {Schwab}},
  \bibinfo {author} {\bibfnamefont {R.}~\bibnamefont {Raimondi}},\ and\
  \bibinfo {author} {\bibfnamefont {A.~L.}\ \bibnamefont {Shelankov}},\
  }\bibfield  {title} {\bibinfo {title} {Non-abelian gauge fields in the
  gradient expansion: Generalized boltzmann and eilenberger equations},\ }\href
  {https://doi.org/10.1103/PhysRevB.82.195316} {\bibfield  {journal} {\bibinfo
  {journal} {Phys. Rev. B}\ }\textbf {\bibinfo {volume} {82}},\ \bibinfo
  {pages} {195316} (\bibinfo {year} {2010})}\BibitemShut {NoStop}%
\bibitem [{\citenamefont {Raimondi}\ \emph {et~al.}(2012)\citenamefont
  {Raimondi}, \citenamefont {Schwab}, \citenamefont {Gorini},\ and\
  \citenamefont {Vignale}}]{Raimondi2012}%
  \BibitemOpen
  \bibfield  {author} {\bibinfo {author} {\bibfnamefont {R.}~\bibnamefont
  {Raimondi}}, \bibinfo {author} {\bibfnamefont {P.}~\bibnamefont {Schwab}},
  \bibinfo {author} {\bibfnamefont {C.}~\bibnamefont {Gorini}},\ and\ \bibinfo
  {author} {\bibfnamefont {G.}~\bibnamefont {Vignale}},\ }\bibfield  {title}
  {\bibinfo {title} {Spin-orbit interaction in a two-dimensional electron gas:
  A su(2) formulation},\ }\href
  {https://doi.org/https://doi.org/10.1002/andp.201100253} {\bibfield
  {journal} {\bibinfo  {journal} {Ann. Phys. (Berlin)}\ }\textbf {\bibinfo
  {volume} {524}},\ \bibinfo {pages} {153} (\bibinfo {year}
  {2012})}\BibitemShut {NoStop}%
\bibitem [{\citenamefont {Bi}\ \emph {et~al.}(2013)\citenamefont {Bi},
  \citenamefont {He}, \citenamefont {Hankiewicz}, \citenamefont {Winkler},
  \citenamefont {Vignale},\ and\ \citenamefont {Culcer}}]{Xintao2013}%
  \BibitemOpen
  \bibfield  {author} {\bibinfo {author} {\bibfnamefont {X.}~\bibnamefont
  {Bi}}, \bibinfo {author} {\bibfnamefont {P.}~\bibnamefont {He}}, \bibinfo
  {author} {\bibfnamefont {E.~M.}\ \bibnamefont {Hankiewicz}}, \bibinfo
  {author} {\bibfnamefont {R.}~\bibnamefont {Winkler}}, \bibinfo {author}
  {\bibfnamefont {G.}~\bibnamefont {Vignale}},\ and\ \bibinfo {author}
  {\bibfnamefont {D.}~\bibnamefont {Culcer}},\ }\bibfield  {title} {\bibinfo
  {title} {Anomalous spin precession and spin hall effect in semiconductor
  quantum wells},\ }\href {https://doi.org/10.1103/PhysRevB.88.035316}
  {\bibfield  {journal} {\bibinfo  {journal} {Phys. Rev. B}\ }\textbf {\bibinfo
  {volume} {88}},\ \bibinfo {pages} {035316} (\bibinfo {year}
  {2013})}\BibitemShut {NoStop}%
\bibitem [{\citenamefont {Shen}\ \emph
  {et~al.}(2014{\natexlab{b}})\citenamefont {Shen}, \citenamefont {Raimondi},\
  and\ \citenamefont {Vignale}}]{Shen2014b}%
  \BibitemOpen
  \bibfield  {author} {\bibinfo {author} {\bibfnamefont {K.}~\bibnamefont
  {Shen}}, \bibinfo {author} {\bibfnamefont {R.}~\bibnamefont {Raimondi}},\
  and\ \bibinfo {author} {\bibfnamefont {G.}~\bibnamefont {Vignale}},\
  }\bibfield  {title} {\bibinfo {title} {Theory of coupled spin-charge
  transport due to spin-orbit interaction in inhomogeneous two-dimensional
  electron liquids},\ }\href {https://doi.org/10.1103/PhysRevB.90.245302}
  {\bibfield  {journal} {\bibinfo  {journal} {Phys. Rev. B}\ }\textbf {\bibinfo
  {volume} {90}},\ \bibinfo {pages} {245302} (\bibinfo {year}
  {2014}{\natexlab{b}})}\BibitemShut {NoStop}%
\bibitem [{\citenamefont {Szolnoki}\ \emph {et~al.}(2017)\citenamefont
  {Szolnoki}, \citenamefont {D\'ora}, \citenamefont {Kiss}, \citenamefont
  {Fabian},\ and\ \citenamefont {Simon}}]{Szolnoki2017}%
  \BibitemOpen
  \bibfield  {author} {\bibinfo {author} {\bibfnamefont {L.}~\bibnamefont
  {Szolnoki}}, \bibinfo {author} {\bibfnamefont {B.}~\bibnamefont {D\'ora}},
  \bibinfo {author} {\bibfnamefont {A.}~\bibnamefont {Kiss}}, \bibinfo {author}
  {\bibfnamefont {J.}~\bibnamefont {Fabian}},\ and\ \bibinfo {author}
  {\bibfnamefont {F.}~\bibnamefont {Simon}},\ }\bibfield  {title} {\bibinfo
  {title} {Intuitive approach to the unified theory of spin relaxation},\
  }\href {https://doi.org/10.1103/PhysRevB.96.245123} {\bibfield  {journal}
  {\bibinfo  {journal} {Phys. Rev. B}\ }\textbf {\bibinfo {volume} {96}},\
  \bibinfo {pages} {245123} (\bibinfo {year} {2017})}\BibitemShut {NoStop}%
\bibitem [{\citenamefont {Gorini}\ \emph {et~al.}(2017)\citenamefont {Gorini},
  \citenamefont {Maleki~Sheikhabadi}, \citenamefont {Shen}, \citenamefont
  {Tokatly}, \citenamefont {Vignale},\ and\ \citenamefont
  {Raimondi}}]{Gorini2017}%
  \BibitemOpen
  \bibfield  {author} {\bibinfo {author} {\bibfnamefont {C.}~\bibnamefont
  {Gorini}}, \bibinfo {author} {\bibfnamefont {A.}~\bibnamefont
  {Maleki~Sheikhabadi}}, \bibinfo {author} {\bibfnamefont {K.}~\bibnamefont
  {Shen}}, \bibinfo {author} {\bibfnamefont {I.~V.}\ \bibnamefont {Tokatly}},
  \bibinfo {author} {\bibfnamefont {G.}~\bibnamefont {Vignale}},\ and\ \bibinfo
  {author} {\bibfnamefont {R.}~\bibnamefont {Raimondi}},\ }\bibfield  {title}
  {\bibinfo {title} {Theory of current-induced spin polarization in an electron
  gas},\ }\href {https://doi.org/10.1103/PhysRevB.95.205424} {\bibfield
  {journal} {\bibinfo  {journal} {Phys. Rev. B}\ }\textbf {\bibinfo {volume}
  {95}},\ \bibinfo {pages} {205424} (\bibinfo {year} {2017})}\BibitemShut
  {NoStop}%
\bibitem [{\citenamefont {Sheikhabadi}\ \emph {et~al.}(2018)\citenamefont
  {Sheikhabadi}, \citenamefont {Miatka}, \citenamefont {Sherman},\ and\
  \citenamefont {Raimondi}}]{Sheikhabadi2018}%
  \BibitemOpen
  \bibfield  {author} {\bibinfo {author} {\bibfnamefont {A.~M.}\ \bibnamefont
  {Sheikhabadi}}, \bibinfo {author} {\bibfnamefont {I.}~\bibnamefont {Miatka}},
  \bibinfo {author} {\bibfnamefont {E.~Y.}\ \bibnamefont {Sherman}},\ and\
  \bibinfo {author} {\bibfnamefont {R.}~\bibnamefont {Raimondi}},\ }\bibfield
  {title} {\bibinfo {title} {Theory of the inverse spin galvanic effect in
  quantum wells},\ }\href {https://doi.org/10.1103/PhysRevB.97.235412}
  {\bibfield  {journal} {\bibinfo  {journal} {Phys. Rev. B}\ }\textbf {\bibinfo
  {volume} {97}},\ \bibinfo {pages} {235412} (\bibinfo {year}
  {2018})}\BibitemShut {NoStop}%
\bibitem [{\citenamefont {Tkach}(2021)}]{Tkach2021}%
  \BibitemOpen
  \bibfield  {author} {\bibinfo {author} {\bibfnamefont {Y.~Y.}\ \bibnamefont
  {Tkach}},\ }\bibfield  {title} {\bibinfo {title} {Specific features of the
  conductivity and spin susceptibility tensors of a two-dimensional electron
  gas with rashba and dresselhaus spin-orbit interactions},\ }\href
  {https://doi.org/10.1103/PhysRevB.104.085413} {\bibfield  {journal} {\bibinfo
   {journal} {Phys. Rev. B}\ }\textbf {\bibinfo {volume} {104}},\ \bibinfo
  {pages} {085413} (\bibinfo {year} {2021})}\BibitemShut {NoStop}%
\bibitem [{\citenamefont {Tkach}(2022)}]{Tkach2022}%
  \BibitemOpen
  \bibfield  {author} {\bibinfo {author} {\bibfnamefont {Y.~Y.}\ \bibnamefont
  {Tkach}},\ }\bibfield  {title} {\bibinfo {title} {Identification of a state
  of persistent spin helix in a parallel magnetic field, and exploration of its
  transport properties},\ }\href {https://doi.org/10.1103/PhysRevB.105.165409}
  {\bibfield  {journal} {\bibinfo  {journal} {Phys. Rev. B}\ }\textbf {\bibinfo
  {volume} {105}},\ \bibinfo {pages} {165409} (\bibinfo {year}
  {2022})}\BibitemShut {NoStop}%
\bibitem [{\citenamefont {Suzuki}\ and\ \citenamefont
  {Kato}(2023)}]{Suzuki2023}%
  \BibitemOpen
  \bibfield  {author} {\bibinfo {author} {\bibfnamefont {Y.}~\bibnamefont
  {Suzuki}}\ and\ \bibinfo {author} {\bibfnamefont {Y.}~\bibnamefont {Kato}},\
  }\bibfield  {title} {\bibinfo {title} {Spin relaxation, diffusion, and
  edelstein effect in chiral metal surface},\ }\href
  {https://doi.org/10.1103/PhysRevB.107.115305} {\bibfield  {journal} {\bibinfo
   {journal} {Phys. Rev. B}\ }\textbf {\bibinfo {volume} {107}},\ \bibinfo
  {pages} {115305} (\bibinfo {year} {2023})}\BibitemShut {NoStop}%
\bibitem [{\citenamefont {Yama}\ \emph
  {et~al.}(2023{\natexlab{a}})\citenamefont {Yama}, \citenamefont {Matsuo},\
  and\ \citenamefont {Kato}}]{Yama2023a}%
  \BibitemOpen
  \bibfield  {author} {\bibinfo {author} {\bibfnamefont {M.}~\bibnamefont
  {Yama}}, \bibinfo {author} {\bibfnamefont {M.}~\bibnamefont {Matsuo}},\ and\
  \bibinfo {author} {\bibfnamefont {T.}~\bibnamefont {Kato}},\ }\bibfield
  {title} {\bibinfo {title} {Effect of vertex corrections on the enhancement of
  gilbert damping in spin pumping into a two-dimensional electron gas},\ }\href
  {https://doi.org/10.1103/PhysRevB.107.174414} {\bibfield  {journal} {\bibinfo
   {journal} {Phys. Rev. B}\ }\textbf {\bibinfo {volume} {107}},\ \bibinfo
  {pages} {174414} (\bibinfo {year} {2023}{\natexlab{a}})}\BibitemShut
  {NoStop}%
\bibitem [{\citenamefont {T\"olle}\ \emph {et~al.}(2017)\citenamefont
  {T\"olle}, \citenamefont {Eckern},\ and\ \citenamefont {Gorini}}]{Tolle2017}%
  \BibitemOpen
  \bibfield  {author} {\bibinfo {author} {\bibfnamefont {S.}~\bibnamefont
  {T\"olle}}, \bibinfo {author} {\bibfnamefont {U.}~\bibnamefont {Eckern}},\
  and\ \bibinfo {author} {\bibfnamefont {C.}~\bibnamefont {Gorini}},\
  }\bibfield  {title} {\bibinfo {title} {Spin-charge coupled dynamics driven by
  a time-dependent magnetization},\ }\href
  {https://doi.org/10.1103/PhysRevB.95.115404} {\bibfield  {journal} {\bibinfo
  {journal} {Phys. Rev. B}\ }\textbf {\bibinfo {volume} {95}},\ \bibinfo
  {pages} {115404} (\bibinfo {year} {2017})}\BibitemShut {NoStop}%
\bibitem [{\citenamefont {Dey}\ \emph {et~al.}(2018)\citenamefont {Dey},
  \citenamefont {Prasad}, \citenamefont {Register},\ and\ \citenamefont
  {Banerjee}}]{Dey2018}%
  \BibitemOpen
  \bibfield  {author} {\bibinfo {author} {\bibfnamefont {R.}~\bibnamefont
  {Dey}}, \bibinfo {author} {\bibfnamefont {N.}~\bibnamefont {Prasad}},
  \bibinfo {author} {\bibfnamefont {L.~F.}\ \bibnamefont {Register}},\ and\
  \bibinfo {author} {\bibfnamefont {S.~K.}\ \bibnamefont {Banerjee}},\
  }\bibfield  {title} {\bibinfo {title} {Conversion of spin current into charge
  current in a topological insulator: Role of the interface},\ }\href
  {https://doi.org/10.1103/PhysRevB.97.174406} {\bibfield  {journal} {\bibinfo
  {journal} {Phys. Rev. B}\ }\textbf {\bibinfo {volume} {97}},\ \bibinfo
  {pages} {174406} (\bibinfo {year} {2018})}\BibitemShut {NoStop}%
\bibitem [{\citenamefont {Fleury}\ \emph {et~al.}(2023)\citenamefont {Fleury},
  \citenamefont {Barth},\ and\ \citenamefont {Gorini}}]{Fleury2023}%
  \BibitemOpen
  \bibfield  {author} {\bibinfo {author} {\bibfnamefont {G.}~\bibnamefont
  {Fleury}}, \bibinfo {author} {\bibfnamefont {M.}~\bibnamefont {Barth}},\ and\
  \bibinfo {author} {\bibfnamefont {C.}~\bibnamefont {Gorini}},\ }\bibfield
  {title} {\bibinfo {title} {Tunneling anisotropic spin galvanic effect},\
  }\href {https://doi.org/10.1103/PhysRevB.108.L081402} {\bibfield  {journal}
  {\bibinfo  {journal} {Phys. Rev. B}\ }\textbf {\bibinfo {volume} {108}},\
  \bibinfo {pages} {L081402} (\bibinfo {year} {2023})}\BibitemShut {NoStop}%
\bibitem [{\citenamefont {Yama}\ \emph
  {et~al.}(2023{\natexlab{b}})\citenamefont {Yama}, \citenamefont {Matsuo},\
  and\ \citenamefont {Kato}}]{Yama2023}%
  \BibitemOpen
  \bibfield  {author} {\bibinfo {author} {\bibfnamefont {M.}~\bibnamefont
  {Yama}}, \bibinfo {author} {\bibfnamefont {M.}~\bibnamefont {Matsuo}},\ and\
  \bibinfo {author} {\bibfnamefont {T.}~\bibnamefont {Kato}},\ }\bibfield
  {title} {\bibinfo {title} {Theory of inverse rashba-edelstein effect induced
  by spin pumping into a two-dimensional electron gas},\ }\href
  {https://doi.org/10.1103/PhysRevB.108.144430} {\bibfield  {journal} {\bibinfo
   {journal} {Phys. Rev. B}\ }\textbf {\bibinfo {volume} {108}},\ \bibinfo
  {pages} {144430} (\bibinfo {year} {2023}{\natexlab{b}})}\BibitemShut
  {NoStop}%
\bibitem [{Note1()}]{Note1}%
  \BibitemOpen
  \bibinfo {note} {We further assumed that an exchange bias due to the
  mean-field term, $\langle {\protect \bm {S}}_i \rangle \cdot {\protect \bm
  {s}}_j$, is sufficiently small.}\BibitemShut {Stop}%
\bibitem [{\citenamefont {Wilson}(1953)}]{Wilson1953}%
  \BibitemOpen
  \bibfield  {author} {\bibinfo {author} {\bibfnamefont {A.~H.}\ \bibnamefont
  {Wilson}},\ }\href@noop {} {\emph {\bibinfo {title} {The Theory of Metals}}}\
  (\bibinfo  {publisher} {Cambridge University Press, Cambridge, UK},\ \bibinfo
  {year} {1953})\BibitemShut {NoStop}%
\bibitem [{\citenamefont {Ziman}(1960)}]{Ziman1960}%
  \BibitemOpen
  \bibfield  {author} {\bibinfo {author} {\bibfnamefont {J.~M.}\ \bibnamefont
  {Ziman}},\ }\href@noop {} {\emph {\bibinfo {title} {Electrons and Phonons:
  The Theory of Transport Phenomena in Solids}}}\ (\bibinfo  {publisher}
  {Clarendon Press, Oxford},\ \bibinfo {year} {1960})\BibitemShut {NoStop}%
\bibitem [{\citenamefont {Lundstrom}(2000)}]{Lundstrom2000}%
  \BibitemOpen
  \bibfield  {author} {\bibinfo {author} {\bibfnamefont {M.}~\bibnamefont
  {Lundstrom}},\ }\href@noop {} {\emph {\bibinfo {title} {Fundamentals of
  Carrier Transport}}}\ (\bibinfo  {publisher} {Cambridge University Press,
  Cambridge},\ \bibinfo {year} {2000})\BibitemShut {NoStop}%
\bibitem [{Note2()}]{Note2}%
  \BibitemOpen
  \bibinfo {note} {For example, see Eq.~(2) in Ref.~\cite {Ganichev2004}. In
  previous work, two types of current were derived for $\Gamma \gg k_{\protect
  \rm F}\alpha , k_{\protect \rm F}\beta $ depending on the different spin
  relaxation mechanisms, that is, the Elliott-Yafet mechanism and the
  Dyakonov–Perel mechanism~\cite {Ivchenko2008}. However, we should note that
  this separation by the spin relaxation mechanisms is not possible in our
  calculation, which treats the opposite condition, $\Gamma \ll k_{\protect \rm
  F}\alpha , k_{\protect \rm F}\beta $~\cite
  {Szolnoki2017,Suzuki2023}.}\BibitemShut {Stop}%
\bibitem [{\citenamefont {Varotto}\ \emph {et~al.}(2022)\citenamefont
  {Varotto}, \citenamefont {Johansson}, \citenamefont {G{\"o}bel},
  \citenamefont {Vicente-Arche}, \citenamefont {Mallik}, \citenamefont
  {Br{\'e}hin}, \citenamefont {Salazar}, \citenamefont {Bertran}, \citenamefont
  {F{\`e}vre}, \citenamefont {Bergeal}, \citenamefont {Rault}, \citenamefont
  {Mertig},\ and\ \citenamefont {Bibes}}]{Varotto2022}%
  \BibitemOpen
  \bibfield  {author} {\bibinfo {author} {\bibfnamefont {S.}~\bibnamefont
  {Varotto}}, \bibinfo {author} {\bibfnamefont {A.}~\bibnamefont {Johansson}},
  \bibinfo {author} {\bibfnamefont {B.}~\bibnamefont {G{\"o}bel}}, \bibinfo
  {author} {\bibfnamefont {L.~M.}\ \bibnamefont {Vicente-Arche}}, \bibinfo
  {author} {\bibfnamefont {S.}~\bibnamefont {Mallik}}, \bibinfo {author}
  {\bibfnamefont {J.}~\bibnamefont {Br{\'e}hin}}, \bibinfo {author}
  {\bibfnamefont {R.}~\bibnamefont {Salazar}}, \bibinfo {author} {\bibfnamefont
  {F.}~\bibnamefont {Bertran}}, \bibinfo {author} {\bibfnamefont {P.~L.}\
  \bibnamefont {F{\`e}vre}}, \bibinfo {author} {\bibfnamefont {N.}~\bibnamefont
  {Bergeal}}, \bibinfo {author} {\bibfnamefont {J.}~\bibnamefont {Rault}},
  \bibinfo {author} {\bibfnamefont {I.}~\bibnamefont {Mertig}},\ and\ \bibinfo
  {author} {\bibfnamefont {M.}~\bibnamefont {Bibes}},\ }\bibfield  {title}
  {\bibinfo {title} {{Direct visualization of Rashba-split bands and
  spin/orbital-charge interconversion at KTaO${}_3$ interfaces}},\ }\href
  {https://doi.org/10.1038/s41467-022-33621-1} {\bibfield  {journal} {\bibinfo
  {journal} {Nat. Commun.}\ }\textbf {\bibinfo {volume} {13}},\ \bibinfo
  {pages} {6165} (\bibinfo {year} {2022})}\BibitemShut {NoStop}%
\bibitem [{\citenamefont {Passell}\ \emph {et~al.}(1976)\citenamefont
  {Passell}, \citenamefont {Dietrich},\ and\ \citenamefont
  {Als-Nielsen}}]{Passell76}%
  \BibitemOpen
  \bibfield  {author} {\bibinfo {author} {\bibfnamefont {L.}~\bibnamefont
  {Passell}}, \bibinfo {author} {\bibfnamefont {O.~W.}\ \bibnamefont
  {Dietrich}},\ and\ \bibinfo {author} {\bibfnamefont {J.}~\bibnamefont
  {Als-Nielsen}},\ }\bibfield  {title} {\bibinfo {title} {{Neutron scattering
  from the Heisenberg ferromagnets EuO and EuS. I. The exchange
  interactions}},\ }\href {https://doi.org/10.1103/PhysRevB.14.4897} {\bibfield
   {journal} {\bibinfo  {journal} {Phys. Rev. B}\ }\textbf {\bibinfo {volume}
  {14}},\ \bibinfo {pages} {4897} (\bibinfo {year} {1976})}\BibitemShut
  {NoStop}%
\bibitem [{\citenamefont {Hasegawa}(2012)}]{Hasegawa12}%
  \BibitemOpen
  \bibfield  {author} {\bibinfo {author} {\bibfnamefont {Y.}~\bibnamefont
  {Hasegawa}},\ }\bibfield  {title} {\bibinfo {title} {{Magnetic Semiconductor
  EuO, EuS, and EuSe Nanocrystals for Future Optical Devices}},\ }\href
  {https://doi.org/10.1246/cl.2013.2} {\bibfield  {journal} {\bibinfo
  {journal} {Chem. Lett.}\ }\textbf {\bibinfo {volume} {42}},\ \bibinfo {pages}
  {2} (\bibinfo {year} {2012})}\BibitemShut {NoStop}%
\bibitem [{\citenamefont {Olejn\'{\i}k}\ \emph {et~al.}(2012)\citenamefont
  {Olejn\'{\i}k}, \citenamefont {Wunderlich}, \citenamefont {Irvine},
  \citenamefont {Campion}, \citenamefont {Amin}, \citenamefont {Sinova},\ and\
  \citenamefont {Jungwirth}}]{Olejnik12}%
  \BibitemOpen
  \bibfield  {author} {\bibinfo {author} {\bibfnamefont {K.}~\bibnamefont
  {Olejn\'{\i}k}}, \bibinfo {author} {\bibfnamefont {J.}~\bibnamefont
  {Wunderlich}}, \bibinfo {author} {\bibfnamefont {A.~C.}\ \bibnamefont
  {Irvine}}, \bibinfo {author} {\bibfnamefont {R.~P.}\ \bibnamefont {Campion}},
  \bibinfo {author} {\bibfnamefont {V.~P.}\ \bibnamefont {Amin}}, \bibinfo
  {author} {\bibfnamefont {J.}~\bibnamefont {Sinova}},\ and\ \bibinfo {author}
  {\bibfnamefont {T.}~\bibnamefont {Jungwirth}},\ }\bibfield  {title} {\bibinfo
  {title} {{Detection of Electrically Modulated Inverse Spin Hall Effect in an
  $\mathrm{Fe}/\mathrm{GaAs}$ Microdevice}},\ }\href
  {https://doi.org/10.1103/PhysRevLett.109.076601} {\bibfield  {journal}
  {\bibinfo  {journal} {Phys. Rev. Lett.}\ }\textbf {\bibinfo {volume} {109}},\
  \bibinfo {pages} {076601} (\bibinfo {year} {2012})}\BibitemShut {NoStop}%
\bibitem [{\citenamefont {Bozorth}(1951)}]{Bozorth1951}%
  \BibitemOpen
  \bibfield  {author} {\bibinfo {author} {\bibfnamefont {R.~M.}\ \bibnamefont
  {Bozorth}},\ }\href@noop {} {\emph {\bibinfo {title} {{Ferromagnetism}}}}\
  (\bibinfo  {publisher} {Princeton, New Jersey, D. Van Nostrand Company
  Inc.},\ \bibinfo {year} {1951})\BibitemShut {NoStop}%
\bibitem [{\citenamefont {Lynn}(1975)}]{Lynn75}%
  \BibitemOpen
  \bibfield  {author} {\bibinfo {author} {\bibfnamefont {J.~W.}\ \bibnamefont
  {Lynn}},\ }\bibfield  {title} {\bibinfo {title} {{Temperature dependence of
  the magnetic excitations in iron}},\ }\href
  {https://doi.org/10.1103/PhysRevB.11.2624} {\bibfield  {journal} {\bibinfo
  {journal} {Phys. Rev. B}\ }\textbf {\bibinfo {volume} {11}},\ \bibinfo
  {pages} {2624} (\bibinfo {year} {1975})}\BibitemShut {NoStop}%
\end{thebibliography}%
	
\clearpage

\end{document}